\documentclass[11pt]{report}

\usepackage[margin=1in]{geometry}
\usepackage{setspace}
\usepackage[hyperfootnotes=false]{hyperref}
\hypersetup{ 		
    pdfauthor={Aidar Zhetessov},        
    colorlinks=true,			
    linkcolor=NavyBlue,
    urlcolor=NavyBlue,
    citecolor=NavyBlue		
}
\usepackage{afterpage}
\newcommand\blankpage{
	\null
	\thispagestyle{empty}
	\addtocounter{page}{-1}
	\newpage
}

\usepackage[pdftex]{graphicx}
\usepackage{float}
\graphicspath{{./fig/}}
\usepackage[usenames,dvipsnames]{color}
\usepackage{epstopdf}
\usepackage{graphicx}
\usepackage{wrapfig}
\usepackage{lscape}
\usepackage{rotating}
\usepackage[final]{pdfpages}
\usepackage{subcaption}
\usepackage[export]{adjustbox}
\usepackage{array}
\newcolumntype{P}[1]{>{\centering\arraybackslash}p{#1}}

\usepackage{fancyhdr}
\usepackage{gensymb}
\usepackage{amsmath,amsthm,amssymb,mathrsfs,dsfont,amsmath}

\begin{document}

\lhead[<even output>]{}
\chead[<even output>]{}
\rhead[<even output>]{\thepage}
\cfoot{} 

\title{Design and Realization of a Novel Buck-Boost	Phase-Modular Three-Phase AC/DC Converter System with Low Component Number}
\date{}
\begin{titlepage}

\newcommand{\HRule}{\rule{\linewidth}{0.5mm}} 



\begin{figure}
	\begin{subfigure}{0.5\textwidth}
		\includegraphics[width=7cm,left]{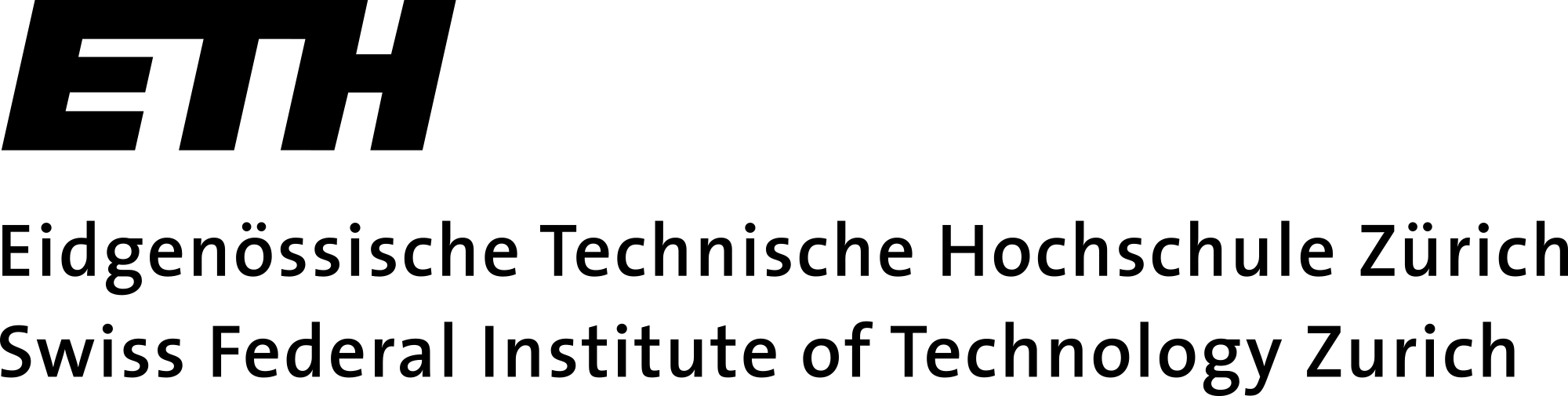}
	\end{subfigure}
	\begin{subfigure}{0.5\textwidth}
		\includegraphics[width=7cm,right]{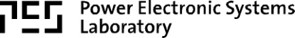}
	\end{subfigure}
\end{figure}
 

\center 

\makeatletter
\HRule \\[0.5cm]
{ \huge \bfseries \@title}\\[1cm] 

\includegraphics[width=0.95\textwidth,center]{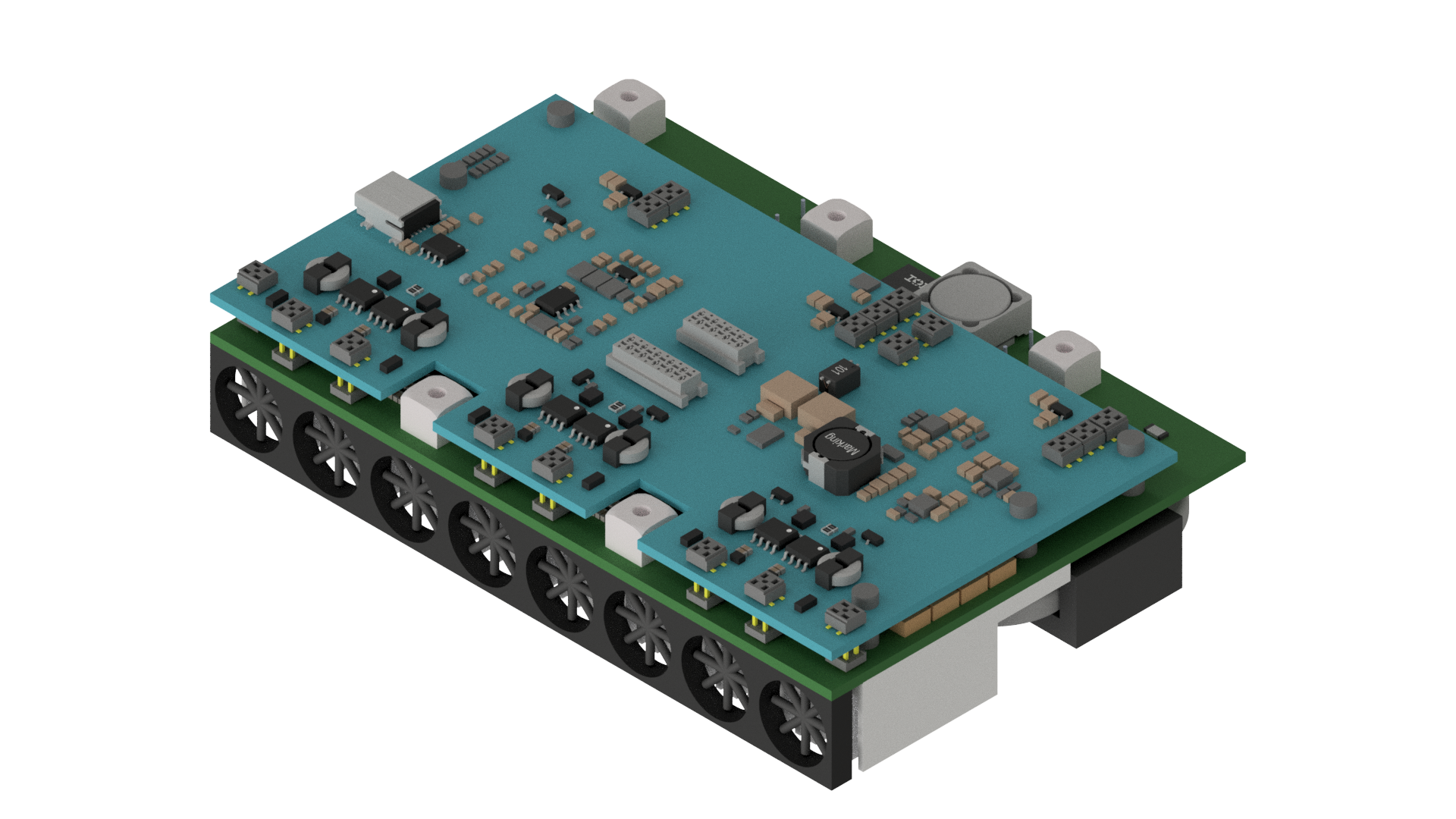}
\\[1cm]
\textsc{\LARGE Master Thesis}\\[0.2cm] 
\textsc{\Large Spring Semester 2019}\\[0.2cm] 
\HRule \\[0.5cm]


\begin{minipage}{0.4\textwidth}
\begin{flushleft} \large
\emph{Author:}\\
Aidar Zhetessov \\
ID: 15-948-268
\end{flushleft}
\end{minipage}
~
\begin{minipage}{0.4\textwidth}
\begin{flushright} \large
\emph{Supervisor:} \\
David Menzi \\[1.2em] 
\emph{Professor:} \\
Prof. Dr. J. W. Kolar 
\end{flushright}
\end{minipage}\\[1cm]
\makeatother



{\large July 30, 2019}\\[2cm] 

\vfill 

\afterpage{\blankpage}

\end{titlepage}
\afterpage{\blankpage}
\begin{abstract}
Scalability and modularity are key features for future power converters, such that these systems can be easily employed in many applications with different electrical specifications. In this thesis, the potential of a new bidirectional phase-modular three-phase AC/DC converter with buck-boost capability is evaluated by means of studying two potential application cases and developing a hardware prototype for one of them.

The DC-DC inverting buck-boost converter is a well known and established topology. By connecting three such systems in parallel, a phase-modular bidirectional buck-boost DC-AC converter employing a minimum number of active components results, where for given AC voltage amplitudes, an arbitrary DC voltage can be generated and vice versa. Such a three-phase converter was not yet described in literature and this project aims at investigating the fundamental topology properties, as well as its performance limits. A hardware demonstrator is designed for one potential application in order to verify the basic operation and the expected high performance in terms of efficiency and power density.
\end{abstract}

\includepdf[pages=-]{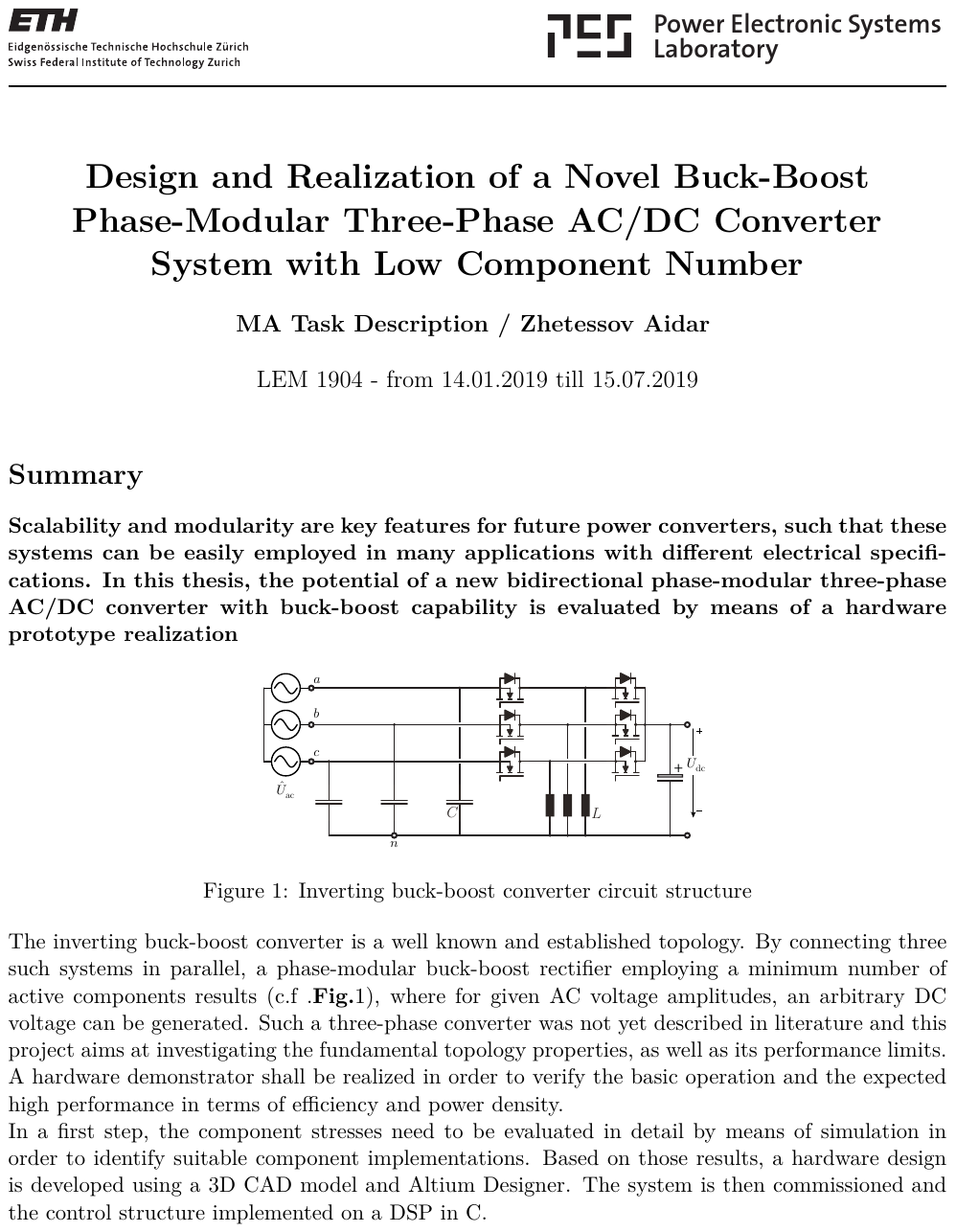}
\afterpage{\blankpage}
\tableofcontents
\chapter{Introduction}

\section{Background and Motivation}

With the recent proliferation of power electronics in e-mobility and distributed renewable energy generation the demand for more efficient, plug-and-play, modular and compact bidirectional converters with voltage buck-boost capability is intensified. Example emerging applications are the Electric Vehicles (EV), their charging stations, grid-feeding converters with variable input voltage in microgrids, and some specific applications in a More Electric Aircraft. In case of Fuel Cell Vehicles (FCV), for instance, one wants to have full control over the machine torque irrespective of the fuel cell output voltage as long as the output power can be provided. The same holds true not only for the main motor, but also for an auxiliary one, which provides the compressed air needed for the Fuel Cell chemical reaction \cite{Antivachis}. For the EV charging stations, the established standards require very wide range of both DC and AC voltage to be covered. The idea is to ensure compatibility between various EV manufacturers that employ different technologies. For microgrids, having a PV panel connected to the DC-link capacitor in a standard inverter setup, one obviously cannot ensure constant DC voltage at a certain level due to illumination variations. And even if an Energy Storage System (ESS) is there to reduce the DC voltage variations, a control system could require the converter to step up or down the AC bus voltage with respect to the input DC voltage in case of some microgrid transients. These again demand voltage buck-boost capability of the inverter.

Apart from buck-boost capability, other important performance metrics for the aforementioned applications can be listed as follows:\\

$\bullet$ \ Power Factor Correction (PFC) - the capability of the converter to absorb/feed an in-phase and clean AC currents from/to the mains/load.\\

$\bullet$ \ Operational Safety - the capability of delivering the required performance in presence of grid imbalance and faults. Safe two- and single-phase operation should also be considered here \cite{Gong}-\cite{DirectPFC}.\\

$\bullet$ \ Reliability - the capability of handing a high peak power, direct start up without pre-charging \cite{Gong}-\cite{DirectPFC}, as well as low complexity and thermal stress to ensure a long Mean Time To Failure (MTTF).\\

$\bullet$ \ Realization Effort - this could include manufacturing cost (phase-modularity), complexity (number of active components), control effort (number of sensors, gate drivers, computational burden), overvoltage-limiting effort as well as wiring and mounting effort (isolated mounting of active devices) \cite{Gong}-\cite{DirectPFC}.\\

$\bullet$ \ Power Density - an important parameter for a space-constrained application, such as in e-mobility.\\

Regarding the Realization Effort, \cite{Parallellism} highlights the importance of phase modularity. According to \cite{Parallellism}, the industry-preferred approach favors the paralleling of single-phase converter units to form a three-phase high-power converter systems. Such a modular development approach has several advantages: \\

$\bullet$ \ Well-proven and reliable single-phase converter technology can be used immediately. \\

$\bullet$ \ No major change of the existing production line is required. \\

$\bullet$ \ The power expandability offers great flexibility in the development of power converter products for different power levels. \\

$\bullet$ \ There are fewer requirements for the maintenance and repair of power converter modules because standard single-phase converter units are used. \\

$\bullet$ \ Standard single-phase converter units do not require high-voltage devices that are normally needed in specially designed three-phase converters. \\

Bearing this in mind, phase modularity (both in terms of topology and associated control) is beneficial concerning manufacturing and maintenance aspects.

Conventional solutions for such buck-boost application would employ a cascaded converter system, e.g. a boost DC-DC converter with a buck-type Voltage Source Inverter (VSI). Essentially this cascade simply boosts the input DC voltage $V_{dc}$ through the boost DC-DC converter to increase the achievable output voltage range for a VSI. Considering the performance metrics stated above, there are many disadvantages associated with the cascaded converter structures in terms of realization effort (phase modularity) and power density. Indeed, a cascaded structure implies more components, cost, volume, losses, design effort etc. Moreover, in many particular cases cascaded structures imply more time-scale separation steps in the control system, which imposes other control-related complications and stability concerns. Therefore, for the mentioned applications it is desired to have a single-stage (bidirectional) buck-boost AC-DC converter, playing the role of a generalized AC-DC interface.

Having observed the aforementioned deficiencies of a conventional solution, several single-stage buck-boost three-phase converter concepts can be found in the literature. A thorough analysis and comparison of three converter concepts was presented in \cite{Gong}. Also there is a buck-boost three-phase AC-DC converter with inherent PFC capability and simple control presented in \cite{PanChen}, its single-switch counterpart from \cite{Correspondence} as well as phase-modular and direct three-phase single-stage flyback-type power factor correctors from \cite{DirectPFC}.

Apart from these literature solutions, more options of such three-phase topologies could be generated by paralleling three single-phase DC-DC converters with buck-boost capability, as it was stated in \cite{NgoCuk83}. In buck-boost DC-DC converters operated in DC-AC mode, usually the DC link voltage midpoint is considered the AC reference point (open star point of the machine for instance). By paralleling DC-DC converters, the AC voltages are referenced to negative DC rail, thus one introduces a CM voltage shift. This CM voltage degree of freedom can be used to realize a three-phase DC-AC converter using three DC-DC converter (phase) modules \cite{NgoCuk83}.

\begin{figure}[H]
	\centering
	\includegraphics[width=0.66\textwidth]{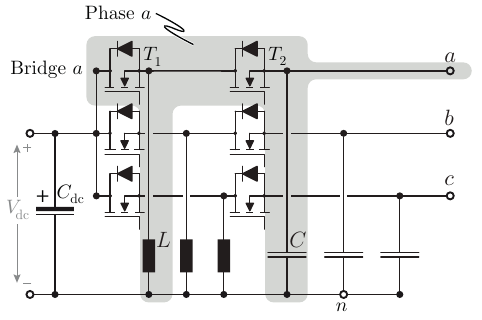}
	\caption{Inverting Buck-Boost Converter Topology}
	\label{fig:Inverswandler}
\end{figure}

Partially inspired by \cite{Antivachis}, \cite{DirectPFC} and \cite{NgoCuk83}, a new topology, the three-phase inverting buck-boost converter, is presented and analyzed in this thesis (Fig. \ref{fig:Inverswandler}). Essentially, this solution results from paralleling three inverting buck-boost DC-DC converters. This novel topology is an interesting alternative to the mentioned solutions, as it is proven to have all the required performance metrics above: it has a low number of semiconductors and passives, provides voltage buck-boost capability, single-stage high-frequency energy conversion and is phase-modular both in terms of topology and control. Apart from these advantages, it has a mentioned Common Mode (CM) AC voltage degree of freedom, which can be used to adapt the modulation and, consequently, improve converter losses and semiconductor utilization. Together, these factors are in favor of a power dense design, thus making the investigation of this topology interesting. Therefore, the objective of this thesis is to explore the applicability of a three-phase inverting buck-boost converter topology in the mentioned emerging applications. This is done through the design of a prototype for a selected application. Particularly considered applications comprise (but are not limited to) a rectifier for an Electro Hydrostatic Actuator (EHA) in a More Electric Aircraft and an auxiliary motor's inverter in a Fuel Cell Vehicle (FCV).

\section{Outline}
This thesis has five chapters, the first one being an introduction. The \textbf{chapter 2} provides a general information of interest regarding the topology, such as topology derivation, operating principle, review of hard- and soft-switched operation and applicable modulation schemes. \textbf{Chapter 3} explores the applicability potential of the proposed topology in a rectifier application within the context of a More Electric Aircraft. This is done by designing appropriate modulation and associated control for the topology in rectifier mode. \textbf{Chapter 4} does the same for inverter application within the context of a Fuel Cell Vehicle. In addition, the prototype design process is thoroughly explained and a virtual prototype is presented. The thesis is concluded in \textbf{chapter 5}. As a side project, a calorimetric switching loss measurement setup for a GaN GIT transistor is addressed in \textbf{Appendix A}.

\afterpage{\blankpage}
\chapter{Converter Fundamentals and Modulation Scheme Options} \label{TheoryRectifier}

This chapter is dedicated to the theoretical background of the three-phase inverting buck-boost converter. The information of this chapter lays the background on the converter topology and operation. All subsequent chapters of the thesis will build up on this background. The chapter consists of four sections. In the first section we discuss the derivation of the three-phase inverting buck-boost converter topology. The second section reveals the basic operating principle of the topology, starting from a DC-DC inverting buck-boost phase module and proceeding to the three-phase DC-AC operation. The third section is a brief review of a standard half-bridge hard- and soft-switching operation. This information is useful because, as it was already shown in Fig. \ref{fig:Inverswandler}, the half-bridge constitutes a basic building block of our converter. Moreover, understanding hard- and soft-switched operation will be useful in subsequent evaluation and comparison of modulation schemes. The fourth section proposes several modulation scheme candidates and discusses respective operating principles.

\section{Derivation of the Three-Phase Inverting Buck-Boost Converter Topology}

\begin{figure}[h]
	\centering
	\includegraphics[width=\textwidth]{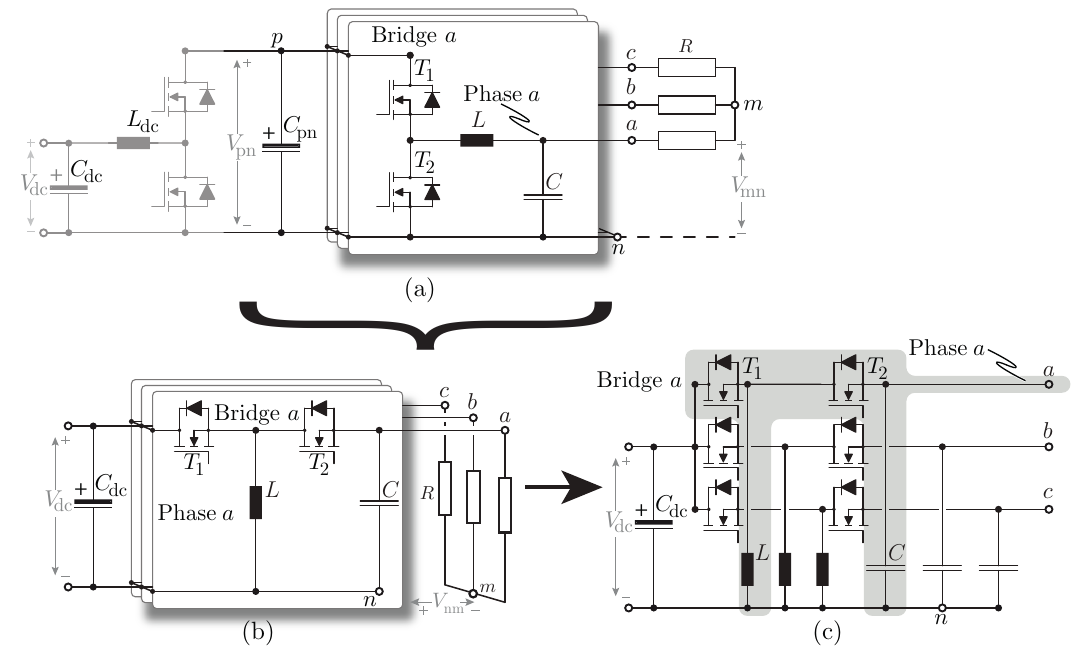}
	\caption{Inverting buck-boost converter justification. Conventional solution: DC-DC boost + VSI (a); Replacing DC-DC buck stages with DC-DC buck-boost stages (b); and three-phase inverting buck-boost converter topology (c)}
	\label{fig:InverswandlerOrigin}
\end{figure}

Before studying the operation of the converter in more detail, it is worth justifying the use of a three-phase inverting buck-boost converter topology first. To do that, one starts with the aforementioned conventional cascaded solution - DC-DC boost converter and Voltage-Source Inverter (VSI, Fig. \ref{fig:InverswandlerOrigin}a). The VSI can also be seen as three paralleled DC-DC buck converters that can generate output voltages between [0..$V_{pn}$], where $V_{pn}$ is the intermediate DC link voltage. Provided unipolarity of the output voltages ([0..$V_{pn}$] $\geq$ 0) three DC-DC buck converters can drive a three-phase AC load only with a positive Common Mode (CM) voltage offset between the load star point and the DC- rail. Moreover, the maximum AC output phase voltage amplitude is limited to $\hat{V}_{AC,max} = V_{pn}/2$ due to buck converter upper limit of $V_{pn}$. In these conditions to provide the overall buck-boost capability one has to boost up the DC input voltage $V_{dc}$ by employing the preceding DC-DC boost stage (Fig. \ref{fig:InverswandlerOrigin}a).

On the other hand, instead of paralleling three DC-DC buck stages one could parallel three DC-DC inverting buck-boost stages to result in a three-phase inverting buck-boost converter instead of a VSI (Fig. \ref{fig:InverswandlerOrigin}b). This proposal is advantageous because the three-phase inverting buck-boost converter provides the voltage buck-boost capability without the preceding DC-DC boost stage, while keeping the number of active/passive components the same as in the VSI stage. It is worth mentioning that the output voltage is unipolar and inverted (as the name suggests), meaning that the converter features a negative CM voltage offset between the load star point and the DC- rail. This is reflected in polarity change of $V_{mn}$ in Fig. \ref{fig:InverswandlerOrigin}b compared to Fig. \ref{fig:InverswandlerOrigin}a.

\section{Inverting Buck-Boost Converter Operating Principle} \label{InverswandlerOperatingPrinciple}

This section deals with the basic operating principle of the three-phase inverting buck-boost converter, starting from a DC-DC inverting buck-boost stage and proceeding to the three-phase DC-AC operation. As it was mentioned, this procedure is inspired by \cite{Antivachis}, \cite{DirectPFC} and \cite{NgoCuk83}.

\subsection{DC-DC operation of one inverting buck-boost stage}

\begin{figure}[h]
	\centering
	\begin{subfigure}[H]{0.49\textwidth}
		\includegraphics[width=\textwidth]{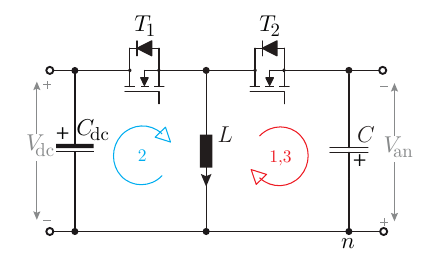}
		\caption{}
		\label{fig:FlybackConverterTopology}
	\end{subfigure}
	~
	\begin{subfigure}[H]{0.48\textwidth}
		\includegraphics[width=\textwidth]{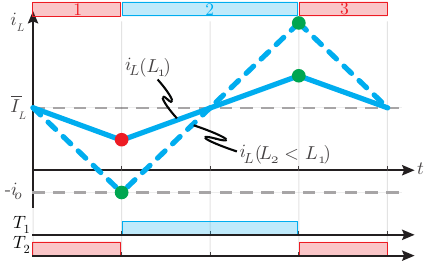}
		\caption{}
		\label{fig:FlybackConverterWaveform}
	\end{subfigure}
	\caption{DC-DC inverting buck-boost stage. Topology (a); Steady state current waveforms in DC-DC operation $i_L(L_1)$ and $i_L(L_2)$ (b)}
	\label{fig:FlybackConverter}
\end{figure}

The well-known DC-DC inverting buck-boost topology \ref{fig:FlybackConverterTopology} has a full control over its output DC voltage. To understand its operation, consider	 steady state with $V_{dc} = |V_{an}|$. In that case, possible steady state buck-boost inductor current waveforms for two values of inductance are shown in Fig. \ref{fig:FlybackConverterWaveform}. There are three distinct time intervals within the depicted switching period:

$\bullet$ \ First, the period starts with the transistor $T_2$ turned on and the inductor current equal to its average value $\bar{I}_L$ as shown in Fig. \ref{fig:FlybackConverterWaveform}. The inductor is demagnetized by the output capacitor $C$ during the first interval (number $1$ red demagnetization loop in Fig.  \ref{fig:FlybackConverterTopology}).

$\bullet$ \ Second, after some time (equal to the quarter of the switching period in this case) transistor $T_2$ is turned off and the transistor $T_1$ is turned on, signifying the start of the second interval. The buck-boost inductor is now magnetized by the input DC capacitor $C_{dc}$. Second interval corresponds to number $2$ blue magnetization loop in Fig.  \ref{fig:FlybackConverterTopology}.

$\bullet$ \ Third, after the second interval (which is equal to half of the switching period in this case) transistor $T_1$ is turned off and the transistor $T_2$ is turned on back again, signifying the start of the third interval. The buck-boost inductor is now demagnetized back by the output capacitor $C$. This interval corresponds to number $3$ red demagnetization loop in Fig.  \ref{fig:FlybackConverterTopology}.

During the first and the third (red) time intervals, the inductor demagnetization is described by the following equation: $-|V_{an}| = \frac{L di_L}{dt}$. Similarly, for the second (blue) time interval the inductor magnetization is described by: $V_{dc} = \frac{L di_L}{dt}$. Defining the duty cycle $d$ to be a fraction of the switching period $T_s$ corresponding to the turn-on state of $T_1$ (time interval $2$), and writing the steady state inductor volt-second balance we arrive at the following equations:

\begin{equation}
\label{eq:VSBalance_Ipkpk}
\begin{split}
&V_{dc} d T_s - |V_{an}| (1-d) T_s = 0
\\
&V_{dc} d T_s = L \Delta I_{pkpk}
\\
\end{split}
\end{equation}

From (\ref{eq:VSBalance_Ipkpk}) the converter duty cycle and the peak-to-peak inductor current ripple $I_{pkpk}$ can be derived as a function of input and output voltages, inductance value and the switching frequency $f_s$ as follows:

\begin{equation}
\label{eq:Duty_Ipkpk}
\begin{split}
d &= \frac{|V_{an}|}{V_{dc}+|V_{an}|}
\\
I_{pkpk} &= \frac{|V_{an}| V_{dc}}{L f_s (|V_{an}| + V_{dc})}
\\
\end{split}
\end{equation}

From the equation (\ref{eq:Duty_Ipkpk}) it is seen that in case $V_{dc}=|V_{an}|$ the duty cycle $d$ equals $0.5$, which coincides with the interval durations in Fig. \ref{fig:FlybackConverterWaveform}. Equation (\ref{eq:Duty_Ipkpk}) with its \textit{non-linear} input and output voltage dependent duty cycle and current ripple, combined with Fig. \ref{fig:FlybackConverter}, provide a comprehensive overview of the DC-DC inverting buck-boost converter steady state operation.

\subsection{DC-AC operation of one inverting buck-boost stage}

\begin{figure}[h]
	\centering
	\includegraphics[width=\textwidth]{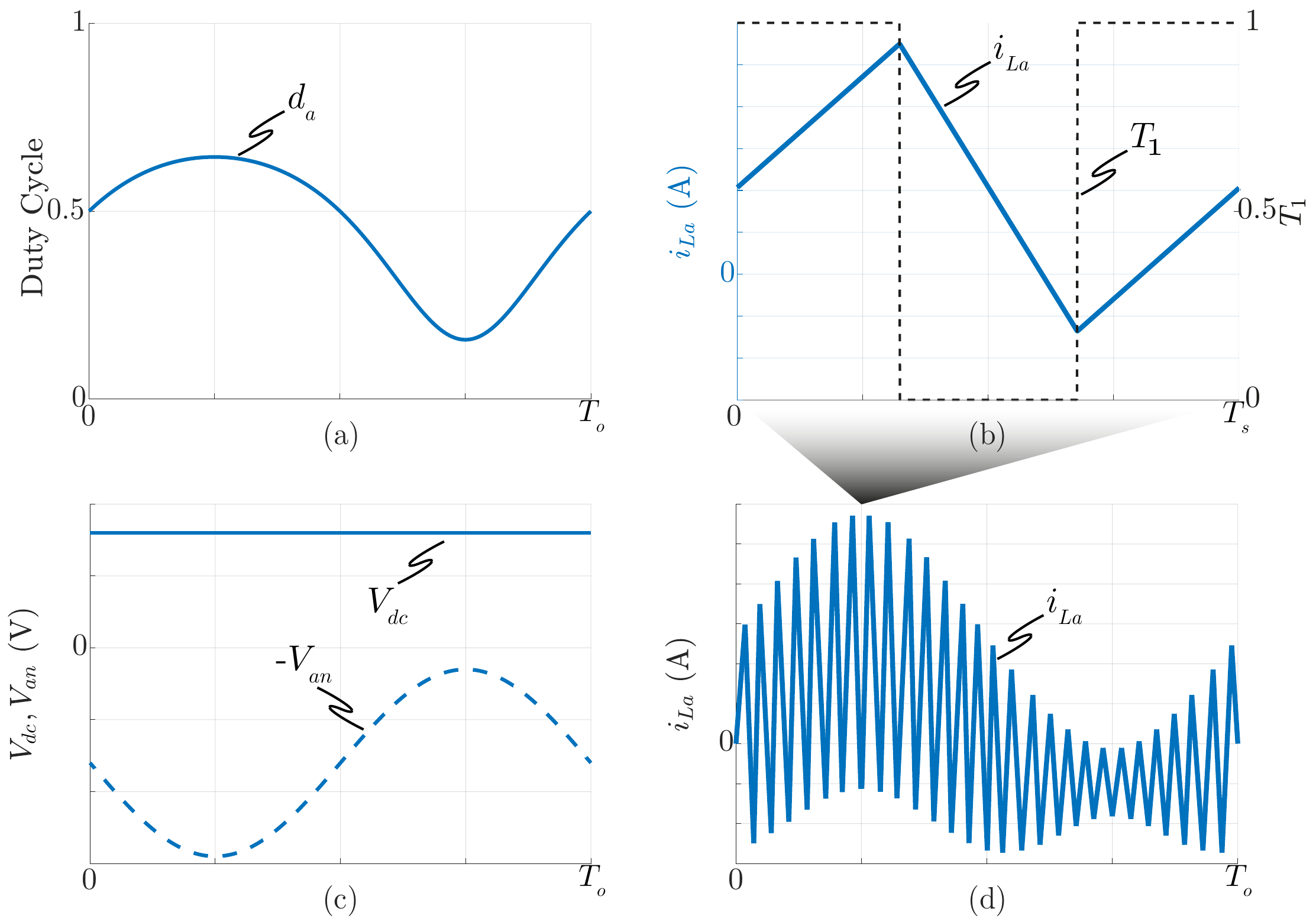}
	\caption{Theoretical inverting buck-boost stage DC-AC waveforms. Duty cycle over fundamental period (a); Inductor current over one switching period (b); Converter voltages over fundamental period (note $V_{an} \leq 0$)(c); Inductor current over fundamental period (d)}
	\label{fig:FlybackDCACWaveforms}
\end{figure}

To operate the inverting buck-boost stage from Fig. \ref{fig:FlybackConverterTopology} in a DC-AC mode, one could zoom out in time scale and slowly vary the steady state duty cycle from (\ref{eq:Duty_Ipkpk}) in a non-linear manner, similar to \cite{Antivachis}. For that though, the fundamental frequency $f_o$ of the converter AC operation should be much slower than the switching frequency ($f_o<<f_s$) to ensure a quasi-DC-DC operation within any given switching period. Potential steady state waveforms of an inverting buck-boost stage DC-AC operation are depicted in Fig. \ref{fig:FlybackDCACWaveforms}.

In Fig. \ref{fig:FlybackDCACWaveforms} we can see that provided the duty cycle variation within one switching period is negligible, the converter always operates in quasi-DC-DC mode as it can be seen in the magnification from Fig. \ref{fig:FlybackDCACWaveforms}d to Fig. \ref{fig:FlybackDCACWaveforms}b. Moreover, Fig. \ref{fig:FlybackDCACWaveforms}c shows a nice sinusoidal output voltage $-V_{an}$ with a negative DC offset. In this case the offset is equal to $-V_{dc}$, which can also be seen from the Fig. \ref{fig:FlybackDCACWaveforms}a, because the duty cycle equals $0.5$ once in every half-fundamental period interval. Finally, Fig. \ref{fig:FlybackDCACWaveforms}a illustrates the \textit{non-linear} relation between duty cycle and the output voltage, since the duty cycle is not sinusoidal for an offsetted sinusoidal output voltage.

\subsection{DC-AC operation of three paralleled inverting buck-boost stages}
\begin{figure}[h]
	\centering
	\begin{subfigure}[H]{0.55\textwidth}
		\includegraphics[width=\textwidth]{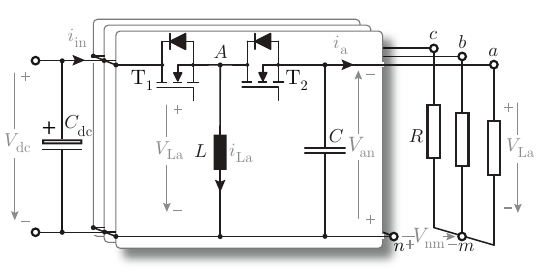}
		\caption{}
		\label{fig:InverswandlerTopology}
	\end{subfigure}
	~
	\begin{subfigure}[H]{0.42\textwidth}
		\includegraphics[width=\textwidth]{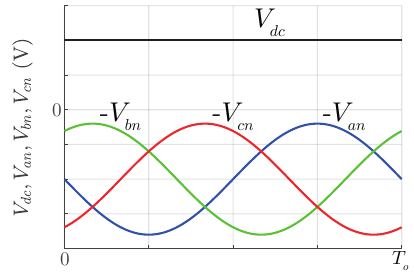}
		\caption{}
		\label{fig:Inverswandler_DCAC_V}
	\end{subfigure}
	\caption{A three-phase inverting buck-boost converter realized as a paralleling of three DC-DC inverting buck-boost stages. Topology (a); Steady state voltage waveforms (b)}
	\label{fig:InverswandlerDCAC}
\end{figure}

Following the idea from \cite{NgoCuk83}, also shown in Fig. \ref{fig:InverswandlerOrigin}c, a three-phase inverting buck-boost converter can be assembled from three DC-AC operated inverting buck-boost stages. In this case, unlike the single-phase case, only three AC terminals $a,b,c$ (without negative DC terminal $n$) are provided from AC side as it is shown in Fig. \ref{fig:InverswandlerDCAC}.

Considering the example case depicted in Fig. \ref{fig:InverswandlerDCAC}, three-phase converter can be operated as an inverter feeding a resistive load. Fig. \ref{fig:Inverswandler_DCAC_V} shows the input DC voltage and the output capacitor voltages when each paralleled inverting buck-boost stage is operated with a standard PWM and a $120$ degree phase shifted duty cycle from Fig. \ref{fig:FlybackDCACWaveforms}a for each phase. It can be seen that $-V_{an,bn,cn}$ are always less than zero due to the nature of the topology. In a three-phase context this property can be referred to as an already mentioned \textit{Common Mode} offset $V_{nm}$ (Fig. \ref{fig:InverswandlerTopology}).



\section{Review of a Hard- and Soft-Switching Operation}

\begin{figure}[h]
	\centering
	\begin{subfigure}[H]{0.49\textwidth}
		\includegraphics[width=\textwidth]{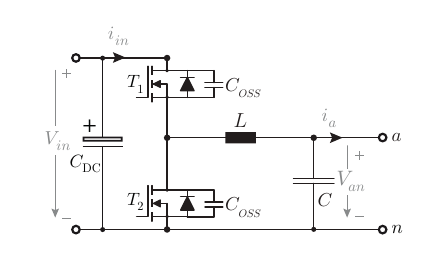}
		\caption{}
		\label{fig:BuckConverterTopology}
	\end{subfigure}
	~
	\begin{subfigure}[H]{0.48\textwidth}
		\includegraphics[width=\textwidth]{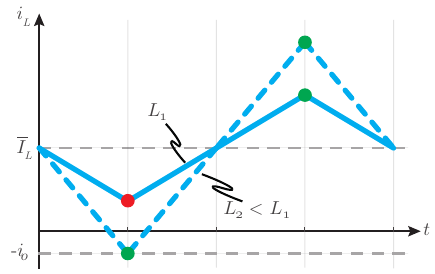}
		\caption{}
		\label{fig:BuckConverterWaveform}
	\end{subfigure}
	\caption{A buck converter used for analysis of soft-switching. Topology (a); Waveforms for two inductance values(b)}
	\label{fig:BuckConverter}
\end{figure}

Before diving into the various modulation options, a brief review of hard- and soft-switched operation would be useful, as the modulations will be evaluated with respect to the impact on the switching losses. To do that, consider an example of a DC-DC buck converter (Fig. \ref{fig:BuckConverterTopology}) in steady state operation with $V_{in} > V_{an}$. 

Possible steady state inductor current waveforms are provided in Fig. \ref{fig:BuckConverterWaveform} for two inductance values. The duty cycle and the inductor current ripple for a buck converter are given by:

\begin{equation}
\label{eq:Buck_Duty_DeltaI}
\begin{split}
d &= V_{an}/V_{in}
\\
\Delta I_L &= (1-d) T_s V_{an} / L = (V_{in} - V_{an}) T_s V_{an} / V_{in} L
\\
\end{split}
\end{equation}

From the equation for the inductor current ripple $\Delta I_L$ it is seen that the ripple is proportional to the switching period $T_s$ and inversely proportional to the inductance value $L$. 

Consider the waveform corresponding to the inductance value $L_1$. From Fig. \ref{fig:BuckConverterWaveform} it is seen that the half of the inductor current ripple $\Delta I_L$ is less than the average value of the inductor current $\bar{I}_L$ such that both switching transitions occur at positive instantaneous inductor current values. If one now analyzes the first switching transition, depicted by a red dot, in more detail, it corresponds to the turn-off of the transistor T2 and subsequent turn-on of the transistor T1 after a dead time interval. Bearing in mind the positive instantaneous inductor current (directed from the half-bridge to the output capacitor), after turn-off of the transistor T2 the inductor current continues flowing through it using the respective body diode. As soon as the dead time interval is finished, the transistor T1 is turned on and the inductor current commutates from the body diode of the transistor T2 to the channel of the transistor T1. The energy stored in the output capacitance of the transistor T1, which is equal to $C_{OSS} V_{in}^2/2$, is dissipated in the transistor T1 at the instant of turn-on, thus forming a hard switched transition of the half-bridge (a red dot in Fig. \ref{fig:BuckConverterWaveform}).

Regarding the second transition of the switching period under consideration, it is seen that the half of the inductor current ripple $\Delta I_L$ and the average value of the inductor current $\bar{I}_L$ add up to form a highly positive instantaneous inductor current. If one now analyzes the second switching transition, depicted by a green dot, in more detail, it corresponds to the turn-off of the transistor T1 and subsequent turn-on of the transistor T2 after a dead time interval. Bearing in mind the positive instantaneous inductor current (directed from the half-bridge to the output capacitor), after the turn-off of the transistor T1 the inductor current starts charging up the output capacitance $C_{OSS}$ of the transistor T1 and discharging the output capacitance $C_{OSS}$ of the transistor T2. Provided the dead time interval is long enough and the inductor current is large enough, the charge-discharge procedure of the output capacitances is finished before the end of the dead time interval. After the end of the charge-discharge procedure, the inductor current commutated to the body diode of the transistor T2. Corresponding voltages of the output capacitances of the transistors T1 and T2 at the end of the dead time interval are $V_{in}$ and $0$ respectively, so that there is no energy stored in the output capacitance of the transistor T2 at the instant of turning on after the dead time. This forms a soft-switched transition of the half-bridge (a green dot in Fig. \ref{fig:BuckConverterWaveform}).

The derivations for the dead time interval and the minimum inductor current requirements can be obtained from literature. For example, in \cite{Waffler09} $i_o$ is defined as

\begin{equation}
\label{eq: io}
i_o \geq \text{max}\big(\text{max}(V_{in}),\text{max}(V_{an})\big) \sqrt{\frac{C_{OSS}}{L}}
\end{equation}

Now consider the second inductor current waveform in Fig. \ref{fig:BuckConverterWaveform}. It corresponds to an inductance value $L_2 < L_1$ so that $\Delta I_L$ is increased (dashed line). From Fig. \ref{fig:BuckConverterWaveform} it is seen that the half of the inductor current ripple $\Delta I_L$ is more than the average value of the inductor current $\bar{I}_L$ so that two switching transitions occur at negative and positive instantaneous inductor current values, respectively. If one now analyzes the first switching transition, depicted by a green dot at $-i_o$ current value, in more detail, it again corresponds to the turn-off of the transistor T2 and subsequent turn-on of the transistor T1 after the dead time interval. Bearing in mind the negative instantaneous inductor current (directed from the output capacitor to the half-bridge), after the turn-off of the transistor T2 the inductor current starts charging the output capacitance of the transistor T2 and discharging that of T1. Provided the dead time interval is long enough and the inductor current is large enough (which is true for the waveform under consideration), the charge-discharge procedure of the output capacitances is finished before the end of the dead time interval. After the end of the charge-discharge procedure, the inductor current commutated to the body diode of the transistor T1. The corresponding voltages of the output capacitances of the transistors T1 and T2 at the end of the dead time interval are $0$ and $V_{in}$, respectively, so that there is no energy stored in the output capacitance of the transistor T1 at the instant of turning on after the dead time. This forms a soft switched transition of the half-bridge (a green dot in Fig. \ref{fig:BuckConverterWaveform}).

Regarding the second transition of the switching period under consideration, it is seen that the half of the inductor current ripple $\Delta I_L$ and the average value of the inductor current $\bar{I}_L$ add up to form a highly positive instantaneous inductor current, which is even more than that of the first waveform, corresponding to $L_1$. Therefore the analysis that we've done for the second transition of the first waveform ($L_1$), also applies for the second transition of the second (dashed waveform, $L_2$), yielding a soft-switched transition.

In this section we've reviewed a basic model for hard- and soft-switching operation of one half-bridge. We've seen that it is possible to operate a half-bridge in a soft-switched manner, thus resulting in a significantly reduced switching losses. The main idea is to have an inductor-impressed current of a sufficient magnitude flowing into the half-bridge (in case of transition from low to high) or out of the half-bridge (in case of transition from high to low) within the dead time interval of sufficient length. The same considerations can also be applied to the half-bridge of the DC-DC inverting buck-boost converter. The only difference is in the blocking voltage of the half-bridge. While in DC-DC buck converter the blocking voltage is equal to $V_{in}$, in DC-DC inverting buck-boost converter (a single phase in Fig. \ref{fig:InverswandlerOrigin}c) the blocking voltage is equal to $V_{in}+V_{an}$, which is the sum of the input and output voltages according to Kirchoff's Voltage Law applied to the outer loop through $T_1,T_2,C$ and $C_{dc}$ in Fig. \ref{fig:InverswandlerOrigin}c.

\subsection{MOSFET semiconductor losses}

\begin{figure}[h]
	\centering
	\includegraphics[width=0.65\textwidth]{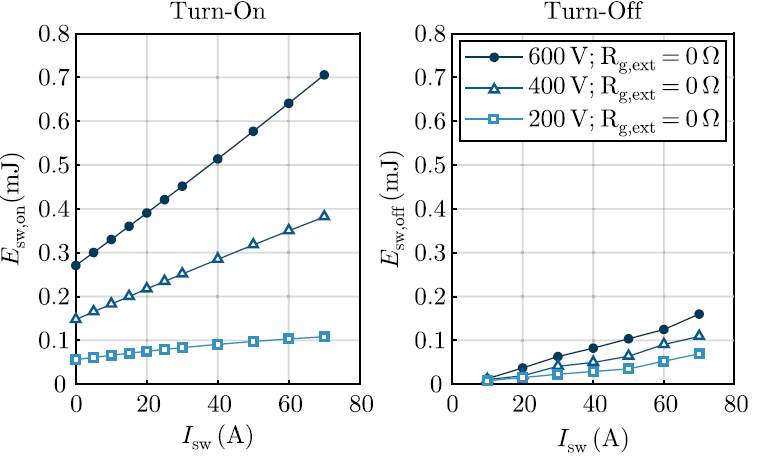}
	\caption{The experimentally obtained loss map for a 900 V SiC MOSFET based half-bridge \cite{C3M0030090K}}
	\label{fig:LossMap1}
\end{figure}

The equation \ref{eq: io} provides only a theoretical estimation of the minimum soft-switching current. On a separate note, there is a comprehensive experimental evaluation of the semiconductor losses provided in \cite{Azurza17}. In that work, a loss map of a 900 V SiC MOSFET half-bridge was experimentally obtained (c.f. Fig. \ref{fig:LossMap1}). The loss map for this device is of particular interest in this chapter due to the applicability of the device in an Electro Hydrostatic Actuator rectifier application, which will be discussed in the next chapter.

Here the turn-on graph on the left provides the half-bridge switching losses in case of a hard-switched transition (e.g. the red dot in Fig. \ref{fig:BuckConverterWaveform}). From this graph we can see that the switching loss is proportional to the half-bridge voltage and the switched current. Even in case of zero switched current (Zero Current Switching) there is still a substantial loss originating from the energy stored in the output capacitance of the transistor $E_{OSS} = C_{OSS} V_{in}^2 / 2$, which motivates the soft-switching.

The turn off graph on the right provides the half-bridge switching losses in case of a soft transition (e.g. a green dot in Fig. \ref{fig:BuckConverterWaveform}). From this graph we can see that the switching loss is still proportional to the half-bridge voltage and the switched current, although the losses are considerably lower than those of the turn on graph. These dependencies are expected, since both hard- and soft-switching losses are essentially the voltage-current overlap areas in the channel of the switch. At low values of the switched current the switching loss becomes negligible as there is almost no voltage-current overlap area. This is an example of a complete soft-switching. Note that the data for a very low values of the switched current are not shown because at those values the secondary effects, which change based on the application, such as the duration of the dead time, $L / C_{OSS}$ ratio etc., start affecting the measurements. The respective transitions become partially hard and at zero current the loss figures coincide with those of ZCS, ensuring the continuity of the loss map.

\section{Modulation Schemes for the Three-Phase Inverting Buck-Boost Converter}

Having understood the basic operating principle of the three-phase topology at hand and revisited the hard- and soft-switching mechanisms, this section introduces three modulation scheme candidates - standard Pulse Width Modulation (PWM), Discontinuous Pulse Width Modulation (DPWM) and a Boundary Conduction Mode (BCM). For all three candidates the steady state theoretical three-phase converter waveforms are presented. In these waveforms any effect of the converter reactive components was neglected. In particular, the voltage drops over the buck-boost inductances and reactive currents consumed by AC-side capacitors were neglected due to the fact that respective voltage/current amplitudes are small compared to the nominal output voltage/current amplitudes. Moreover, for each modulation the envelopes of the buck-boost inductor currents are also depicted as a green/red lines, corresponding to soft-/hard-switched transitions respectively.

\subsection{Standard Pulse Width Modulation (PWM)}

\begin{figure}[H]
	\centering
	\includegraphics[width=\textwidth]{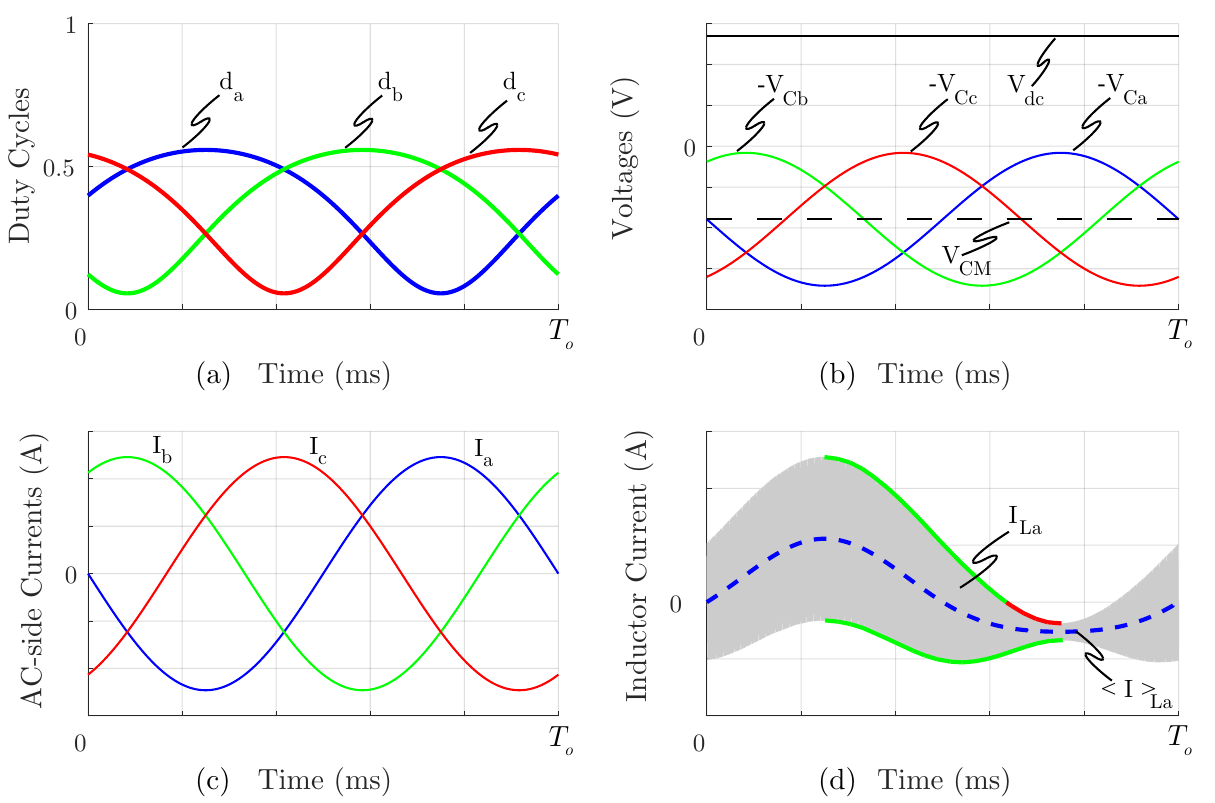}
	\caption{Three-phase inverting buck-boost converter theoretical steady state PWM waveforms over fundamental period. Phase duty cycles (a); DC- and AC-side voltages (b); AC-side currents (c); and phase $a$ buck-boost inductor current (d)}
	\label{fig:PWMwaveforms}
\end{figure}

The first modulation candidate is standard Pulse Width Modulation (PWM). The concept of PWM is simple: provide a large enough constant CM offset $V_{CM}$, such that the AC-side voltages $V_{Ca}, V_{Cb}, V_{Cc}$ are strictly unipolar throughout the fundamental period $T_o$, and apply the non-linear duty cycles (\ref{eq:Duty_Ipkpk}) in each phase to yield the sinusoidal Differential Mode (DM) voltages. Fig. \ref{fig:PWMwaveforms} shows the steady state theoretical PWM waveforms of the three-phase inverting buck-boost converter.

Fig. \ref{fig:PWMwaveforms}a-b tell us that the duty cycles below/above 0.5 correspond to buck/boost phase operation respectively. Fig. \ref{fig:PWMwaveforms}c depicts the AC-side load currents. Fig. \ref{fig:PWMwaveforms}d shows the phase $a$ buck-boost inductor current without taking into account the reactive capacitive currents of the AC-side. Although, strictly speaking, these reactive currents have to be incorporated for a proper modulation evaluation, the effect of these currents is relatively small, especially when AC-side capacitances are selected such that the reactive currents remain small compared to the load currents. Thus, for evaluation and comparison purposes, capacitive current and inductive voltage drop contributions are neglected in the presented steady state theoretical waveforms.


\subsection{Discontinuous Pulse Width Modulation (DPWM)}
\label{DPWM}

\begin{figure}[H]
	\centering
	\includegraphics[width=\textwidth]{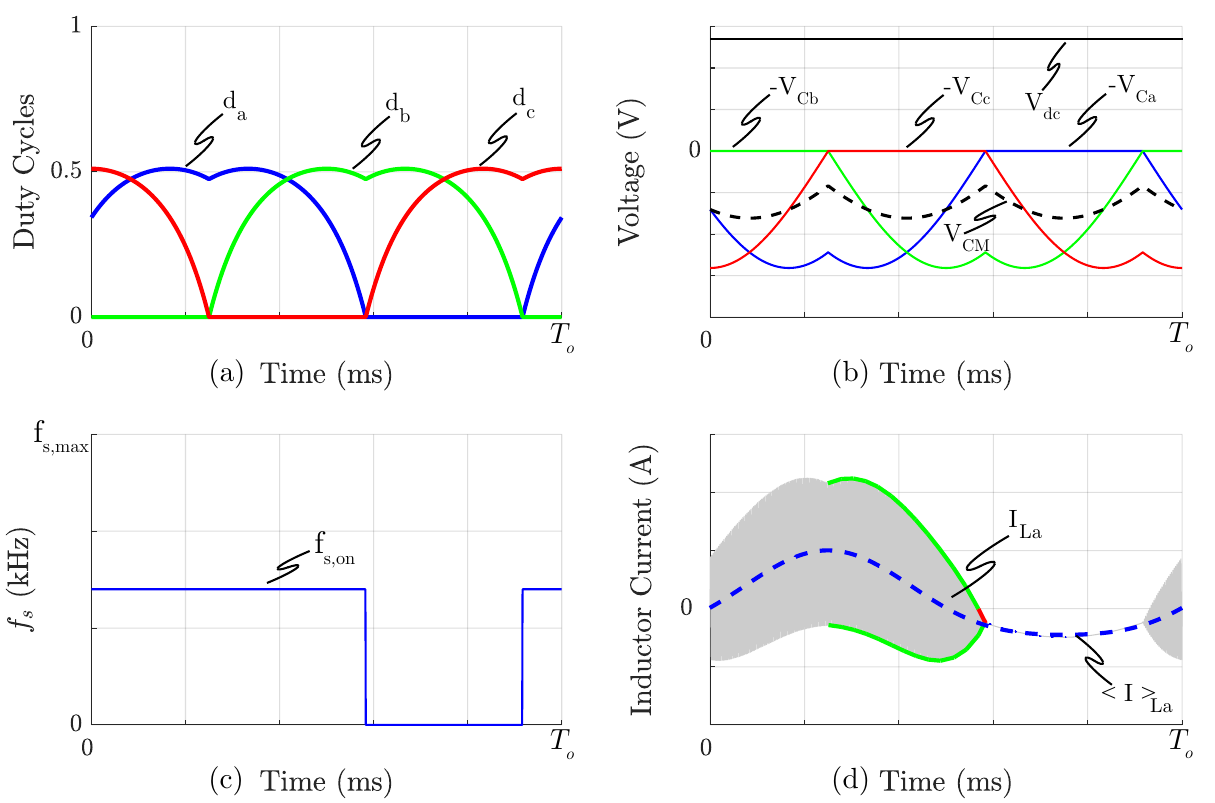}
	\caption{Three-phase inverting buck-boost converter theoretical steady state DPWM waveforms over fundamental period. Phase duty cycles (a); DC- and AC-side voltages (b); phase $a$ switching frequency (c); and phase $a$ buck-boost inductor current (d)}
	\label{fig:DPWMwaveforms}
\end{figure}

The second modulation candidate is Discontinuous Pulse Width Modulation (DPWM), inspired by \cite{Antivachis}. The concept of DPWM is as follows: use the CM offset $V_{CM}$ as a degree of freedom (DOF) and shape it in a way that results in a reduced blocking voltage per switch and switching actions compared to standard PWM (compare Fig. \ref{fig:PWMwaveforms}b, \ref{fig:PWMwaveforms}d to Fig. \ref{fig:DPWMwaveforms}b, \ref{fig:DPWMwaveforms}d). Fig. \ref{fig:DPWMwaveforms} shows the steady state theoretical DPWM waveforms of the three-phase inverting buck-boost converter.

As it can be seen from Fig. \ref{fig:DPWMwaveforms}a, the duty cycles are clamped to zero for one third of the fundamental period. This can also be seen from Fig. \ref{fig:DPWMwaveforms}c, where phase $a$ switching frequency equals zero (no switching) for one-third of the fundamental period. Fig. \ref{fig:DPWMwaveforms}d shows the phase $a$ buck-boost inductor current without taking into account the reactive capacitive currents of the AC-side. In this idealized case one would expect the inductor current to be exactly equal to the sinusoidal phase current for the duration of clamping, which can actually be observed in the figure. Moreover, phases are almost purely soft-switched due to DPWM as it can be seen from Fig. \ref{fig:DPWMwaveforms}d. Regarding the phase voltages in Fig. \ref{fig:DPWMwaveforms}b, they are obviously far from sinusoidal when referenced to the negative DC rail of the converter (point $n$ in Fig. \ref{fig:InverswandlerRectifier}). However, these voltages are purely sinusoidal when referenced to the star point of the grid (point $m$ in Fig. \ref{fig:InverswandlerRectifier}). Indeed, consider phase $a$ voltage relations (Fig. \ref{fig:DPWMPhaseAVoltageRelation}). From the schematic one can see that:

\begin{figure}[h]
	\centering
	\includegraphics[width=\textwidth]{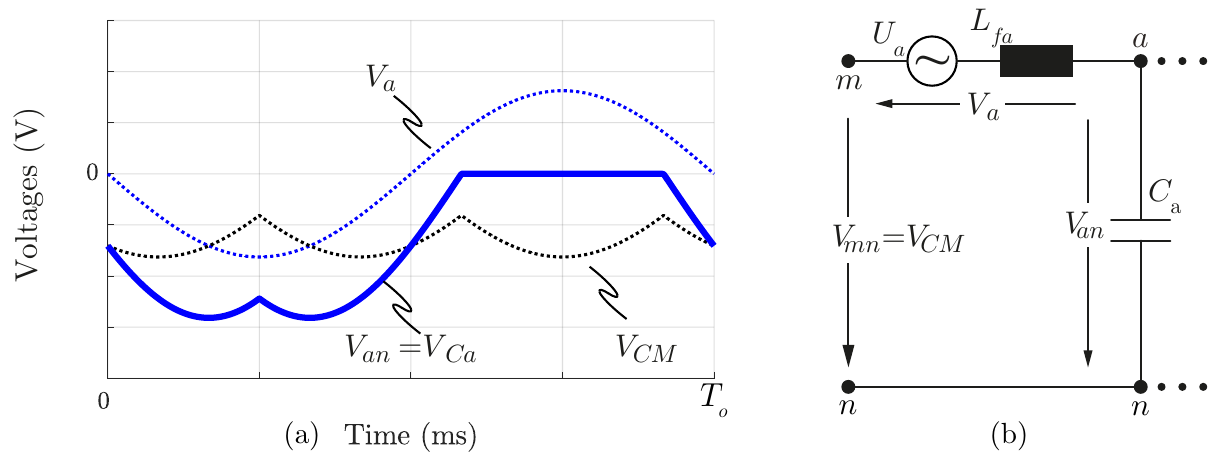}
	\caption{DPWM phase $a$ voltage relation. Phase $a$ voltages with respect to DC- and star point (a); phase $a$ AC-side circuit(b).}
	\label{fig:DPWMPhaseAVoltageRelation}
\end{figure}

\begin{equation}
\label{eq:PhaseAVoltageRelation}
V_{Ca} = V_{an} = V_{a} + V_{mn} = V_{a} + V_{CM} = V_{Ca,DM} + V_{CM}
\end{equation}

The relation $V_{an} = V_a + V_{CM}$ is easily seen in Fig. \ref{fig:DPWMPhaseAVoltageRelation}a, thus proving that AC-side voltage is purely sinusoidal when referenced to the grid star point.

\subsection{Boundary (Triangular) Conduction Mode (BCM or TCM)}

\begin{figure}[H]
	\centering
	\includegraphics[width=\textwidth]{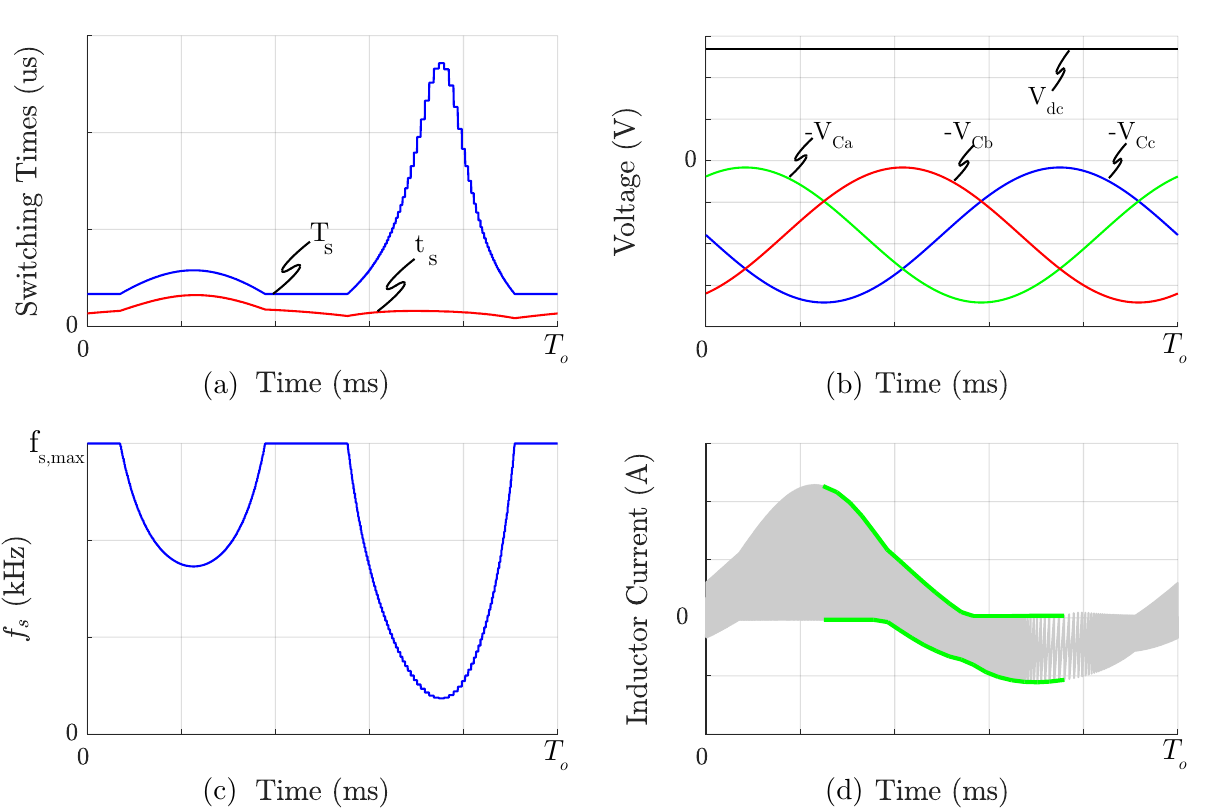}
	\caption{Three-phase inverting buck-boost converter theoretical steady state BCM waveforms over fundamental period. Phase $a$ switching times (a); DC- and AC-side voltages (b); phase $a$ switching frequency (c); and phase $a$ buck-boost inductor current (d)}
	\label{fig:BCMwaveforms}
\end{figure}

The third modulation candidate is Boundary Condution Mode (BCM), inspired by \cite{Knecht18}. The concept of BCM is as follows: use the per-phase switching frequency $f_s$ as a degree of freedom (DOF) and shape it in a way that results in a pure soft-switched operation throughout the fundamental period $T_o$. Fig. \ref{fig:BCMwaveforms} shows the steady state theoretical BCM waveforms of the three-phase inverting buck-boost converter.

Fig. \ref{fig:BCMwaveforms}a immediately reveals the qualitative difference between BCM and PWM/DPWM. Indeed, for BCM the duty cycle $d = f(f_s)$ as this modulation scheme utilizes the variable switching frequency DOF to facilitate a soft-switched operation throughout the fundamental period. Instead of the duty cycles one plots the switching times $t_s$ and $T_s$, where $t_s$ stands for the switching instance of the respective phase within the switching period and $T_s$ stands for the final switching transition of the period. In this context, the duty cycle can be defined as $d = \frac{t_s}{T_s}$. Fig. \ref{fig:BCMwaveforms}d shows the phase $a$ buck-boost inductor current without taking into account the reactive capacitive currents of the AC-side. It can be seen that BCM ensures soft switching throughout the fundamental period. Also the peak/ripple inductor currents are slightly reduced compared to DPWM/PWM (compare Fig. \ref{fig:PWMwaveforms}d, \ref{fig:DPWMwaveforms}d to Fig. \ref{fig:BCMwaveforms}d), thus potentially improving the switching losses as well as the conduction losses both in semiconductors and inductors. Fig. \ref{fig:BCMwaveforms}c depicts the switching frequency variation over the fundamental period. Two clear valleys below the maximum switching frequency of 300 kHz are observed. Taking into account these valleys, the input filter has to be designed for the minimal frequency of the CISPR band (150 kHz for rectifier applications), thus resulting in elevated filtering effort.

\subsection{Conclusion}
To sum up, remember that all three modulations are applicable to the Inverswandler irrespective of the application. Thus, the choice of modulation has to be done based on the quantitative evaluation and comparison, which takes into account the electrical specifications of an application at hand. Up to this point though, all the analysis was generic. In the next chapter, mentioned evaluation will be performed.

\chapter{Modulation, Control and Evaluation of the Three-Phase Inverting Buck-Boost Converter as a Rectifier}

This chapter deals with the evaluation and selection of a modulation scheme, and control system design for the three-phase inverting buck-boost converter as a rectifier for an Electro-Hydrostatic Actuator (EHA) application in a More Electric Aircraft. Section one introduces the topology as a three-phase rectifier, and the electrical specifications. Also it discusses the component design, since not only the electrical specifications, but also the component designs are important for performance evaluation. Next, section two evaluates and compares the modulation scheme candidates from the previous chapter in the context of application at hand. One modulation is selected to be implemented in a closed-loop control scheme. Rectifier control for the selected modulation is addressed in section three. Challenges in the closed-loop rectifier control implementation are identified, and ameliorating measures (both software and hardware) are proposed for those in section four. Finally, section five summarizes the study of the three-phase inverting buck-boost converter within the rectifier application context.

\section{Topology and Electrical Specifications in Rectifier Mode}

Regarding applications for a More Electric Aircraft, the authors of \cite{Gong} compare various three-phase AC-DC converter concepts for a drive train of an Electro-Hydrostatic Actuator (EHA). EHA is an actuator system that uses a high-power stepper motor to precisely control the pump position and pressure inside the hydraulic cylinder, which in turn is used to control the plate positions on the wings for aircraft positioning. Conventional designs used mechanical valve control to change the pressure in the hydraulic cylinder manually. The EHA provides redundancy (robustness), simplicity in system design and operation, energy economy and weight reduction. In this application, analyzed AC-DC converters are used to feed the DC link of the motor-driving inverter from the aircraft on-board mains. Here the advantage of having a controllable buck-boost output voltage is the capability of handling a wide input voltage range from the mains while maintaining the favorable output voltage.

The three-phase inverting buck-boost converter topology and the electrical specifications for a rectifier operation within the EHA application are shown in Fig. \ref{fig:InverswandlerRectifier} and Table \ref{t:SpecsRectifier} below. The first set of specifications in Table \ref{t:SpecsRectifier} (starting with power rating and ending with THD) are the direct application requirements. Note that the rectifier has to provide a constant DC voltage and output power in the conditions of voltage- and frequency-oscillating grid, while maintaining a very high power factor and grid current quality. Although the grid voltage fluctuations are not that wide, the DC voltage is defined such that the buck-boost capability is required. The load resistor $R_{dc}$ is used to model the load behavior and is selected such that 600W of power are consumed at 270V DC link voltage. The second set of specifications in Table \ref{t:SpecsRectifier} (starting with operating switching frequency and ending with maximum switching frequency limit) are the derived electrical specifications. The derivation reasoning behind those is provided below.

\begin{figure}[h]
	\centering
	\includegraphics[width=\textwidth]{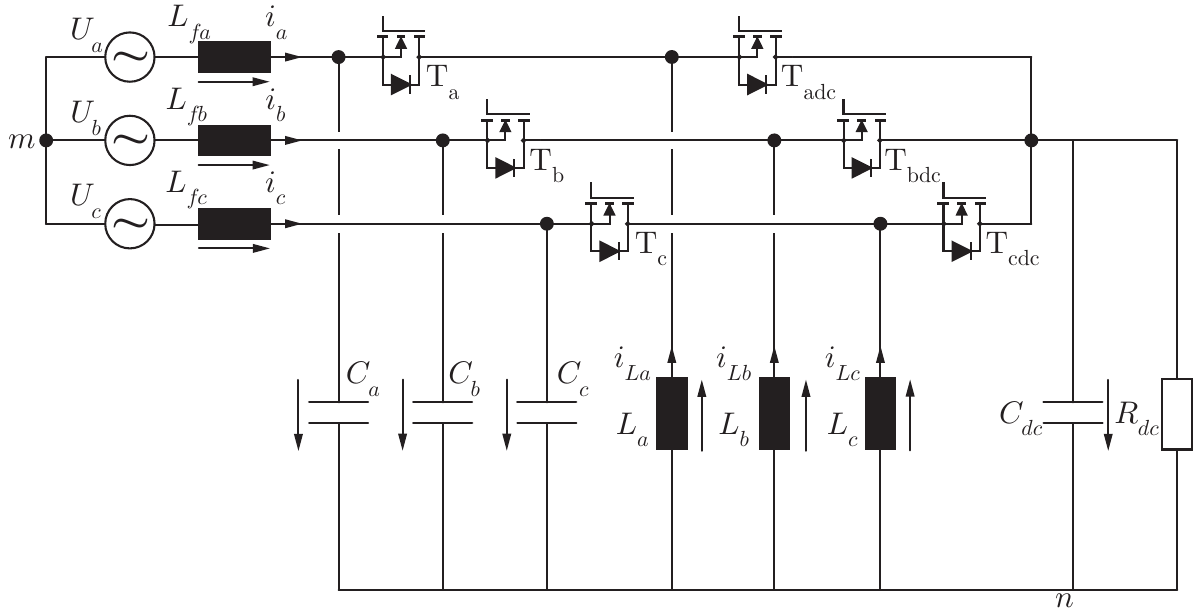}
	\caption{Grid-interfaced three-phase inverting buck-boost converter in a rectifier configuration}
	\label{fig:InverswandlerRectifier}
\end{figure}
\begin{table}[H]
	\centering
	\caption{System specifications for a rectifier application}
	\label{t:SpecsRectifier}
	\begin{tabular}{ | P{2cm} || P{2cm} | P{1.5cm} | P{8cm} | }
		\hline
		Parameter & Value & Unit & Param. Explanation (Given $|$ Derived)\\
		\hline\hline
		$P$ & $600$ & W & Rectifier power rating\\
		$U_{o}$ & $115 \pm 15\%$ & Vrms & Input grid phase voltage rms with tolerance\\
		$V_{dc}$ & $270$ & V & Output DC voltage\\		
		$f_o$ & $360 - 800$ & Hz & Grid frequency range\\
		$PF$ & $>99\%$ & - & Power factor requirement for $[50-100]\%$ load\\
		$THD$ & $<5\%$ & - & Grid phase current THD requirement\\
		\hline
		$f_s$ & $140$ & kHz & Switching frequency\\
		$L$ & $75$ & $\mu$H & Buck-boost inductance\\
		$L_f$ & $75$ & $\mu$H & Grid-side filter inductance\\
		$C$ & $1$ & $\mu$F & Grid-side phase capacitance\\
		$C_{dc}$ & $10$ & $\mu$F & Load-side DC capacitance\\
		Device & SiC 900V & - & Semiconductor: 1 x 900V, 30mOhm SiC \cite{C3M0030090K}\\
		$f_{s,max}$ & $300$ & kHz & Maximum switching frequency (BCM)\\	
		\hline
	\end{tabular}
\end{table}

$\bullet$ First, the switching frequency $f_s$ was selected to be 140kHz because this value is at the lower limit of the regulated frequency band of the DO160 airborne equipment standard (between 150kHz - 30MHz for aviation applications \cite{DO160}). By selecting 140kHz switching frequency we ensure that the first regulated switching frequency harmonic is 280kHz. Since the first regulated harmonic determines the corner frequencies of our input filter stages, the higher the regulated harmonic is, the higher the filter corner frequencies will be. Thus, the size of the filter is expected to reduce considerably. Bearing in mind the stringent application requirements on the grid current quality, having the first regulated harmonic as high is possible is beneficial. From the overall system perspective, of course one could consider switching frequencies above 140kHz. Ideally, in a proper Pareto analysis the switching frequency is a Degree Of Freedom (DOF), but for now the objective is to compare the modulation schemes for a single set of electrical specifications.

$\bullet$ Second, the buck-boost inductance value $L$ was selected to be 75$\mu$H because at this value the inductor current ripple is just large enough such that the Discontinuous Pulse Width Modulation (DPWM) results in complete soft-switching converter throughout the whole fundamental period, while standard PWM is soft-switched over a considerable part of the fundamental period as we will see in the later sections. On the other hand, the selected inductance value precludes the current ripple from becoming too large, which would increase the converter conduction losses. Bearing in mind that BCM employs variable switching frequency to facilitate soft-switching throughout the whole fundamental period, having the other two modulation candidates (PWM and DPWM) predominantly soft-switched provides a fair comparison between the three. In addition, the per-phase inverting buck-boost converter inherently features a high peak-to-average buck-boost inductor current ratio due to the discontinuous input-output power transfer. This inherent disadvantage in terms of converter conduction losses can be used to reduce the corresponding switching losses by allowing large current ripple with consequent inductor current-reversals needed for soft-switching. Regarding the filter inductor value, its value depends on the buck-boost inductance and AC-side capacitance values. The objective here is to filter the grid phase currents to comply with stringent current quality requirement. The filter inductance was iteratively simulated and selected to be 75$\mu$H, but in general this value is not strict and can be subject to change depending on aforementioned inductance and capacitance values.

$\bullet$ Third, the AC-side capacitance per phase was iteratively simulated and selected to be 1$\mu$F because it provides a good compromise between the grid-side voltage ripple and the capacitive current consumption. Indeed, in case AC-side capacitance is too small, the respective voltage ripple will become too large, potentially deteriorating the AC-side current quality (THD, PF) or even closed-loop control stability. On the other hand, in case AC-side capacitance is too large, the reactive current consumed by that capacitor will become large, resulting in elevated converter current values and conduction losses. Regarding the DC-side capacitance, it was iteratively simulated and selected to be 10$\mu$F because this value is enough to suppress the DC-side voltage ripple to $\pm5\%$.

$\bullet$ Finally, the maximum switching frequency for BCM case was selected to be 300kHz. The objective of such a choice is to limit the occurring switching losses near the fundamental zero-crossing of the buck-boost inductor current. At 300kHz the switching losses already dominate the conduction losses of the semiconductor by far, increasing the switching frequency higher could result in too much losses and controllability (gate driver) problems. Based on this reasoning, the limit of 300kHz was selected.


$\bullet$ Capacitor realization. To evaluate the effect of modulation on the converter performance, apart from the electrical specifications of capacitance and inductance, one also needs to determine the respective capacitor and inductor realizations, because the converter efficiency and power density depend on the volumes and losses of the respective components. The designs are determined using existing component scripts. Regarding capacitors, only film capacitors have been considered both for AC and DC-side. There are two reasons for that. First, in a proper converter design with a full Pareto optimization, the capacitor losses tend to be small compared to other component losses and this is well captured in our modulation evaluation, because film capacitors are lossless in a good approximation. Second, film capacitors are the least power dense and the most sensitive to the rated voltage variations. By utilizing them in the modulation evaluation one can more easily distinguish the effects of modulation-dependent parameters on the volume of the capacitive components.

$\bullet$ Inductor realization. Regarding the buck-boost inductors, the component script sweeps through all the possible core-coil combinations that result in the specified inductance value and outputs the losses-volume scatter graph, where each individual inductor design corresponds to a point. For a proper converter design, factors like inductor shape, fill-factor, manufacturing process etc. would have to be considered. For a qualitative modulation comparison, however, it is enough to pick a simply viable inductor design that comprehensively captures the expected characteristics of a properly designed inductor. Generally, those expected characteristics comprise a small volume and low losses. From these considerations, it was decided to estimate the volume-loss product for each inductor and simply pick the design with minimum product. By doing so we ensure that our modulation comparison outputs a reliable results, because irrespective of the final inductor design, it will have similar performance characteristics as the design with minimum loss-volume product.

\section{Comparison of Modulation Schemes}

\begin{sidewaysfigure}[h]
	\centering
	\includegraphics[width=\textwidth]{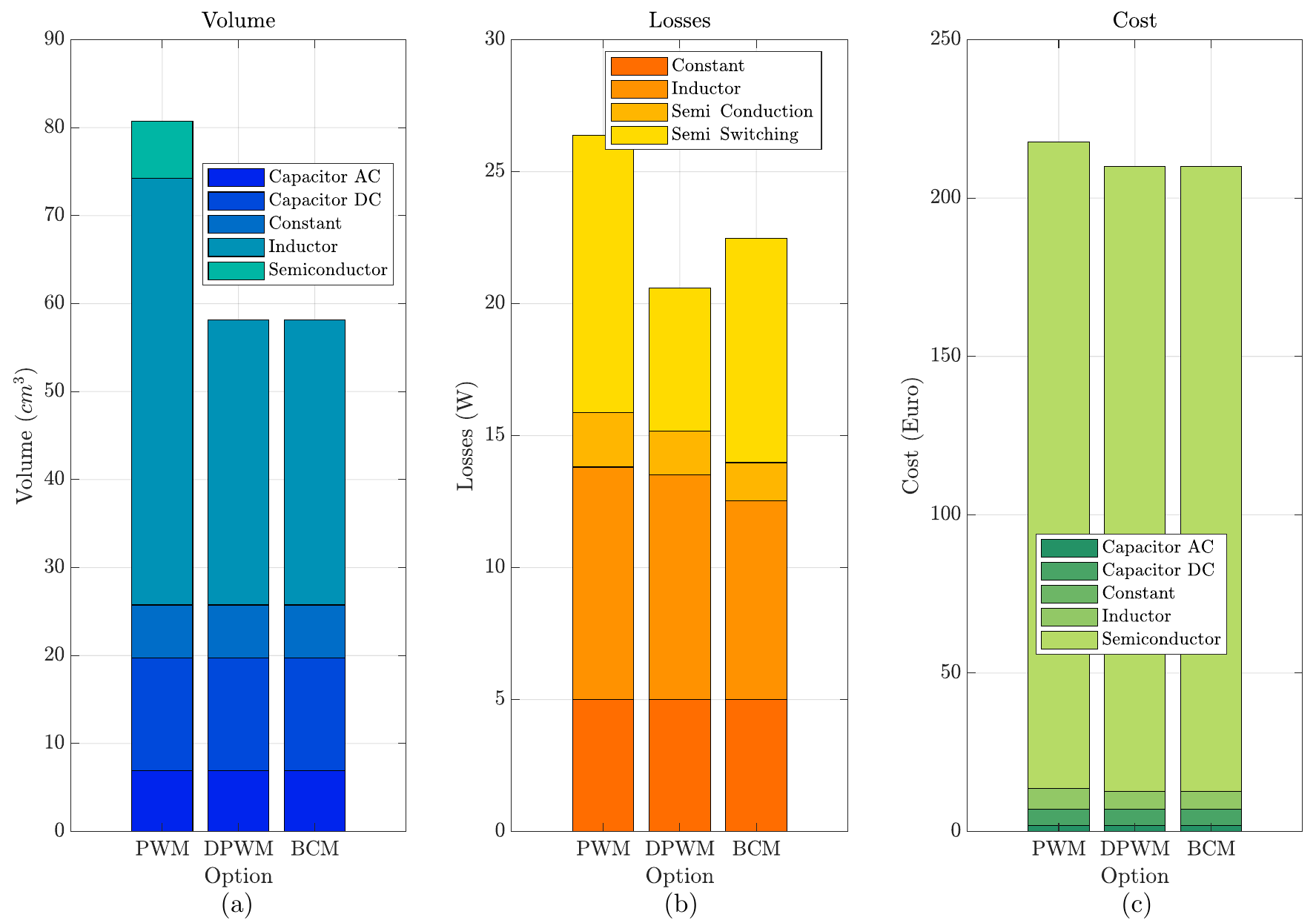}
	\caption{Estimated volume (a), losses (b) and cost (c) of the three-phase inverting buck-boost converter employing the considered modulation schemes}
	\label{fig:ModulationsComparison}
\end{sidewaysfigure}

Having presented the electrical specifications and the converter design guidelines, this section deals with the modulation comparison. Evaluated performance metrics comprise theoretical converter volume, losses and cost. Volume is estimated as a sum of boxed component volumes. The passives (capacitors, inductors) are designed based on the procedures above and volumes/losses are estimated. For example, the blue line in Fig. \ref{fig:PWMwaveforms}d is the buck-boost inductor current average over the fundamental period. From this line as well as the instantaneous ripple one can estimate the inductor current and, consequently, losses. Regarding semiconductors, loss estimation is done using the instantaneous voltage/currents across the devices at switching instances and the loss map (shown in Fig. \ref{fig:LossMap1}). From these, estimated performances are presented in Fig. \ref{fig:ModulationsComparison} and Table \ref{t:CompareModulations}.

\begin{table}[h]
	\centering
	\caption{Comparison of modulation performances for nominal operating conditions}
	\label{t:CompareModulations}
	\begin{tabular}{ | P{2cm} || P{1.5cm} | P{1.5cm} | P{1.5cm} | P{1cm} | P{6cm} | }
		\hline
		Parameter & PWM & DPWM & BCM & Unit & Param. explanation\\
		\hline\hline
		$I_{L,rms}$ & 4.3 & 3.8 & 3.5 & A & Buck-boost inductor rms current\\
		$I_{L,rms}^2$ & 18.2 & 14.4 & 12.2 & A$^2$ & Buck-boost inductor rms$^2$ current\\
		$I_{L,max}$ & 12.7 & 11.2 & 11.3 & A & Buck-boost inductor max. current\\
		\hline
		$f_{sw,avg}$ & 140 & 91 & 200 & kHz & Average switching frequency\\
		$f_{sw,max}$ & 140 & 140 & 300 & kHz & Maximum switching frequency\\
		$f_{sw,min}$ & 140 & 0 & 37 & kHz & Minimum switching frequency\\
		\hline
		$V_{ss,max}$ & 611.5 & 551.6 & 611.5 & V & Maximum soft-switched voltage\\        
		$I_{ss,max}$ & 12.7 & 11.2 & 11.3 & A & Maximum soft-switched current\\        
		$V_{ss,avg}$ & 498.8 & 404.0 & 448.9 & V & Average soft-switched voltage\\
		$I_{ss,avg}$ & 5.6 & 3.4 & 3.2 & A & Average soft-switched current\\        
		\hline        
		$V_{hs,max}$ & 319.9 & - & - & V & Maximum hard-switched voltage\\
		$I_{hs,max}$ & 1.9 & - & - & A & Maximum hard-switched current\\        
		$V_{hs,avg}$ & 299.1 & - & - & V & Average hard-switched voltage\\        
		$I_{hs,avg}$ & 1.2 & - & - & A & Average hard-switched current\\
		\hline
	\end{tabular}
\end{table}

\clearpage

From the comparison it is seen that both DPWM and BCM are expected to show a comparable performance in terms of volume, losses and cost, which is better than that of standard PWM. Indeed, the comparison reveals the same potential converter volume and cost for DPWM/BCM, meaning that the small differences in the respective buck-boost inductor current waveforms are not significant enough to opt for a different inductors/capacitors realization. Nevertheless, small waveform differences are reflected in the loss distribution, where DPWM slightly outperforms the BCM. This result, however, cannot be generalized, as for a slightly different inductor realization or a varying rectifier grid frequency/voltage, the result can be different.

Finally, the decisive factors in modulation selection are the control complexity and filtering considerations. BCM features a very complex control with its current zero-crossing detection techniques. Apart from that, the BCM filter is expected to be over-sized as it was mentioned earlier. These complications are not present in DPWM, where the control is relatively straightforward and the filter can be designed for 280kHz $>$ 150kHz. From these considerations, the DPWM was selected to be further investigated.

\section{Rectifier Control for DPWM}

In this section the control structure is designed for the Discontinuous PWM. A standard rectifier control with its dq-transformations and Phase Locked Loop (PLL) is adapted to account for the specific three-phase inverting buck-boost converter topology and the selected modulation. The control system of this section is depicted in Fig. \ref{fig:ControllerDPWM}. For convenience purposes, the topology under consideration is again depicted below:

\begin{figure}[h!]
	\centering
	\includegraphics[width=0.96\textwidth]{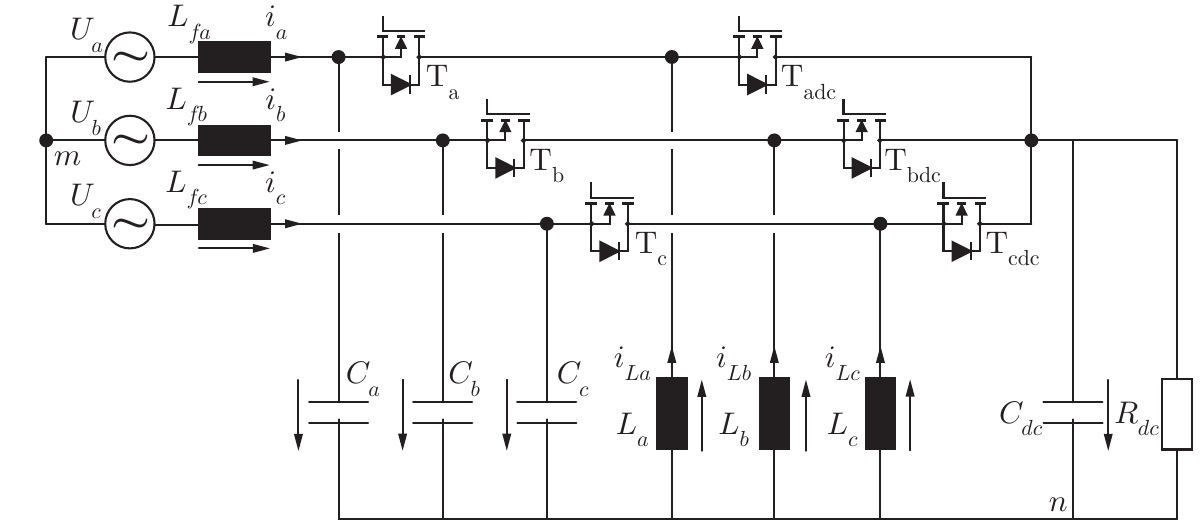}
	\caption{Grid-interfaced three-phase inverting buck-boost converter in a rectifier configuration}
	\label{fig:InverswandlerRectifierNoDiodes}
\end{figure}

\subsection{DC-side voltage controller}
First, the control system starts with the outermost controller, which is a DC-side voltage regulator. DC-side controller is used to track the reference of the desired DC link voltage. As an output, it provides the required DC link current that, in addition to the DC-side load current measurement, results in a reference for the sum of the DC-side switch current averages $i_{Tdc}^*$. The product of $i_{Tdc}^*$ and $V_{dc}^*$, in turn, sets the required power demand of the converter. Assuming that the inner controllers in dq-frame as well as the PLL itself are operating properly and that unity power factor is tracked (which should be true due to time scale separation of the controllers), active power reference can immediately provide the d-component of the AC-side grid current reference in dq-frame through the well-known three-phase active power relation. Given the requirement of unity power factor, q-component reference is set to zero. These dq-components are the final outputs of the outermost DC-side voltage controller.

\subsection{Grid current controllers in dq-frame}
Next, the dq-components of the grid current reference are compared to the grid current measurements. The difference is fed to the second cascaded controllers, responsible for grid current control over the filter inductors in dq-frame. These controllers output the voltage drop required over the filter inductors to draw the required current from the grid. Of course, the resulting dq-components of voltage reference do not carry any information about the CM voltage in converter phases, since dq-transformation does not capture the CM voltage. However, in DPWM we also want to shape the CM voltage as a function of the instantaneous DM voltage references. Therefore to control both DM and CM components of the AC-side capacitor voltages, from here on one has to move back to the abc-frame.

\subsection{CM voltage and capacitor reference derivation}
The dq-abc transformation gives us the respective voltage references across the filter inductors in abc-frame. By adding these voltages to the respective grid voltages, one gets the required AC-side capacitor voltages referenced to the grid star point or, in other words, DM parts of the AC-side capacitor voltage references $V_{Ci,DM}^*, i \in {a,b,c}$. In DPWM, the required CM voltage reference $V_{CM}^*$ is a function of $V_{Ci,DM}^*$. The relation is as follows (Fig. \ref{fig:DPWMPhaseAVoltageRelation}):

\begin{equation}
\label{eq:CMtoDMrelation}
V_{CM}^* = -\underset{i=a,b,c}{max}(V_{Ci,DM}^*)
\end{equation}

From (\ref{eq:PhaseAVoltageRelation}) and (\ref{eq:CMtoDMrelation}) the total AC-side voltage references can be derived as a sum of DM and CM references, and yielding one phase with zero voltage reference (clamped):

\begin{equation}
\setlength{\jot}{10pt} 
\label{eq:CMplusDM}
\begin{split}
&V_{Cx}^* = V_{Cx,DM}^* + V_{CM}^* = V_{Cx,DM}^* -\underset{i=a,b,c}{max}(V_{Ci,DM}^*) = 0
\\
&V_{Cy}^* = V_{Cy,DM}^* + V_{CM}^* = V_{Cy,DM}^* -\underset{i=a,b,c}{max}(V_{Ci,DM}^*)
\\
&V_{Cz}^* = V_{Cz,DM}^* + V_{CM}^* = V_{Cz,DM}^* -\underset{i=a,b,c}{max}(V_{Ci,DM}^*)
\\
\end{split}
\end{equation}

Indeed, from equation (\ref{eq:CMtoDMrelation}) we know that at any instant in time the CM voltage is equal to the negative DM AC-side capacitor voltage reference of one phase, meaning that the sum of CM and DM references is always equal to zero for one phase. This zero-reference phase changes three times every fundamental period. Thus, for inner controllers instead of any particular phase designator $a, b$ or $c$, one uses the designators $x, y$ and $z$, where $x$ always refers to the clamped phase (can be $a, b$ or $c$ at any particular instant). While the phase with $x$ designator is clamped, the references of the remaining phases $y$ and $z$ are fed to the inner controllers, addressed below.

\subsection{Capacitor AC-side voltage controllers in abc-frame}
As it was already said, the phase with $x$ designator is clamped, meaning that its duty cycle is forced to zero and no further calculations are made for this phase. The remaining phase voltages ($y$ and $z$ designators) are then compared to the respective measurements and the errors are fed to the AC-side voltage controllers - a third stage controllers, responsible for the tracking of the AC-side capacitor voltages. These controllers output the AC capacitor current references. According to KCL (Fig. \ref{fig:InverswandlerRectifierNoDiodes}) the local average AC-side capacitor current consists of two terms:

\begin{equation}
\label{eq:ACsideSwitchKCL}
i_{Cy,Cz} = C \frac{dV_{Cy,Cz}}{dt} = i_{y,z} + i_{Ty,Tz,avg}
\end{equation}

Average buck-boost inductor current, in turn, can be derived from the average AC-side switch current according to the following relation:

\begin{equation}
\setlength{\jot}{10pt} 
\label{eq:InductorACsideSwitchRelation}
\begin{split}
&i_{Ly,avg} = \frac{i_{Ty,avg}}{(1-d_y)} = \Big(C \frac{dV_{Cy}}{dt} - i_y\Big) \Big(1 + \frac{|V_{Cy}|}{V_{dc}}\Big) \\
&i_{Lz,avg} = \frac{i_{Tz,avg}}{(1-d_z)} = \Big(C \frac{dV_{Cz}}{dt} - i_z\Big) \Big(1 + \frac{|V_{Cz}|}{V_{dc}}\Big) \\
\end{split}
\end{equation}

From (\ref{eq:ACsideSwitchKCL} - \ref{eq:InductorACsideSwitchRelation}) table
the buck-boost inductor current references are derived:

\begin{equation}
\label{eq:InductorACsideSwitchRefRelation}
i_{Ly,avg}^* = \frac{i_{Ty,avg}^*}{(1-d_y)} = \Big(i_{Cy}^* - i_y\Big) \Big(1 + \frac{|V_{Cy}|}{V_{dc}}\Big) \text{ }\text{ }\text{ } i_{Lz,avg}^* = \frac{i_{Tz,avg}^*}{(1-d_z)} = \Big(i_{Cz}^* - i_z\Big) \Big(1 + \frac{|V_{Cz}|}{V_{dc}}\Big)
\end{equation}

\subsection{Buck-boost inductor current controllers in abc-frame}
Next, buck-boost inductor current references $i_{Ly,avg}^*$ and $i_{Lz,avg}^*$ are compared to the respective current measurements. The errors are fed to the buck-boost inductor current controllers - a forth stage (innermost, fastest), responsible for the inductor currents of the two active phases. These controllers result in the respective voltage references across the buck-boost inductors $V_{Ly}^*$ and $V_{Lz}^*$.

\subsection{Phase modulators}
Finally, from $V_{Ly}^*$ and $V_{Lz}^*$ the respective duty cycles can be derived through phase modulators. Considering the defined voltage polarities (Fig. \ref{fig:InverswandlerRectifierNoDiodes}), the modulator equation can be written as:

\begin{equation}
\label{eq:Modulator}
\begin{split}
V_{Ly} &= -V_{dc} d_y - V_{Cy} (1-d_y) \Rightarrow d_y = \frac{V_{Ly} + V_{Cy}}{V_{Cy} - V_{dc}} = \frac{-|V_{Cy}|+V_{Ly}}{-|V_{Cy}| - V_{dc}} = \frac{|V_{Cy}|-V_{Ly}}{|V_{Cy}| + V_{dc}}
\\
V_{Lz} &= -V_{dc} d_z - V_{Cz} (1-d_z) \Rightarrow d_z = \frac{V_{Lz} + V_{Cz}}{V_{Cz} - V_{dc}} = \frac{-|V_{Cz}|+V_{Lz}}{-|V_{Cz}| - V_{dc}} = \frac{|V_{Cz}|-V_{Lz}}{|V_{Cz}| + V_{dc}}
\\
\end{split}
\end{equation}

\begin{equation}
\label{eq:DtoTacTdc}
d \geq \text{Carrier} \rightarrow {T_{ac} = 0, T_{dc} = 1} \text{ }\text{ }\text{ } d < \text{Carrier} \rightarrow {T_{ac} = 1, T_{dc} = 0}
\end{equation}

\begin{sidewaysfigure}[h]
	\centering
	\includegraphics[width=\textwidth]{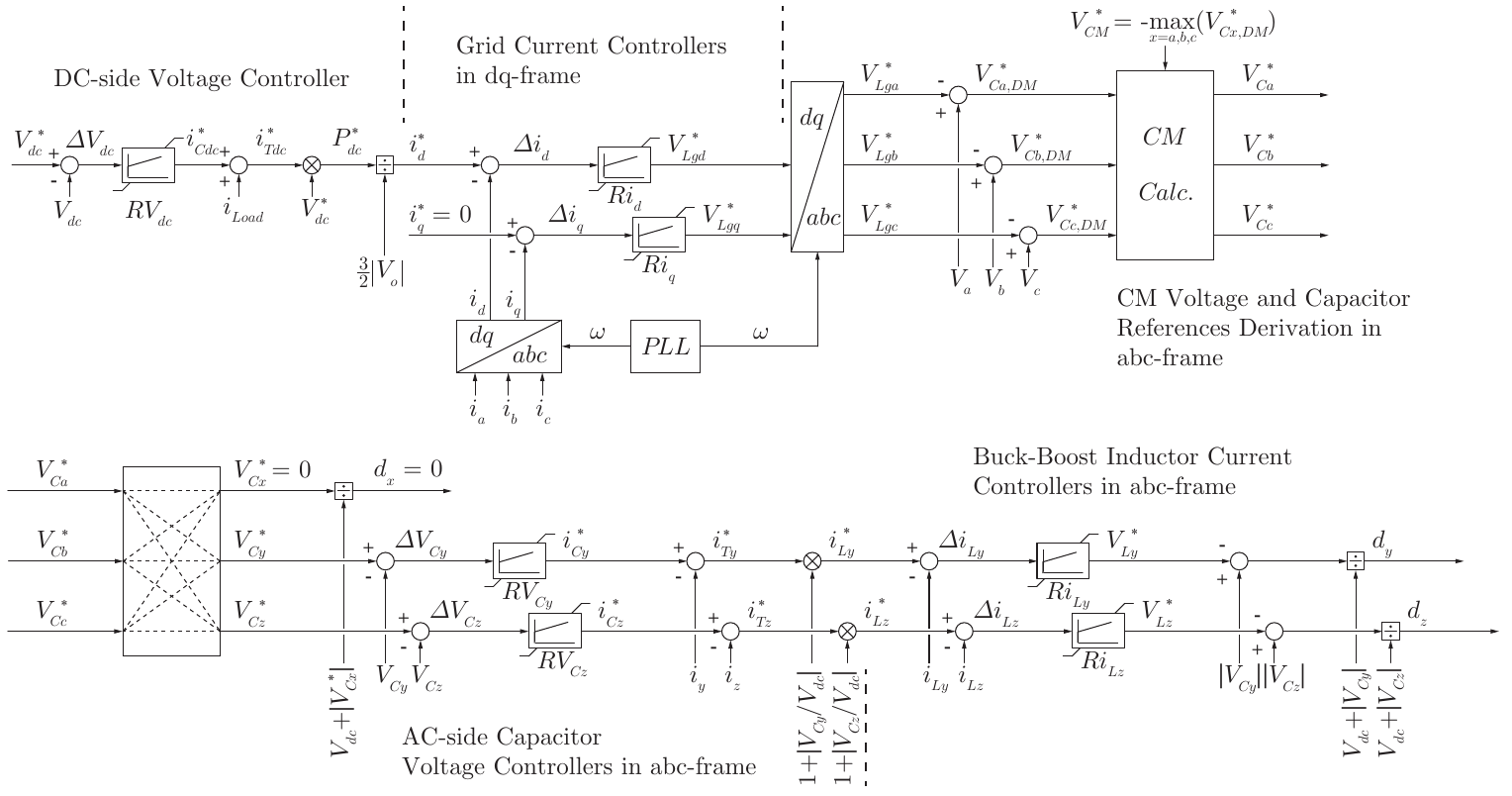}
	\caption{Proposed rectifier control system employing DPWM with four cascaded controllers, where two work in dq- and two in abc-frame, where one phase is clamping at a time}
	\label{fig:ControllerDPWM}
\end{sidewaysfigure}

From (\ref{eq:Modulator}) the duty cycles of the un-clamped phases can be derived. Summarizing block diagram of the control system is depicted in Fig. \ref{fig:ControllerDPWM}.

\clearpage

\subsection{Gecko-Simulink results}

\begin{figure}[h!]
	\centering
	\includegraphics[width=0.7\textwidth]{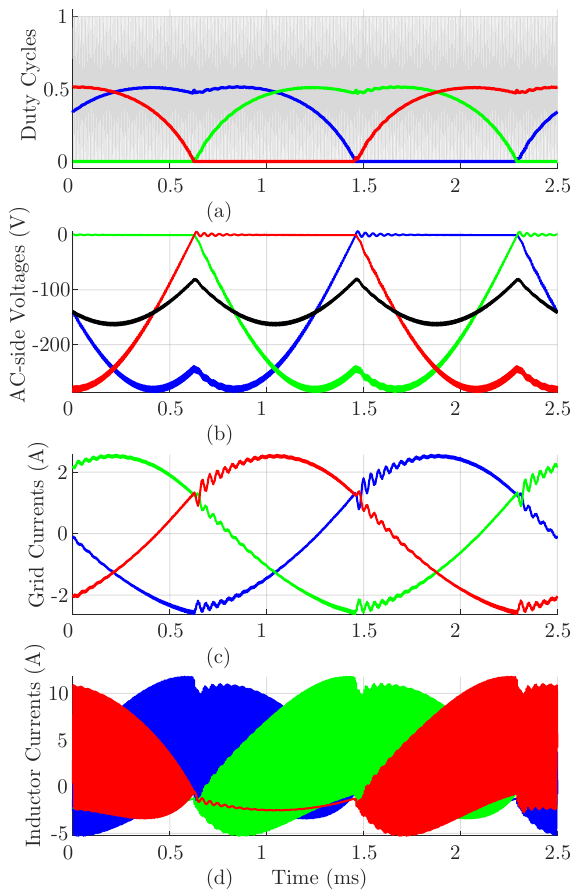}
	\caption{Gecko-simulated steady-state rectifier waveforms with proposed controller}
	\label{fig:DPWMwaveformsGecko}
\end{figure}

In Fig. \ref{fig:DPWMwaveformsGecko} one can observe the simulation results of the implemented control system (Fig. \ref{fig:ControllerDPWM}) applied to the three-phase inverting buck-boost converter (Fig. \ref{fig:InverswandlerRectifierNoDiodes}) with the specifications from Table \ref{t:SpecsRectifier}. Phase duty cycles, AC-side voltages, grid currents and buck-boost inductor currents are shown in graphs ($a$) to ($d$) respectively. From the graphs one can immediately observe the problem of a high frequency ringing in the clamped phases. Apart from that, a Fourier analysis of the grid currents has shown that the major contribution to the grid current THD comes from the second grid frequency harmonic, indicating potential errors in the grid current controller. Therefore, further measures have to be undertaken to eliminate the ringing at clamping instance and to comply with grid current quality requirements.

\section{Further Control Improvements}

Previous section dealt with the design of a control system for a three-phase inverting buck-boost converter employing DPWM. After applying the designed control in the simulation, the problems of high frequency ringing and grid current quality have been identified. This section addresses some further control improvements needed to ameliorate those.

\subsection{High frequency ringing and Gecko-Simulink results}
Fig. \ref{fig:DPWMwaveformsGecko} reveals that the ringing occurs predominantly in the phase that was just clamped. This ringing is expected, since phase clamping essentially means shorting the AC-side capacitor and the buck-boost inductor. Indeed, consider Fig. \ref{fig:InverswandlerRectifierNoDiodes} and the point in time when phase \textit{a} has just been clamped (this also applies to the other phases as well). The switch $T_a$ is permanently turned on, meaning that $C_a$ and $L_a$ are permanently connected to each other. This configuration can be subject to ringing. Moreover, phase \textit{a} grid current is also tied to this ringing due to KCL at the node between $T_a$, $C_a$ and $L_{fa}$. Finally, since the grid currents are tied to each other through the grid star point, phase \textit{a} grid current ringing is translated to the other phases and vice versa, creating the disturbances over those as well (grid currents graph in Fig. \ref{fig:DPWMwaveformsGecko}).

From the considerations above it can be deduced that in order to have a smooth clamping transition, one condition needs to be satisfied. The capacitor current $i_{Ca}$ just before the clamping has to be zero in order not to cause a subsequent change of capacitor voltage. This also implies that the grid and buck-boost inductor currents match perfectly just before the clamping. Obviously, this condition cannot be met within the standard DPWM, because in DPWM the capacitor current $i_{Ca} = C \frac{dV_{Ca}}{dt}$ is not zero before clamping (AC-side voltages graph in Fig. \ref{fig:DPWMwaveformsGecko}). To resolve this issue, the CM-smoothing is proposed. The idea is to smoothen the sharp edge of the CM voltage (black curve in AC-side voltages graph in Fig. \ref{fig:DPWMwaveformsGecko}) in order to provide a smooth transition from one phase clamping to the other. As a result, the capacitor voltages will be reshaped such that $i_{Ca} = C \frac{dV_{Ca}}{dt} \rightarrow 0$ just before the clamping, thus the condition above will be satisfied.

\begin{sidewaysfigure}[h]
	\centering
	\includegraphics[width=\textwidth]{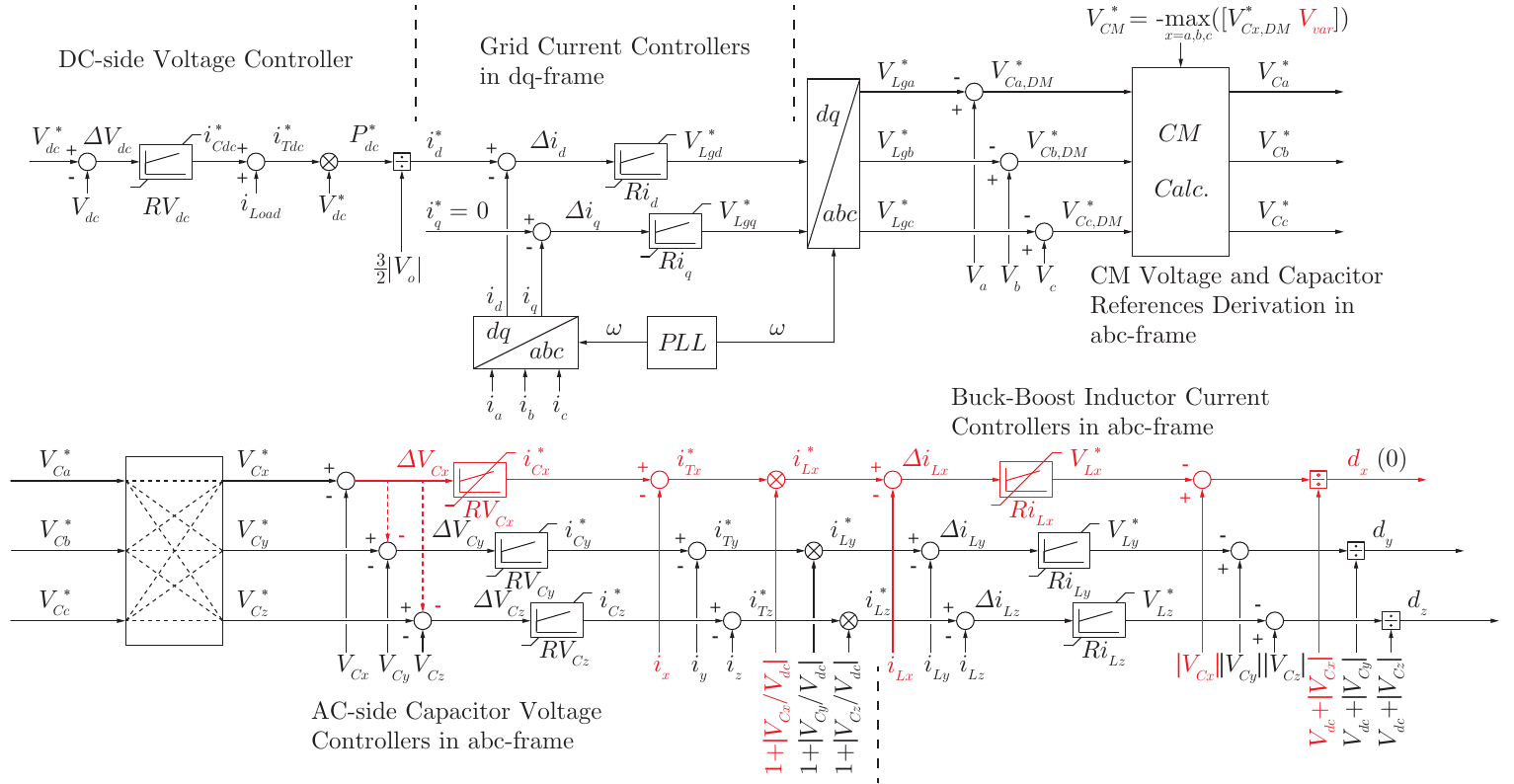}
	\caption{Updated rectifier control system employing DPWM with four cascaded controllers, where two in dq- and two in abc-frame, where one phase is clamping at a time. CM-smoothing algorithm applied}
	\label{fig:ControllerDPWMupdated}
\end{sidewaysfigure}

For a CM-smoothing to take place, the control structure in Fig. \ref{fig:ControllerDPWM} has to be slightly modified. These modifications are depicted in Fig. \ref{fig:ControllerDPWMupdated} (red highlights). First, the smoothing threshold \textcolor{red}{$V_{var}$} is defined such that the CM voltage near the sharp edge is adapted. In fact, the CM voltage near the sharp edge is just fed forward from the predefined reference shape. This CM feedforward is applied for a short time interval corresponding to the transition from one phase clamping to the other. During this time interval all three phases are operated, which can also be seen from the subsequent control structure in red. Eventually, after the phase clamping transition ends and the CM threshold is deactivated, the control structure returns back to standard DPWM operation. This can also be seen from the \textcolor{red}{$d_x$} (0) sign at the output of the modulator, where \textcolor{red}{$d_x$} and (0) correspond to the transition and clamped intervals respectively.

\begin{figure}[h]
	\centering
	\includegraphics[width=0.7\textwidth]{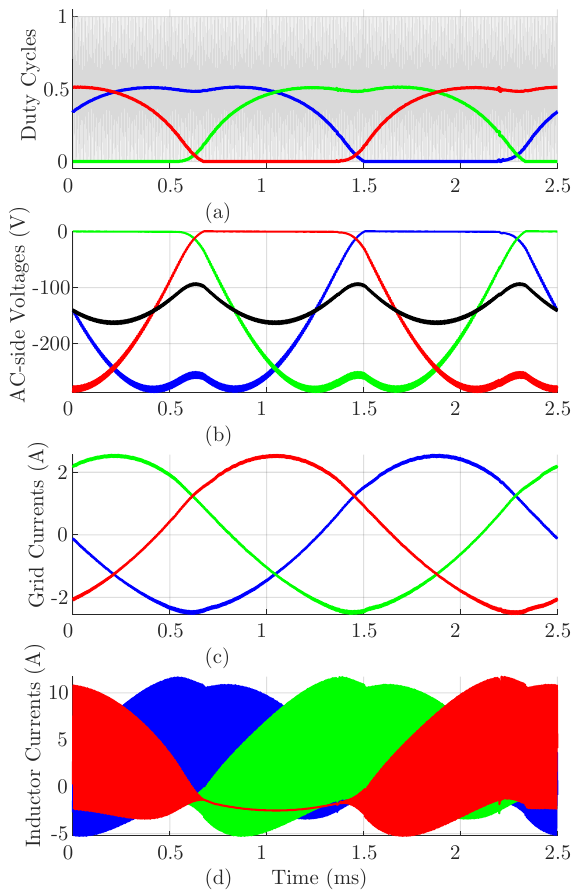}
	\caption{Gecko-simulated steady-state rectifier waveforms with updated controller}
	\label{fig:DPWMwaveformsGeckoUpdated}
\end{figure}

Applying the modified control structure in Gecko-Simulink simulation results in a steady state three-phase inverting buck-boost converter waveforms shown in Fig. \ref{fig:DPWMwaveformsGeckoUpdated}. One can immediately see that the high frequency ringing issue has vanished due to CM-smoothing, which results in a smooth transition from one phase clamping to the other. This, however, comes at the price of a short time intervals, where all three phases are operated. Taking into account that mentioned interval is very short, the CM-smoothing algorithm offers a good solution.

\clearpage

\begin{figure}[h]
	\centering
	\includegraphics[width=\textwidth]{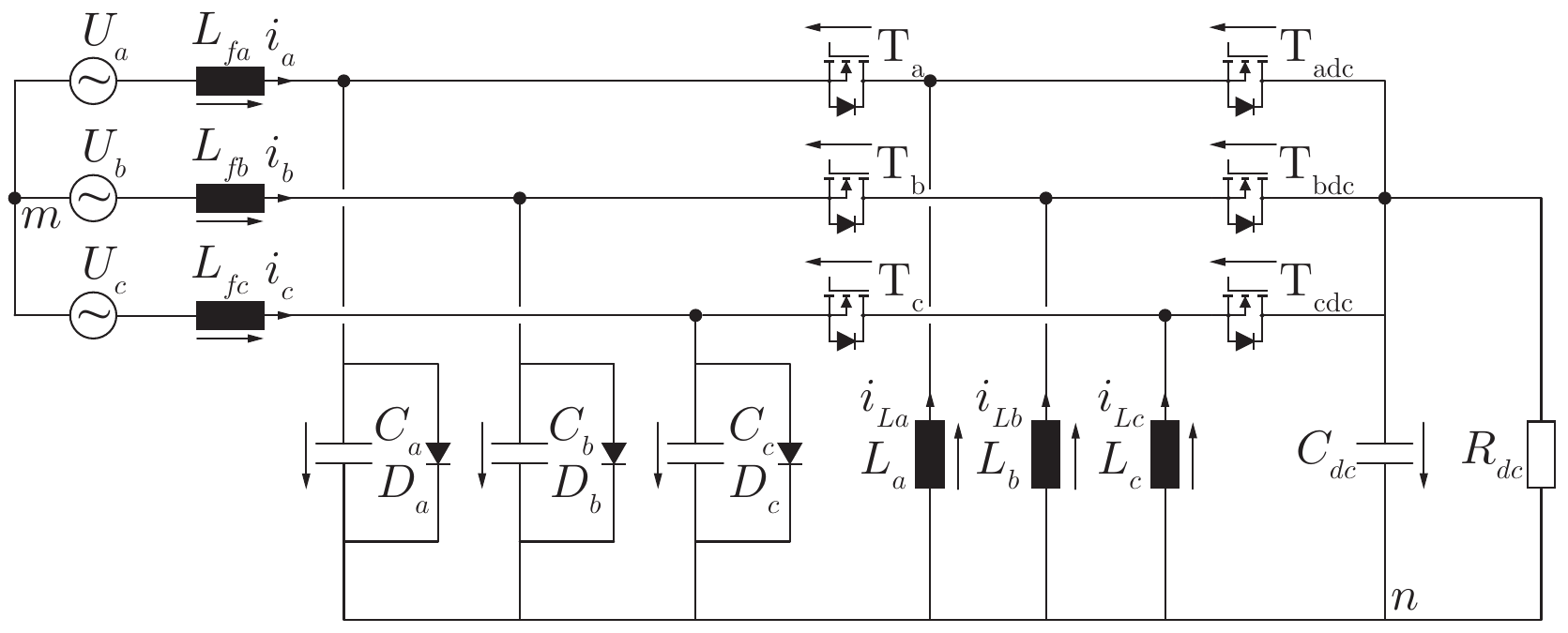}
	\caption{Grid-interfaced three-phase inverting buck-boost converter with anti-parallel diodes in a rectifier configuration}
	\label{fig:InverswandlerRectifierDiodes}
\end{figure}

Apart from that, as a precaution measure, anti-parallel diodes are connected across the AC-side capacitors. By doing so, one ensures that the AC-side capacitors' voltages are kept strictly negative in case the CM-smoothing algorithm fails to operate properly. For instance, in case the aforementioned ringing persists around zero after clamping, the transients could result in positive voltages across the AC-side capacitor of the clamped phase. This positive voltage is then discharged over the anti-parallel diode. The resulting topology modification can be seen in Fig. \ref{fig:InverswandlerRectifierDiodes}.

\section{Rectifier Summary}

To sum up, this chapter explored the applicability potential of the three-phase inverting buck-boost converter topology in an EHA rectifier application within the context of a More Electric Aircraft. First, as the rectifier performance heavily depends on the modulation scheme employed, three modulation candidates have been studied and compared with respect to various performance metrics. As a result, Discontinuous PWM (DPWM) has been selected. Note that the modulation analysis and results are topology- rather than application-specific, meaning that DPWM is expected to outperform its counterparts in inverter application as well.

Second, the control structure has been developed for the three-phase inverting buck-boost converter as a rectifier with DPWM. Several arising problems, such as high frequency ringing and grid current quality have been identified and resolved. Some parts of the control structure, for example the inner cascaded controllers and CM voltage derivation, are applicable in the inverter application as well. This is also true for the CM-smoothing algorithm - a major DPWM enabler. In general, when it comes to control, inverter application cases are usually easier than rectifier ones due to the fact that these are usually inherently stable in open-loop. This topology-specific feature of open-loop stability is very useful for hardware commissioning. Theoretical stability analysis and extensive simulations show that three-phase inverting buck-boost converter also possesses open-loop stability in inverter configuration, meaning that with this topology-specific feature the commissioning of a hardware prototype will be much easier.

Overall, rectifier analysis provided a thorough understanding of the topology, its operation and control. Important topology-specific features have been identified.
\chapter{Three-Phase Inverting Buck-Boost Converter as a Fuel Cell Powered High-Speed Motor Drive}

After finishing the analysis of the three-phase inverting buck-boost converter as a rectifier, this chapter explores a new application, namely a Fuel Cell (FC) powered high-speed motor drive within the context of a Fuel Cell Vehicle (FCV). In section one, a new application and the respective electrical specifications are introduced. Section two deals with the inverter control structure for DPWM. Majority of the results presented in this section are inherited from the rectifier analysis due to latter's applicability. Next, after defining the application specifications, modulation and control, section three starts the hardware discussion by revealing the selection and design procedures of the main converter power components. Section four addresses secondary components, such as measurements and auxiliary supplies. Printed Circuit Board (PCB) layout design is discussed in section five, which finally leads us to the virtual prototype and its performance in section six.

\section{Inverter Application and Electrical Specifications}

\begin{figure}[h]
	\centering
	\includegraphics[width=\textwidth]{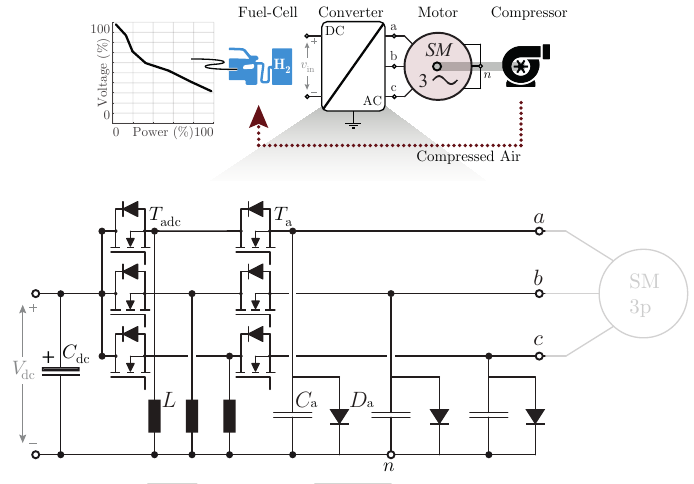}
	\caption{Three-phase inverting buck-boost converter as an inverter in FCV application}
	\label{fig:InverswandlerInverterWithFuelCell}
\end{figure}

Regarding the application for a Fuel Cell Vehicle (FCV), the authors of \cite{Antivachis} briefly explain the operating principle of the FCV and propose the application, where the converters with buck-boost capability could be employed. Fuel Cell Vehicle (FCV) is a vehicle that uses the energy from the chemical reaction in the fuel cell for propulsion. The reaction requires oxygen, which can be extracted from the surrounding air. Thus, there are two distinct motors - main high-power motor for vehicle propulsion and an auxiliary low-power motor that drives the air compressor to sustain the chemical reaction. As it can be seen from Fig. \ref{fig:InverswandlerInverterWithFuelCell}, the output voltage of the fuel cell heavily depends on the extracted power, which results in a very wide input voltage range for both motor drives. In this context, three-phase inverting buck-boost converter topology could be a good choice for a compressor driver setup (Fig. \ref{fig:InverswandlerInverterWithFuelCell}). Indeed, compressor driver is a low-power application with a wide input DC voltage range - exactly the niche of the three-phase inverting buck-boost converter. Low number of active components and power conversion within a single stage could become the competitive advantages over conventional counterparts. Moreover, three-phase inverting buck-boost converter provides an inherent filtering of the output AC voltages through its AC-side capacitors, which is good for reducing the high frequency harmonic content and associated losses in the motor of the compressor.

The three-phase inverting buck-boost converter topology and the electrical specifications for an inverter operation within the FCV application are shown in Fig. \ref{fig:InverswandlerInverterWithFuelCell} and Table \ref{t:SpecsInverter}. Further derivation of the component electrical specifications will be provided in the corresponding sections below.

\begin{table}[H]
	\centering
	\caption{Inverter application specifications for a FC powered high-speed motor drive}
	\label{t:SpecsInverter}
	\begin{tabular}{ | P{2cm} || P{2cm} | P{1.5cm} | P{8cm} | }
		\hline
		Parameter & Value & Unit & Param. Explanation (Application)\\
		\hline\hline
		$P$ & $1000$ & W & Nominal inverter power rating\\
		$\hat{V}_{o}$ & $80$ & V & Nominal output AC phase voltage amplitude\\
		$V_{dc}$ & $[80-240]$ & V & Input DC voltage range\\		
		$f_o$ & $1$ & kHz & Nominal output AC frequency\\
		\hline
	\end{tabular}
\end{table}

\section{Inverter Control for DPWM}

Regarding the three-phase inverting buck-boost converter control for a motor drive application, a standard inverter control is reviewed and updated to account for the specific converter topology and the selected modulation at hand. For instance, recall the example of a VSI directly connected to Permanent Magnet Synchronous Motor (PMSM). There, the control system comprised two cascaded controllers in a rotor-oriented dq-frame. The outer speed controller provided a reference for a q-component of the stator current, as this q-component is responsible for torque generation. Current dq-component controllers, in turn, provided a references for a dq-components of the inverter AC-side voltages. Since in the example at hand the VSI is directly connected to the PMSM, AC-side voltages can be directly generated through the high-frequency operation of semiconductors. Thus, the voltage reference in dq-frame is translated back to abc-frame and fed to the modulators of the respective phases.

As it can be seen from Fig. \ref{fig:InverswandlerInverterWithFuelCell}, in considered application one has a three-phase inverting buck-boost converter instead of a VSI directly connected to the PMSM. No explicit filter is placed, thanks to the structure of the three-phase inverting buck-boost converter topology with its AC-side capacitors smoothing the motor voltages and providing inherent filtering. This filtering, however, comes at the price of an additional control effort. Surely, motor nodes $a$, $b$ and $c$ in Fig. \ref{fig:InverswandlerInverterWithFuelCell} are not directly connected to the switched nodes of the three-phase inverting buck-boost converter, as it was the case in the example of VSI. Thus, at least one additional control stage for the AC-side voltages is needed. With this being said and bearing in mind aforementioned discussions on outer speed and motor current controllers, this section addresses only the inner controller stages, specific to the three-phase inverting buck-boost converter. Outer controllers are similar and therefore not addressed here.

\begin{figure}[h]
	\centering
	\includegraphics[width=\textwidth]{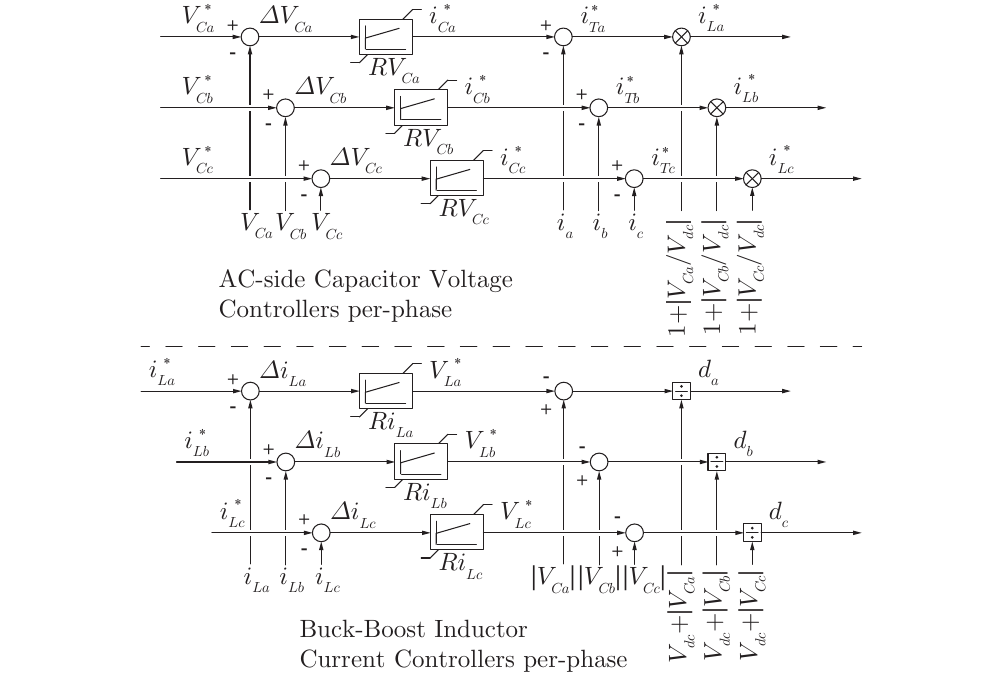}
	\caption{Inner structure of the proposed inverter control system employing DPWM}
	\label{fig:InverterControllerDPWM}
\end{figure}

When it comes to the inner controller stages, those can be inherited from the rectifier analysis, since the control flow in both cases is the same. Surely, in both cases two outermost controllers provide the AC-side voltage references to be tracked. Since in rectifier case it was done by employing two cascaded controllers for AC-side voltages and buck-boost inductor currents, the same is applicable for inverter. Fig. \ref{fig:InverterControllerDPWM} depicts these two cascaded controllers. For details refer to previous chapter.

\section{Component Selection and Design}

This section addresses the selection and design procedures of the main three-phase inverting buck-boost converter components. Given system specifications, relevant for component design, are provided in Table \ref{t:SpecsInverterGiven}.

\begin{table}[H]
	\centering
	\caption{Given system specifications for the inverter application}
	\label{t:SpecsInverterGiven}
	\begin{tabular}{ | P{2cm} || P{2cm} | P{1.5cm} | P{8cm} | }
		\hline
		Parameter & Value & Unit & Param. Explanation (Given)\\
		\hline\hline
		$f_s$ & $300$ & kHz & Switching frequency\\		
		Device & GaN 600V & - & Semiconductor: Infineon GaN 600V, 55mOhm\\
		No. Dev. & 2 & - & Number of parallel devices\\
		$\eta$ & $>95\%$ & - & Minimum efficiency\\
		$\rho$ & - & kW/l & Power density as compact as possible\\
		\hline
	\end{tabular}
\end{table}

\subsection{Semiconductors}

$\bullet$ \textbf{Choice of the device} 

From the Table \ref{t:SpecsInverterGiven} it is known that the switches were predefined for the motor drive application at hand. The selected device is a new generation 600V, 55 mOhm Infineon GaN Gate Injection Transistor (GIT) \cite{IGLD60R070D1}. Previous works have already studied older generations of GaN devices \cite{GaNGIT}, \cite{OriginCoss}, \cite{Heller}. As a result of those works, there are several reasons for the selection, which can be generally combined as follows: 

$a$. New GaN 600V devices provide a cost-effective solution in many 400V key applications \cite{GaNGIT}. 

$b$. New GaN 600V devices help increasing the system performance in terms of achievable efficiencies and power density \cite{GaNGIT}. 

The application at hand lies exactly in the prescribed voltage vicinity with its maximum required blocking voltage being $V_{dc,max} + 2 \times \hat{V}_{o} = 240\text{V} + 2 \times 80\text{V} = 400$V.

$\bullet$ \textbf{Number of parallel devices} 

Table \ref{t:SpecsInverterGiven} also states that the number of parallel devices equals 2. This decision was influenced by practical concerns. First, the upper limit on the number of parallel devices is set by the size concerns of the half-bridge. Indeed, due to non-zero size of each individual device, using many devices in parallel could increase the gate loop inductance (depending on PCB layout), thus decreasing the switching speed, increasing switching losses or even result in non-symmetric operation of parallel devices. Of course one could suggest increasing the number of separate gate drivers, however, this approach would significantly increase the occupied PCB area, thus compromising the power density, which is set as a primary goal of the prototype. Second, the lower limit on the number of parallel devices is set by the current-carrying capability of one device. From the datasheet \cite{IGLD60R070D1} it is seen that the maximum pulsed drain-source current is around 60A, while the maximum DC drain-source current is at most 15A, provided that the junction temperature is as low as 25\degree C. On the other hand, from equations (\ref{eq:Duty_Ipkpk}) and (\ref{eq:InductorACsideSwitchRelation}) one can derive the peak inductor current as a function of input DC voltage and time:

\begin{equation}
\setlength{\jot}{10pt} 
\label{eq:ILpk}
\begin{split}
&I_{La,ripple}(V_{dc},t) = \frac{1}{2 f_s L} \Big(\frac{V_{dc} |V_{Ca}(t)|}{V_{dc} + |V_{Ca}(t)|}\Big)
\\
&I_{La,avg}(V_{dc},t) = \Big(I_a(t) + C \frac{dV_{Ca}(t)}{dt}\Big) \Big(1 + \frac{|V_{Ca}(t)|}{V_{dc}}\Big)
\\
&I_{La,pk}(V_{dc}, t) = \underbrace{\Bigg[\underbrace{\frac{1}{2 f_s L} \Big(\frac{V_{dc} |V_{Ca}(t)|}{V_{dc} + |V_{Ca}(t)|}\Big)}_{\text{Inductor current ripple}} + \underbrace{\Big(I_a(t) + C \frac{dV_{Ca}(t)}{dt}\Big) \Big(1 + \frac{|V_{Ca}(t)|}{V_{dc}}\Big)}_{\text{Inductor current average}}\Bigg]}_{\text{Inductor current envelope}}
\\
\end{split}
\end{equation}

After evaluating this equation for DPWM over a fundamental period for various DC voltages, it was observed that $I_{L,pk}$ varies between 25-32A depending on DC voltage. Based on experience with old generation devices this maximum current range was considered too much for a single device, thus the lower limit on the number of parallel devices was selected to be 2. Overall, selecting two parallel devices is well justified.

$\bullet$ \textbf{Antiparallel diodes} 

Apart from switches, anti-parallel diodes have to be selected and placed near each individual device. The reason is, as it can be seen from typical reverse characteristics in \cite{IGLD60R070D1}, the internal anti-parallel diode of the switch exhibits a very high reverse drain-source voltage as a function of a negative gate-source voltage, meaning that the device will have significant losses in the reverse conduction mode. This problem has also been mentioned in \cite{Heller}, where the explicit SiC 600V Schottky diode from CREE has been used \cite{C3D1P7060Q}. Since both voltage and current ratings of this diode comply with application requirements, it will also be used in present work.

$\bullet$ \textbf{Loss map}

\begin{figure}[h]
	\centering
	\begin{subfigure}[H]{0.48\textwidth}
		\includegraphics[width=\textwidth]{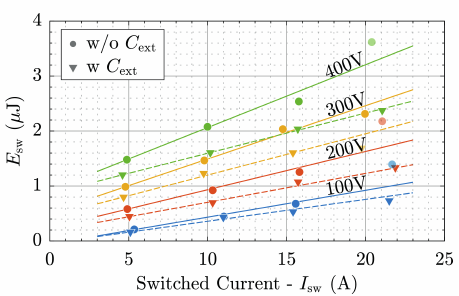}
		\caption{Soft-switching loss map \cite{Heller}}
		\label{fig:SoSLossMap}
	\end{subfigure}
	~
	\begin{subfigure}[H]{0.48\textwidth}
		\includegraphics[width=\textwidth]{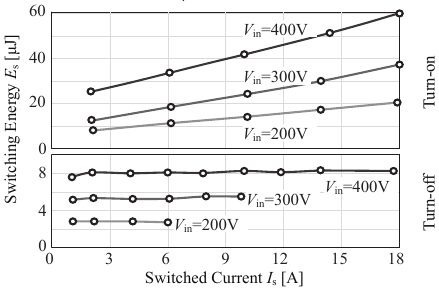}
		\caption{Hard-switching loss map \cite{GaNGIT}}
		\label{fig:HaSLossMap}
	\end{subfigure}
	\caption{Combined loss map based on old measurements}
	\label{fig:SoSHaSLossMap}
\end{figure}

Having selected the switches and diodes, their performance needs to be estimated for a subsequent cooling system design. Expected losses can be estimated using a loss map and on-state resistance for soft-switching and hard-switching respectively. While on-state resistance is readily available in the datasheet, the loss map is usually obtained by measuring hard- and soft-switching losses. For the new devices there exists no loss map yet. To overcome this issue, the loss maps from older generation devices with the same ratings will be applied. Assuming that the performance of a new devices is the same or better, a conservative estimation of the semiconductor losses can be obtained. Soft-switching loss map has been measured in \cite{Heller} using the calorimetric measurement setup. Hard-switching loss map has been measured in \cite{GaNGIT} using double pulse test. Fig. \ref{fig:SoSHaSLossMap} shows the extracted maps. Switched energies are calculated per half-bridge and per transition.

On a separate note, a new measurement board has been designed within the scope of this thesis to measure both hard and soft-switching of the new generation devices. Calorimetric measurement setup was selected due to its higher accuracy. \cite{Heller} has been used as a main reference. The measurement board is addressed in Appendix A.

$\bullet$ \textbf{Semiconductor loss evaluation}

\begin{figure}[H]
	\centering
	\includegraphics[width=0.95\textwidth]{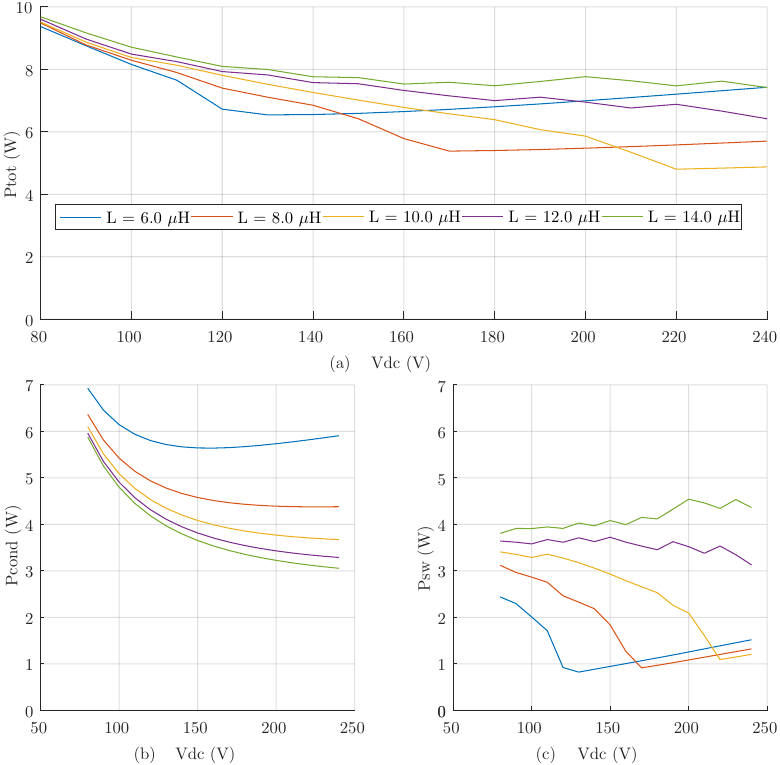}
	\caption{Theoretical semiconductor losses per half-bridge per fundamental period as a function of DC voltage and for various inductances. Total semiconductor losses (a); Conduction losses (b); Switching losses (c)}
	\label{fig:Semi_Dimensioning}
\end{figure}

Using the combined loss map from Fig. \ref{fig:SoSHaSLossMap}, the electrical specifications and the DPWM waveforms from subsection \ref{DPWM}, the semiconductor losses can be estimated for various inductance values. Indeed, as it was shown in subsection \ref{DPWM} and in equation (\ref{eq:ILpk}), the buck-boost inductor current waveform (and, consequently, semiconductor losses) depend on the inductance value, therefore the semiconductor losses are estimated as a function of DC voltage for different inductance values (Fig. \ref{fig:Semi_Dimensioning}). 

The nominal output voltage/power of $\hat{V}_o = 80$V, $P = 1$kW is assumed. In short, at nominal voltage/power both peak and rms inductor currents $I_{pk}$ and $I_{rms}$ reach their maximum levels and since the very same current flows through the half-bridge, one expects the maximum semiconductor losses as well. 

Fig. \ref{fig:Semi_Dimensioning}a, \ref{fig:Semi_Dimensioning}b and \ref{fig:Semi_Dimensioning}c show the semiconductor total, conduction and switching loss components per half-bridge per fundamental period as a function of the DC voltage ranging from 80V to 240V. The considered inductance range is between 6$\mu$H and 14$\mu$H. The same range will be considered in the subsequent inductor design discussion. 

From Fig. \ref{fig:Semi_Dimensioning}b one can see that the semiconductor conduction losses tend to drop as the DC voltage increases. This can be attributed to the fact that the average inductor current drops as a function of DC voltage, as it is shown in equation (\ref{eq:ILpk}). The drop in conduction losses is not steady though. Moreover, it can be seen that for low inductance values the conduction losses tend to grow again after surpassing some DC voltage. This behavior comes from the non-linear contribution of the inductor ripple current, which is also inversely proportional to the inductance value. 

From Fig. \ref{fig:Semi_Dimensioning}c the first thing to be noticed is that with larger inductance the switching loss tends to grow. This can be explained by the reduced inductor current ripple (equation (\ref{eq:ILpk})), which is not enough to achieve the switched node current reversal within every switching period to facilitate pure soft-switching. Another thing to be noticed is that with growing DC voltage the switching loss tends to drop first, followed by a steady rise at lower pace. This behavior is expected, since with growing DC voltage the inductor average current decreases and the ripple current increases, meaning that switched node current reversals within every switching period are becoming more frequent. After reaching the DC voltage of predominantly soft-switching, further increase in DC voltage (and current ripple according to (\ref{eq:ILpk})) results in steady rise of switching losses due to the fact that soft-switching instances occur at higher instantaneous currents and voltages.


\subsection{Capacitors}
$\bullet$ \textbf{AC-side capacitors} 

Regarding AC-side capacitors, a standard design constraint applies for those. On one hand, the maximum capacitance value is limited by the reactive power consumption. On the other hand, minimum capacitance value is limited by the AC-side voltage ripple. Assuming that one wants to limit the capacitive current consumption to below 20\% of the AC-side load current $\hat{I}_o$, maximum capacitance can be found as follows:

\begin{equation}
\label{eq:Cac}
C_{AC,max} = 0.2 \hat{I}_o / \omega_o \hat{V}_o = 3.3 \text{ $\mu$F}
\end{equation}

Here $\hat{I}_o$ and $\hat{V}_o$ correspond to phase current and voltage amplitudes with $\hat{I}_o = 2P/3\hat{V}_o = 8.33$A. Using the obtained value of $C_{AC,max}$ in simulations one can observe a voltage ripple within $\pm3$\%, which is more than enough for the control system to operate properly. Thus, AC-side capacitance was selected to be 3$\mu$F. AC-side capacitors were selected to be ceramic (C5750X7T2W105K250KA). In capacitor selection process several factors have been accounted for, including the fact that for X7T ceramics the capacitance drops significantly with applied voltage.

Apart from the output AC-side capacitors, ceramics were also used as commutation capacitors. Higher voltage rating ceramic capacitors were selected for commutation (KTS501B564M55N0T00), as they should withstand $V_{dc}+2\hat{V}_{o}$, which is greater than $2\hat{V}_{o}$ given for output capacitors.

$\bullet$ \textbf{DC-side capacitors} 

Regarding the DC link design, there are not so many explicit requirements for the DC capacitors. In terms of control, extensive simulations and Matlab theoretical calculations show that DC voltage ripple is kept within $\pm5$\% (which is more than enough for a proper control) already at 30$\mu$F of capacitance. This capacitance can be realized using several small electrolytic capacitors for example. Bering in mind these results as well as already available cooling system and inductor designs, the DC link electrolytics are selected to provide as much capacitance as possible while having a good overall converter fill factor. 


On the other hand, high-frequency ripple in the electrolytic DC link could become a problem due to high ESR in electrolytics at high frequencies. This problem was alleviated by placing DC ceramic capacitors near the DC-side switches of the phases. By doing so one ensures that high-frequency ripple current of the DC link flows predominantly through these ceramic capacitors that, in turn, have much smaller ESR. There are two reasons behind high frequency current redistribution. First, at high frequencies the impedance of ceramic capacitors is much smaller than that of an ESR-dominated electrolytics. Second, since ceramics are much smaller, they can be placed much nearer to the switches on the PCB, meaning that some additional parasitic inductance between electrolytics and switches (due to longer current path) will further increase the impedance of the electrolytics branch of the DC link. With these, DC link values in the final design are as follows: $6$x$10$$\mu$F of electrolytic capacitors (EKXF251ELL100MJC5S) and $10$$\mu$F of ceramic capacitors (C5750X7T2W105K250KA).

\subsection{Inductor dimensioning point}

Regarding the buck-boost inductor, before selection of appropriate inductance value and corresponding design, the dimensioning point within the DC, AC voltage and output power ranges has to be determined, since both inductance sensitivity analysis and design evaluation depend on the operating point. As a good practical rule, the dimensioning point is usually selected based on area product estimation. Two physical limits - core saturation and winding thermal (current density) limit - are accounted for in the area product. Since the inductor core should not saturate and winding should not overheat irrespective of the instantaneous operating point within electrical specifications, the inductor has to be designed for the worst cases both in terms of core and window areas. Consequently, core $A_c$ times winding window $A_w$ area product $A_c A_w$ provides a good volume estimation of the prospective inductor that satisfies both saturation and current density limits. Core and winding window ares $A_c$ and $A_w$, in turn, depend on peak and rms inductor currents respectively. In short, this dependency can be captured by the following equation:

\begin{equation}
\label{eq:AcAw}
A_c A_w \sim L I_{pk} I_{rms}
\end{equation}

Bearing in mind the discussion above, inductor dimensioning point can be determined by evaluating $I_{pk}$ and $I_{rms}$ for various DC voltages and for nominal AC-side voltage/power ($\hat{V}_o = 80$V, $P = 1$kW). The reason for nominal AC-side voltage/power consideration can be seen from equation (\ref{eq:ILpk}). For instance, in equation (\ref{eq:ILpk}) if one analyzes $I_{La,pk}$ as a function of AC-side voltage $V_{Ca}$ and current $I_a$ (proportional to power), one observes that these parameters are in direct relation, meaning that with increasing $V_{Ca}$ and $I_a$ the $I_{La,pk}$ only increases. The same holds true for $I_{La,rms}$. Thus, maximum current product $I_{pk} I_{rms}$ and, consequently, maximum area product $A_c A_w$ occurs at nominal AC-side voltage/power. The same cannot be said about DC voltage relation, since with increasing DC voltage the inductor current ripple term increases, while the average term decreases (\ref{eq:ILpk}). 

\begin{figure}[h]
	\centering
	\includegraphics[width=0.9\textwidth]{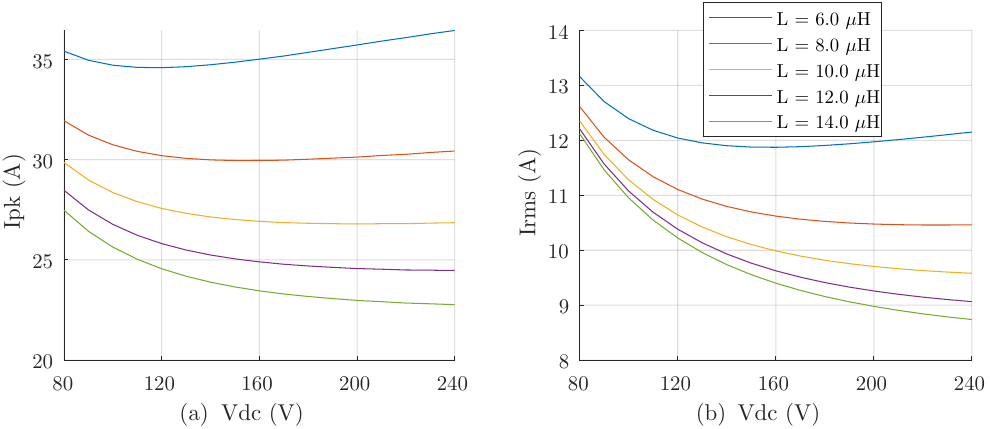}
	\caption{Inductor peak and rms currents versus DC voltage for different inductance values, $\hat{V}_o = 80$V and $P = 1$kW. $I_{pk}$ (a); $I_{rms}$ (b)}
	\label{fig:Ipk_Irms}
\end{figure}

The dimensioning point would then be the nominal AC-side voltage/power and the DC voltage, which results in the largest product. Of course it is possible that $I_{pk}$ and $I_{rms}$ reach their maximums at different DC voltages or that design is compromised by, for example, core losses that were not considered in the area product approach. Nevertheless, the area product approach is a good starting point as it provides a rough estimation of the required inductor dimensions.

In Fig. \ref{fig:Ipk_Irms} one can see $I_{pk}$ and $I_{rms}$ as a function of DC voltage for various inductance values. From Fig. \ref{fig:Ipk_Irms} it can be seen that for most of the inductance values (apart from a very small value of 6 $\mu$H) both $I_{pk}$ and $I_{rms}$ reach their maximums at $V_{dc}=80$V. Thus, 80V DC input voltage is selected as dimensioning point for the subsequent sensitivity analysis and inductor design.

\subsection{Inductor sensitivity analysis, evaluation, design}

$\bullet$ \textbf{Sensitivity analysis} 

\begin{figure}[h]
	\centering
	\captionsetup{justification=centering}
	\includegraphics[width=0.7\textwidth]{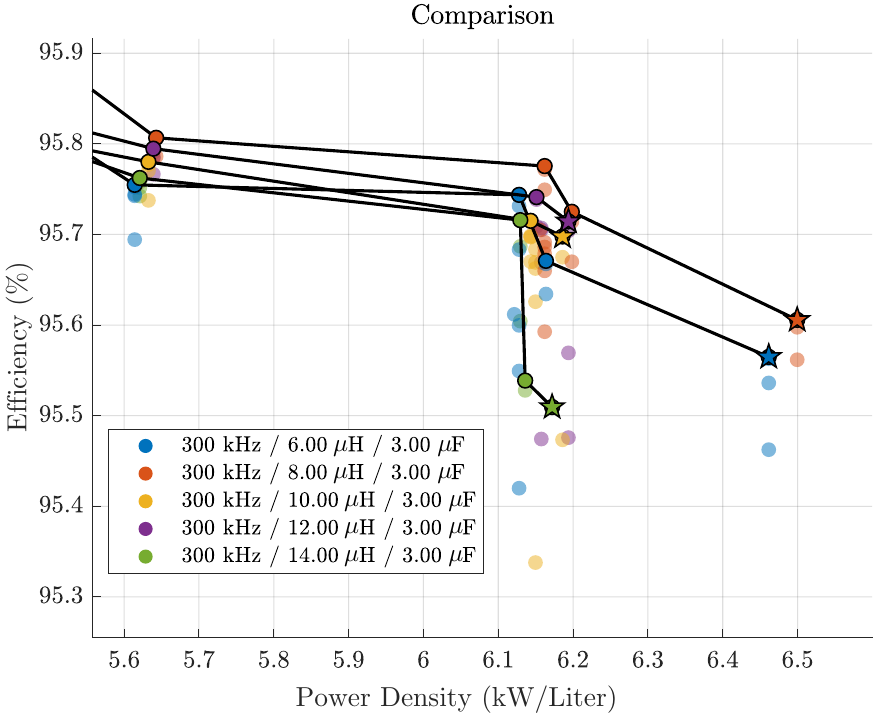}
	\caption{Theoretical converter efficiency - power density graph for $V_{dc} = 80$V, $\hat{V}_o = 80$V, $P = 1$kW. Inductance-based sensitivity analysis}
	\label{fig:SensitivityAnalysis}
\end{figure}

With the dimensioning point the electrical specifications set is now full and one can conduct a small sensitivity analysis to find out which inductance value and inductor design are the most suitable. In this analysis, DC link / AC output capacitances were set to 70$\mu$F / 3$\mu$F  in compliance with the capacitor subsection. Having only inductance as a degree of freedom makes the influence of various inductors on theoretical efficiency and power density figures directly visible. Fig. \ref{fig:SensitivityAnalysis} depicts the sensitivity analysis results for inductance values in the range of [6..14]$\mu$H.

Each point in Fig. \ref{fig:SensitivityAnalysis} corresponds to a different inductor realization with five colors corresponding to the analyzed inductance values. Since one of the objectives of converter design is power density, only high-density realizations are depicted.

From Fig. \ref{fig:SensitivityAnalysis} one could notice that for each realization on a Pareto-front (connected with a line), there are several realizations with same power density, but lower efficiency. These points correspond to the inductor designs with the same core as in Pareto-front design, but with slightly different other parameters, such as number of turns, number of strands, air gap length etc. Variations in these parameters do not affect inductor volume (and hence power density), but reduce efficiency. Therefore sub-Pareto designs can be discarded from further considerations, leaving about 7 to 10 interesting designs remaining in each inductance value group.

$\bullet$ \textbf{Converter evaluation over the whole operating range} 

\begin{figure}[h]
	\centering
	\captionsetup{justification=centering}
	\includegraphics[width=\textwidth]{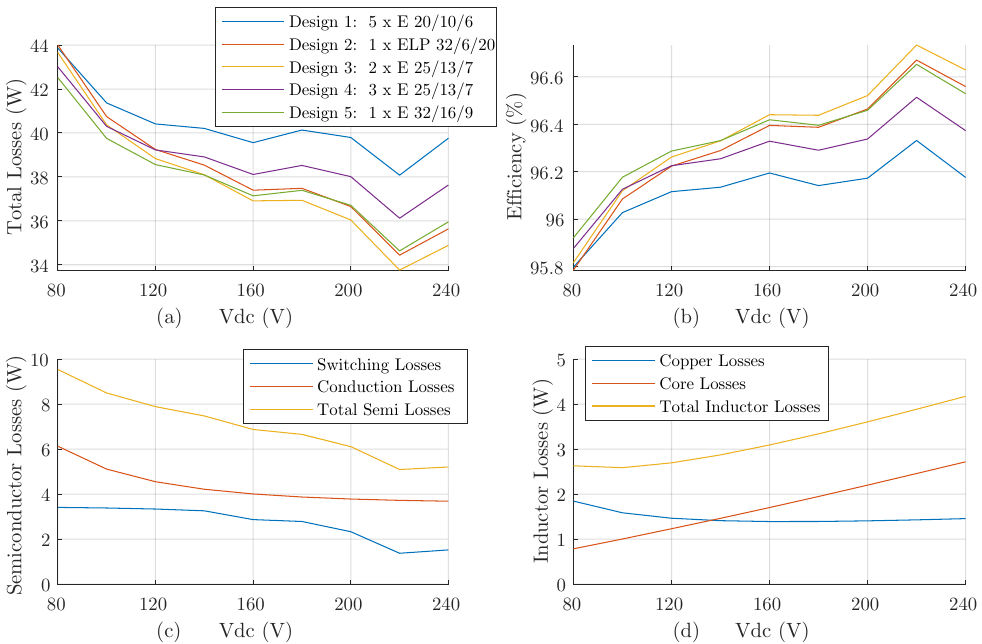}
	\caption{DC voltage sweep of five most compact designs in 10$\mu$H set. Theoretical converter losses (a); Theoretical efficiency (b); Semiconductor losses (independent of inductor design) (c); Inductor losses for Design 2: 1 x ELP 32/6/20 (d)}
	\label{fig:VdcSweep10u}
\end{figure}

After obtaining five sets of possible inductor realizations for each inductance value, these sets have to be evaluated over the whole input voltage range $V_{dc}$ to ensure the inductor realizations do not hit saturation or overheating at some other operating point. Example evaluation graphs for five most compact realizations in the 10$\mu$H set are depicted in Fig. \ref{fig:VdcSweep10u}.

\begin{figure}[h]
	\centering
	\captionsetup{justification=centering}
	\includegraphics[width=\textwidth]{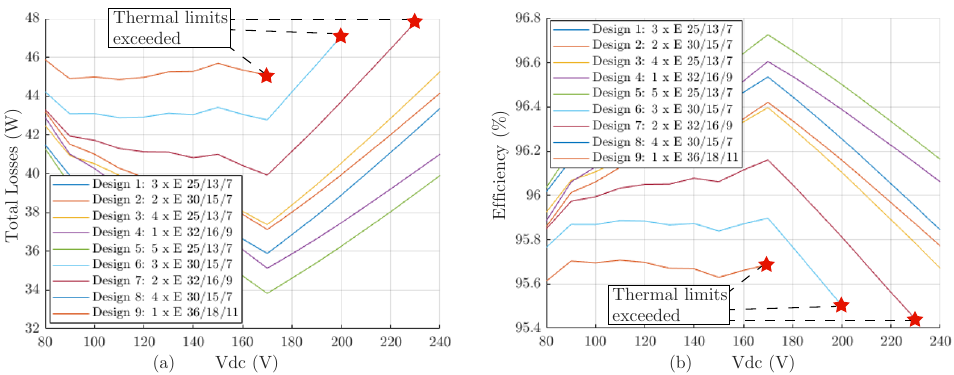}
	\caption{DC voltage sweep of nine most compact designs in 8$\mu$H set. Theoretical converter losses (a); Theoretical efficiency (b)}
	\label{fig:VdcSweep8u}
\end{figure}

From Fig. \ref{fig:VdcSweep10u}a-b it can be seen that none of the considered inductor realizations hits thermal or saturation limits as the graphs are evaluated for the whole DC voltage range. This, however, cannot be said for some inductor realizations in 6$\mu$H and 8$\mu$H sets. For example, in Fig. \ref{fig:VdcSweep8u} one can see that loss graphs are not defined throughout DC voltage range for some realizations. In case of 8$\mu$H this is due to thermal limit.

As noticed from Fig. \ref{fig:VdcSweep10u}d and Fig. \ref{fig:VdcSweep8u}, the electrical operating point of the largest area product ($V_{dc} = 80$V, $V_{o} = 80$V, $P = 1$kW) is not the same as the operating point with the worst case inductor losses. The reason of this discrepancy is the fact that the area product considers the \textit{current density limit of the copper} and the \textit{saturation of the core}, but does not take into account the \textit{thermal limit of the core} (core losses). Indeed, in case of the 10$\mu$H designs at $V_{dc}=80$V both $I_{pk}$ and $I_{rms}$ reach their maximum, therefore aforementioned limits are hit. Nevertheless, regarding core losses, the frequency contributions of the current are important. At $V_{dc}=80$V the inductor current has a large LF component and a moderate HF component. The core loss is then within acceptable limits based on the following equation: 

\begin{equation}
\label{eq:Steinmentz}
P_{core} \sim f^\alpha B^\beta, \alpha, \beta > 1, 2
\end{equation}

Now if one goes to $V_{dc}=240$V, the inductor current has a moderate LF component and a high HF component. While the inductor current rms is still the same or even smaller than for $V_{dc}=80$V, now it has a large HF component that scales exponentially with frequency for core losses (\ref{eq:Steinmentz}). This is the reason of the maximum loss occurrence for $V_{dc}=240$V.

\clearpage

$\bullet$ \textbf{Inductor design}

\begin{figure}[h]
	\centering
	\includegraphics[width=\textwidth]{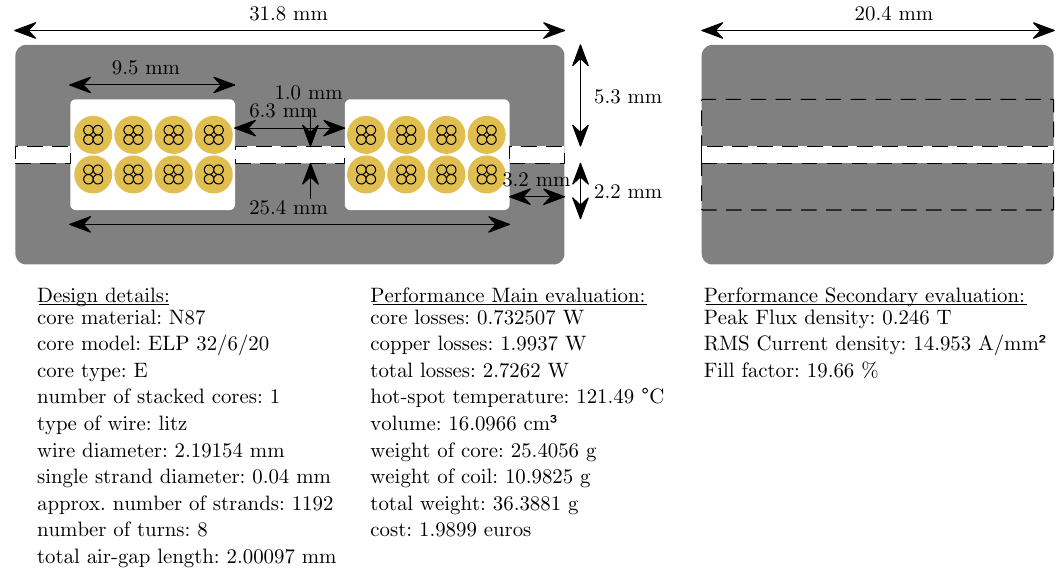}
	\caption{Cross-sectional view, details and performance evaluation of the selected inductor design featuring 10$\mu$H inductance}
	\label{fig:InductorDesign}	
\end{figure}

Regarding the remaining designs that do not hit thermal or saturation limits, preference is given to the designs which are compact and have an appropriate shape. Appropriate shape requirement comes from the objective of a good fill factor in a final converter prototype. After a design process, which took into account the cooling system, capacitors, inductors etc., a second design from 10$\mu$H set (1 x ELP 32/6/20) has been selected (Fig. \ref{fig:InductorDesign}). Performance of the selected design in terms of losses is shown in Fig. \ref{fig:VdcSweep10u}d. Table \ref{t:Passives} summarizes the passive components selected/designed so far.

\begin{table}[H]
	\centering
	\caption{Selected/designed passive components}
	\label{t:Passives}
	\begin{tabular}{ | P{3cm} || P{1.5cm} | P{1.5cm} | P{8cm} | }
		\hline
		Component & Value & Unit & Comment\\
		\hline\hline
		DC link & 70 & $\mu$F & 60$\mu$F - electrolytic, 10$\mu$F - ceramic capacitors\\	
		Inductor & 10 & $\mu$H & Designed buck-boost inductor value\\	
		AC output cap. & 3 & $\mu$F & Per-phase AC-side output ceramic capacitors\\					
		Comm. cap. & 2 & $\mu$F & Per-phase ceramic commutation capacitors\\		
		\hline
	\end{tabular}
\end{table}


\subsection{Cooling system design}

$\bullet$ \textbf{Cooling system dimensioning point}

Recall that Fig. \ref{fig:Semi_Dimensioning}a combines the conduction and switching loss figures. For cooling system dimensioning one should look for the worst case half-bridge losses within the DC voltage range and for 10$\mu$H inductance, selected before. However, as we can see from Fig. \ref{fig:Semi_Dimensioning}, the inductance value (within the defined range [6..14]$\mu$H) does not affect the point of maximum half-bridge losses, for which a perspective cooling system has to be designed. The maximum loss per half-bridge per fundamental period is expected to happen at $V_{dc} = 80$V, $V_{o} = 80$V, $P = 1$kW, and is estimated to be $P_{tot} = 9.5$W. 

From here, if one tries to estimate the losses per single device, one has to account for a non-symmetrical loss distribution between low-side and high-side devices, which can be deduced, for instance, from the fact that only low-side (AC-side) switches experience hard-switching in Fig. \ref{fig:DPWMwaveforms}. This distribution heavily depends on the operating point (DC voltage), thus to be on safe side a 35-65\% worst case relative distribution is assumed. With this, one device will experience maximum per-device power $P_{max,pd}$:

\begin{equation}
\label{eq:Pmax,pd}
P_{max,pd} = P_{tot} \times 0.65/N_{par} = 9.5 \times 0.65/2=3.1 \text{ W}
\end{equation}

Next, in \cite{OriginCoss} it was stated that for the semiconductor package at hand the case-to-ambient thermal resistance $R_{ca}$ is very high compared to the sum of case-to-heatsink and heatsink-to-ambient resistances ($R_{chs}/2 + R_{hsa}$), located in a parallel branch of a thermal model therein \cite{OriginCoss}. In fact, it was stated that $R_{chs}/2 + R_{hsa}$ is negligible compared to $R_{ca}$. The stated thermal resistance values were $R_{chs}/2 + R_{hsa} = 4.8/2 + 13 = 15.4$K/W. Even if we consider $R_{ca}$ to be only four times higher than the stated thermal resistance (which obviously cannot be considered much higher), we obtain the approximate lower limit of possible junction-to-ambient thermal resistance:

\begin{equation}
\label{eq:Rc-a}
R_{ca,min} = 4 \times (R_{chs}/2 + R_{hsa}) \approx 64 \text{ K/W}
\end{equation} 
 
For the device with maximum dissipation that results in a junction temperature of:

\begin{equation}
\label{eq:Rth_ja_max}
T_{j} = T_{amb} + P_{max,pd} \times (R_{jc} + R_{ca,min}) = 25 + 3.1 \times (1 + 64) = 227 \text{\degree C}
\end{equation}

Obtained value is way above the maximum junction temperature of $T_{j,max} = 150$\degree C according to the datasheet \cite{IGLD60R070D1}. Thus one can conclude that a cooling system is necessary.

$\bullet$ \textbf{Cooling requirements, thermal model} 

When it comes to cooling system design, one has to account for several factors and requirements. First, note that selected semiconductor devices are bottom-cooled. They have a large metal pad at the bottom with a good thermal connection to the junction, while top side is covered by a plastic case with a bad thermal conductivity. This device feature has to be accounted for by providing a good thermal connection from the bottom pad of the device to the heatsink. The design with numerous filled vias through PCB should suffice, as it was done in \cite{Heller}. There, numerous filled vias connected top and bottom sides of the PCB just under the device's pad to provide a low-resistance thermal connection between the device and heatsink. Electrical isolation was ensured by placing a thermal pad between PCB's bottom side and the heatsink.

\begin{figure}[h]
	\centering
	\captionsetup{justification=centering}
	\includegraphics[width=0.9\textwidth]{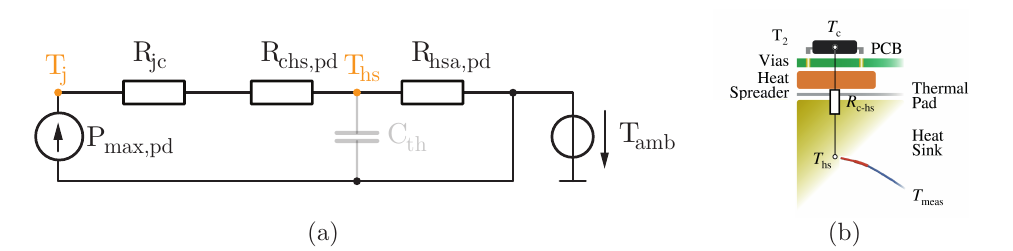}
	\caption{Thermal model used for per-device heatsink-to-ambient thermal resistance $R_{hsa,pd}$ estimation. Thermal model (a); Stack of elements forming the per-device case-to-heatsink thermal resistance $R_{chs,pd}$ \cite{OriginCoss} (b)}
	\label{fig:ThermalModel}
\end{figure}

Second, thermal resistance of the total cooling system should be below the maximum limit, which is determined through thermal model calculations assuming $T_{j,max} \leq 120$\degree C. The applied per-device thermal model is depicted in Fig. \ref{fig:ThermalModel}a. The model features the semiconductor device modeled as a power source, three thermal resistances (junction-to-case $R_{jc}$, per-device case-to-heatsink $R_{chs,pd}$ and per-device heatsink-to-ambient $R_{hsa,pd}$) and ambient, modeled as a constant temperature source. During steady state operation the heatsink thermal capacitance $C_{th}$ does not affect the junction temperature, thus it was grayed out. Regarding transients, those are not that important as long as their time constant $C_{th}R_{hsa}$ is longer than fundamental period, which is definitely the case considering the size of the heatsink. Per-device stack of elements forming $R_{chs,pd}$ is shown in Fig. \ref{fig:ThermalModel}b. All the parameters needed for the calculation of thermal resistances in Fig. \ref{fig:ThermalModel}a are listed in Table \ref{t:ThermalParams}. Note that the PCB from \cite{Heller} was taken as a basis for PCB related parameters in Table \ref{t:ThermalParams}.

\begin{table}[H]
	\centering
	\caption{General and PCB-related thermal model parameters}
	\label{t:ThermalParams}
	\begin{tabular}{ | P{2cm} || P{2cm} | P{1.5cm} | P{8cm} | }
		\hline
		Parameter & Value & Unit & Param. Explanation (General $|$ PCB)\\
		\hline\hline
		$N_{par}$ & $2$ & - & Number of parallel devices\\
		$N_{HB}$ & $4$ & - & Number of devices per half-bridge\\
		$R_{jc}$ & $1$ & K/W & Device junction-to-case thermal resistance \cite{IGLD60R070D1}\\
		$T_{j,max}$ & $120$ & \degree C & Maximum tolerable junction temperature\\
		$T_{amb}$ & $40$ & \degree C & Maximum ambient temperature\\
		$P_{max,pd}$ & $3.1$ & W & Maximum dissipation per device\\
		\hline
		$l$ & $1.7$ & mm & PCB thickness (via length)\\
		$K_{Cu}$ & $385$ & W/K.m & Thermal conductivity of copper\\
		$K_s$ & $60$ & W/K.m & Thermal conductivity of solder\\
		$r_{out}$ & $0.15$ & mm & Outer radius of a single via\\
		$r_{in}$ & $0.10$ & mm & Inner radius of a single via\\		
		$d$ & $0.3$ & mm & Thickness of employed thermal pad\\
		$\lambda_{pad}$ & $17$ & W/K.m & Thermal Conductivity of thermal pad material\\
		$A_{pad}$ & $13.6$ & mm\textsuperscript{2} & Base plate area (used by vias)\\		
		$N_{vias}$ & $36$ & - & Approximate number of vias per device\\		
		\hline
	\end{tabular}
\end{table}

Using these parameters one can estimate the per-device heatsink-to-ambient thermal resistance $R_{hsa,pd}$ based on the following equations:

\begin{equation}
\setlength{\jot}{10pt} 
\label{eq:Rth_hsa_pd}
\begin{split}
&R_{chs,pd} = \frac{d}{\lambda_{pad} A_{pad}} + \frac{1}{N_{vias}} \Big(K_s \pi r_{in}^2 + K_{Cu} \pi (r_{out}^2 - r_{in}^2)\Big)^{-1}
\\
&T_{j,max} \geq T_{amb} + P_{max,pd} \big(R_{jc} + R_{chs,pd} + R_{hsa,pd}\big) \Rightarrow
\\
&R_{hsa,pd} \leq \frac{T_{j,max} - T_{amb}}{P_{max,pd}} - R_{jc} - R_{chs,pd} = 20.84 \text{ K/W} \Rightarrow
\\
&R_{hsa} \leq R_{hsa,pd}/(3N_{HB}) = 1.74 \text{ K/W}
\end{split}
\end{equation}

The last equation is true under assumptions that all devices face the same $R_{hsa,pd}$ and that there is only one heatsink for all three phases. Thus, the upper limit for the heatsink thermal resistance $R_{hsa}$ (heatsink-to-ambient) is $1.74$ K/W.

\begin{figure}[h]
	\centering
	\begin{subfigure}[H]{0.28\textwidth}
		\captionsetup{justification=centering}
		\includegraphics[width=\textwidth]{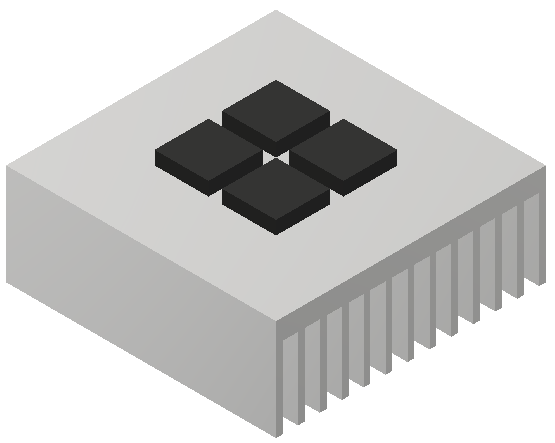}
		\caption{Per-phase pin fin heatsink with half-bridge}
		\label{fig:PerPhaseHeatsink}
	\end{subfigure}
	~
	\begin{subfigure}[H]{0.65\textwidth}
		\includegraphics[width=\textwidth]{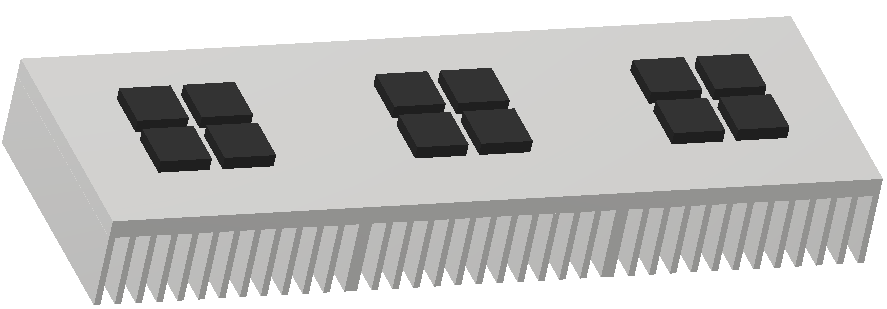}
		\caption{Three-phase stacked pin fin heatsink with three half-bridges}
		\label{fig:ThreePhaseHeatsink}
	\end{subfigure}
	\caption{Prospective per-phase and three-phase heatsink designs}
	\label{fig:PerspectiveHeatsinks}
\end{figure}

Third, in case one heatsink is used per phase, it should accommodate four switches as well as some additional area around for gate driver circuitry and anti-parallel diodes on top side of the PCB. In addition, wide metallic heat spreaders could be used for a better thermal contact (Fig. \ref{fig:ThermalModel}b) on bottom side of the PCB. With these, a per-phase heatsink could look like shown in Fig. \ref{fig:PerPhaseHeatsink}, with black rectangles signifying half-bridge place holders for relative size comparison. A three-phase heatsink, in turn could look like three per-phase ones, provided that phase-modularity is one of the desired features (Fig. \ref{fig:ThreePhaseHeatsink}). Finally, in case active cooling is needed, the requirement would be to provide a non-restricted air intake for the fans.

$\bullet$ \textbf{Cooling system design, evaluation} 

Taking into account the cooling system requirements and thermal model results above, the following cooling system has been designed (Fig. \ref{fig:FinalHeatsink}):

\begin{figure}[h]
	\centering
	\includegraphics[width=0.6\textwidth]{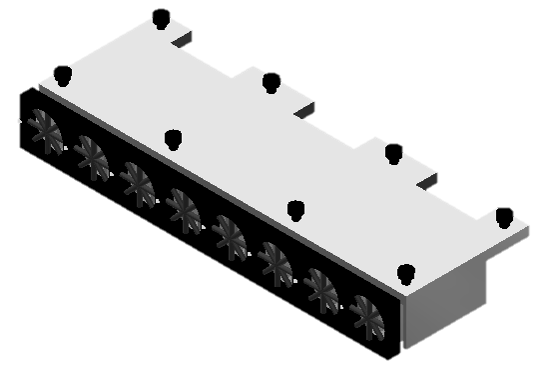}
	\caption{Final cooling system design}
	\label{fig:FinalHeatsink}
\end{figure}

Essentially, final cooling system consists of one large heatsink and eight small fans blowing air through it. The heatsink accommodates all 12 semiconductors, inspired by Fig. \ref{fig:ThreePhaseHeatsink}. There are two reasons for selecting a single heatsink design. First, by having one rigid heatsink the thermal resistance of the whole cooling system is expected to decrease. Second, in a power dense converter design there is only a marginal space left for the screws that connect the heatsink and the PCB, as it will be seen later. Thus, by having a single rigid body one reduces the number of screws without compromising the mechanical pressure between the heatsink and the PCB, which is needed for a good thermal contact of the two. Apart from that, three heatsink cutouts, found between the rear screws, will be necessary for three buck-boost inductor connectors. Final cooling system parameters can be found in Table \ref{t:CoolingParams}. Heatsink parameters were selected in accordance to manufacturing guidelines of the ETH workshop.

\begin{table}[h]
	\centering
	\caption{Cooling system related parameters}
	\label{t:CoolingParams}
	\begin{tabular}{ | P{2cm} || P{2cm} | P{1.5cm} | P{6cm} | }
		\hline
		Parameter & Value & Unit & Param. Explanation\\
		\hline\hline
		$W_{HS}$ & $120$ & mm & Heatsink width\\
		$L_{HS}$ & $40$ & mm & Heatsink length\\
		$d_{HS}$ & $3$ & mm & Heatsink base plate thickness\\
		$h_{HS}$ & $12$ & mm & Heatsink fin height\\
		$t_{HS}$ & $1$ & mm & Heatsink fin thickness\\
		$N_{fins}$ & $40$ & - & Total number of fins\\
		\hline
		$H_{fan}$ & $15$ & mm & Fan height\\
		$L_{fan}$ & $15$ & mm & Fan length\\
		$W_{fan}$ & $4$ & mm & Fan width (or thickness)\\
		$D_{fan}$ & $13$ & mm & Fan blades diameter\\
		$N_{fans}$ & $8$ & - & Total number of fans\\		
		\hline		
	\end{tabular}
\end{table}

To evaluate the cooling system, an existing cooling system evaluation script was used. The script evaluates various fan-heatsink combinations and provides relevant metrics, like Cooling System Performance Index (CSPI), $R_{th}$, volume flow of air etc. Since there was no fan with 15 mm side in the script database, the nearest one with 20 mm side has been evaluated. To account for this discrepancy, the respective heatsink section was updated and the CSPI of the resulting fan-heatsink pair was estimated. Considered heatsink section has a width of 20 mm (same as the fan) and the length of 37 instead of 40 mm to account for the aforementioned cutouts. Other heatsink parameters, namely base plate thickness, fin thickness and the distance between fins, were left unchanged. The resulting figures are as follows: $CSPI = 22.37$ K/W.Liter, $R_{th1} = 2.235$ K/W and $V_{1} = 0.02$ Liters. Assuming that a volume reduction without change of geometry results in the cooling system with similar CSPI, thermal resistance of interest $R_{th2}$, corresponding to one fan segment, can be estimated using the following equation:

\begin{equation}
\setlength{\jot}{10pt} 
\label{eq:CSPI}
\begin{split}
&CSPI = \frac{1}{R_{th1} V_1} = \frac{1}{R_{th2} V_2} = 22.37\text{ K/W.Liter}
\\
&V_2 = L_{eq,HS} H_{fan} L_{fan} = 0.01\text{ Liters} \Rightarrow
\\
&R_{th2} = \frac{1}{CSPI.V_2} = 4.62\text{ K/W} \Rightarrow
\\
&R_{hsa} = R_{th2}/N_{fans} = 0.58\text{ K/W} < 1.74\text{ K/W}
\\
\end{split}
\end{equation}

Equation (\ref{eq:CSPI}) states that our cooling system meets the $R_{th}$ requirement with some margin. Having some margin is useful due to calculation, thermal pad performance as well as loss map uncertainties. As a final remark, proposed cooling system should also provide some active cooling to the components to be placed after the heatsink, for example buck-boost inductors or DC link capacitors.

\section{Measurements and Auxiliaries}
$\bullet$ \textbf{Voltage measurements}

Control system requires measurement of 10 signals in total: DC voltage, 3 AC-side voltages, 3 load currents and 3 buck-boost inductor currents. Regarding voltages, those are measured using a standard resistive divider and a differential operational amplifier (OPV, AD8601) in series. Resistor ratios are selected such that voltage measurement ranges of [0..250]V DC and [+10..-200]V AC are translated to the OPV output voltage range of [0..3]V and [1.12..2.8]V, respectively. The following differential OPV equation and simplified schematic have been used:

\begin{equation}
\label{eq:OpAmp}
V_{out} = \frac{V_p R_{p2}}{(R_{p1} + R_{p2})} \frac{(R_{n1} + R_f)}{R_{n1}} - \frac{V_n R_f}{R_{n1}}
\end{equation}

\begin{figure}[h]
	\centering
	\includegraphics[width=0.9\textwidth]{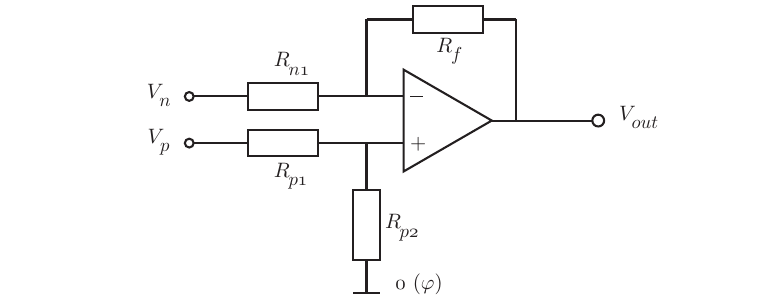}
	\caption{Simplified differential operational amplifier (OPV) circuit}
	\label{fig:OpAmp}	
\end{figure}

\begin{table}[h]
	\centering
	\caption{Operational amplifier (OPV) connections and resistances for DC and AC voltages}
	\label{t:Resistances}
	\begin{tabular}{ | P{2cm} || P{2cm} | P{1.5cm} | P{6cm} | }
		\hline
		Parameter & Value & Unit & Measurement circuit\\
		\hline\hline
		OPV & AD8601 & - & DC, AC voltage measurement\\	
		\hline
		$V_p$ & DC+ & - & DC voltage measurement\\	
		$V_n$ & DC- & - & DC voltage measurement\\					
		$\varphi$ & $0$ & V & DC voltage measurement\\	
		$R_{p1}$ & $500$ & kOhm & DC voltage measurement\\
		$R_{p2}$ & $6$ & kOhm & DC voltage measurement\\
		$R_{n1}$ & $500$ & kOhm & DC voltage measurement\\
		$R_{f}$ & $6$ & kOhm & DC voltage measurement\\
		\hline
		$V_p$ & DC- & - & AC voltage measurement\\	
		$V_n$ & $V_{Ca,Cb,Cc}$ & - & AC voltage measurement\\					
		$\varphi$ & $1.2$ & V & AC voltage measurement\\	
		$R_{p1}$ & $500$ & kOhm & AC voltage measurement\\
		$R_{p2}$ & $4$ & kOhm & AC voltage measurement\\
		$R_{n1}$ & $500$ & kOhm & AC voltage measurement\\
		$R_{f}$ & $4$ & kOhm & AC voltage measurement\\
		\hline
	\end{tabular}
\end{table}

Resistor absolute values are selected such that, on one hand, no excessive losses are generated over those while, on the other, proper OPV operation is not compromised. In case of AC measurements, a small offset is applied to the differential measurements to facilitate the measurement capability of AC signals (potential $\varphi$ in Fig. \ref{fig:OpAmp}). Selected voltage measurement parameters are summarized in the Table \ref{t:Resistances}.

$\bullet$ \textbf{Current measurements}

Regarding current measurements, two types of Allegro ICs have been selected based on current ranges and bandwidths. For the load current, an IC with $\pm20$A measurement capability and 100kHz bandwidth (ACS711KLCTR) has been selected. For buck-boost inductor current, an IC with $\pm60$A measurement capability and 1MHz bandwidth (ACS732/3) has been selected. Among other useful features, these ICs directly provide [0..3]V analog signal as an output, which is directly compatible with DSP logic.

$\bullet$ \textbf{Auxiliary} 

\begin{figure}[h]
	\centering
	\includegraphics[width=\textwidth]{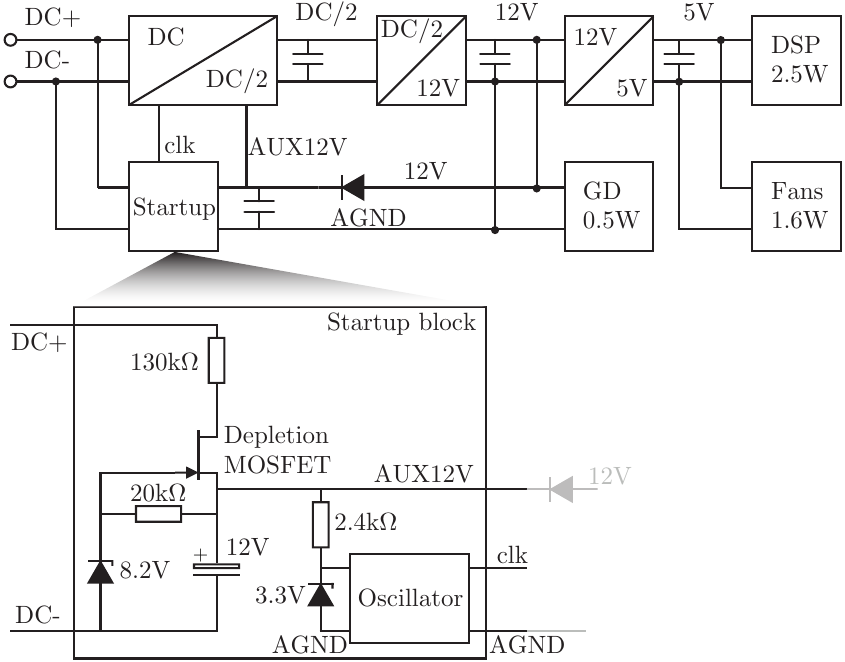}
	\caption{Block diagram of the auxiliary circuit with detailed view of startup block}
	\label{fig:Auxiliary}	
\end{figure}

Auxiliary supply is needed to power the control circuitry, including gate drivers and DSP, as well as secondary loads, such as fans. Also it is responsible for startup of the converter. Taking into account the application specifics, such as a wide DC voltage range, gate driver circuitry, fans etc., the proposed auxiliary supply structure is shown in Fig. \ref{fig:Auxiliary}.

As it can be seen, the auxiliary is used to feed the DSP, fans and Gate Drivers (GD). It starts with an unregulated integrated buck converter (IRSM808-105MH). This IC, as well as some additional components like input, output and bootstrap capacitors as well as the output inductor, are combined to form the first DC-DC/2 conversion stage. This stage requires auxiliary 12V gate driver supply and gate signals (clk) and it is used to half the input DC voltage from [240..80]V to [120..40]V. The second and third stages are independent in the sense that they do not require any additional supplies or gate signals. They are used to further step down the DC/2 to 12V (LTC7138) and 5V (LMZ10501) respectively (Fig. \ref{fig:Auxiliary}). The DSP and fans are powered from 5V supply, while gate drivers are fed from 12V. Of course during converter startup the auxiliary 12V and gate signals for the first DC-DC/2 stage are not available because the down-stream buck stages and DSP are not powered. Thus even if DC voltage is present, the converter cannot start operating. To overcome this problem, the startup circuit has been designed (Startup block in Fig. \ref{fig:Auxiliary}). 

Initially, startup block is fed from the DC terminals through a normally-on MOSFET and a Zener-diode regulation to provide a 50kHz, 3.3V clock and auxiliary 12V signals. The DC-DC/2 stage, fed by DC link and driven by startup block with a constant duty cycle of 0.5, slowly starts charging the output capacitors to DC/2. [Here the underlying startup assumption is that initially the DC link is fed from the on-board accumulator of the car through a DC/DC converter. That DC/DC converter is controlled to ramp up the voltage of the DC link gradually, such that no over-currents/over-voltages occur at the output inductor/ capacitor of the DC-DC/2 stage.] With DC/2 voltage provided, second and third stages directly produce 12V and 5V, powering the GD, fans and DSP. As soon as 12V can be provided from the second stage, the diode turns on and feeds the startup circuits from the second stage as a feedback. The startup circuit was designed such that before diode turn on, AUX12V potential is slightly less than 12V so that when 12V is available, the diode can actually turn on. Apart from that, as soon as AUX12V potential reaches 12V due to diode connection, DC+ supply is disconnected from the startup block. Proposed startup block was verified in Gecko simulation. Regarding the buck stages, those were tested in hardware by some other projects, thus a proper operation in hardware is expected from the proposed auxiliary supply.

\clearpage

\section{Power/Control Board Layout}

$\bullet$ \textbf{General layout description} 

\begin{figure}[h]
	\centering
	\includegraphics[width=0.95\textwidth]{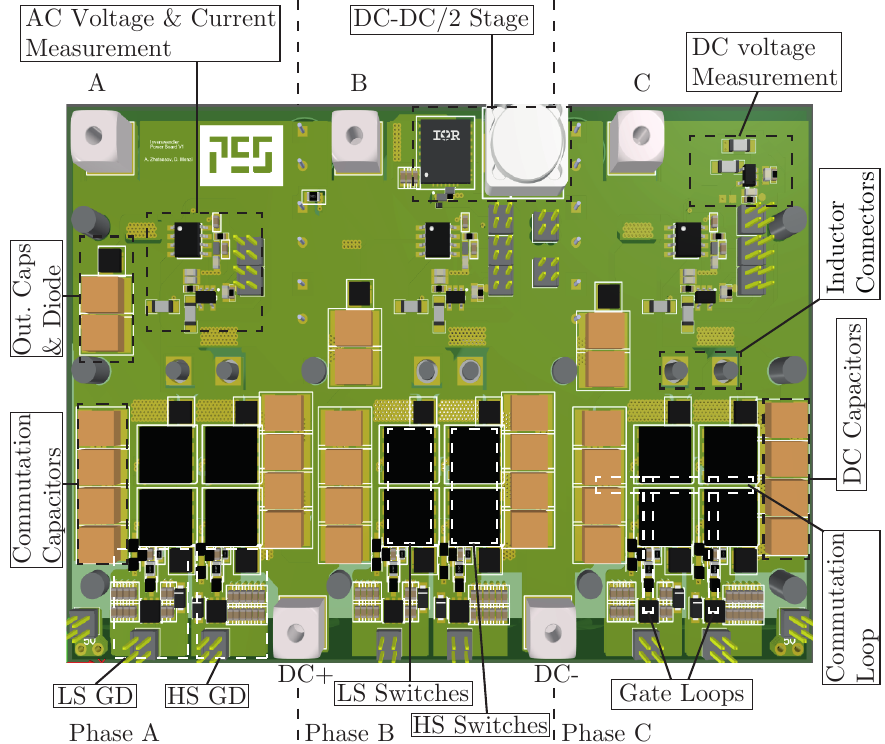}
	\caption{Top view of the power PCB}
	\label{fig:PowerBoardTop}	
\end{figure}

\begin{figure}[h]
	\centering
	\includegraphics[width=0.95\textwidth]{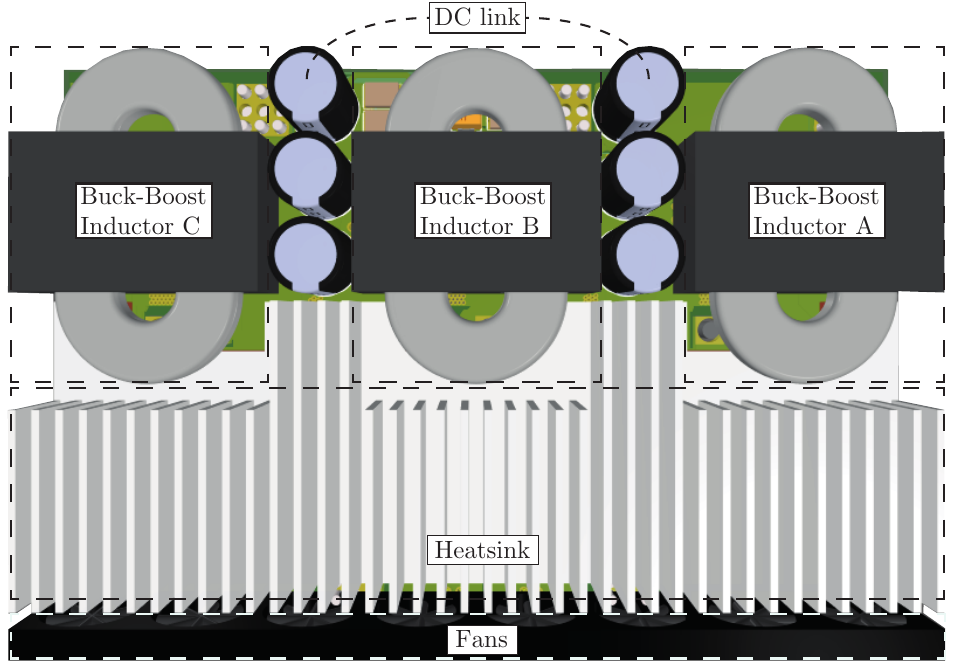}
	\caption{Bottom view of the power PCB}
	\label{fig:PowerBoardBottom}	
\end{figure}

Having defined the design of major components as well as measurements and auxiliaries, Fig. \ref{fig:PowerBoardTop} and Fig. \ref{fig:PowerBoardBottom} demonstrate the positioning of these components on the main PCB, referred to as a power board. Bearing in mind phase modularity of the topology, from general layout it is seen that one tried to preserve phase symmetry as far as possible. Although proposed design is not phase-modular, per-phase symmetry gives an insight about future phase modular design opportunities after the first prototype has been tested.

As it can be seen from Fig. \ref{fig:PowerBoardTop}, each phase features a half-bridge with two parallel devices and anti-parallel diodes, low-side (LS) and high-side (HS) gate drivers (GD), DC and commutation capacitors, output capacitors with anti-parallel diodes (Fig. \ref{fig:InverswandlerInverterWithFuelCell}) as well as AC-side voltage and load current measurements. As far as possible, straight power flow is also preserved with power flowing from DC+/DC- connectors through the half-bridges and output capacitors to the AC connectors A, B and C. Apart from that, the first DC-DC/2 buck stage of the auxiliary supply (Fig. \ref{fig:Auxiliary}) can also be seen on the power board.

From Fig. \ref{fig:PowerBoardBottom} one can see the bottom side of the power board that features fans, heatsink, buck-boost inductors, DC link electrolytic capacitors, inductor current measurements and startup circuitry (concealed by inductors). Components were designed to maximize the fill factor under the power board. In addition, it can be seen that blown air hits the inductors and electrolytic capacitors after passing between heatsink fins, providing some active cooling to those components as well.

The DC link has been realized according to the aforementioned considerations, with ceramic capacitor bank near each half-bridge providing a low-impedance path to a high-frequency current ripple, and a large power-buffering electrolytic capacitors located a bit further away.

$\bullet$ \textbf{Commutation, gate loops}

\begin{figure}[h]
	\centering
	\includegraphics[width=0.9\textwidth]{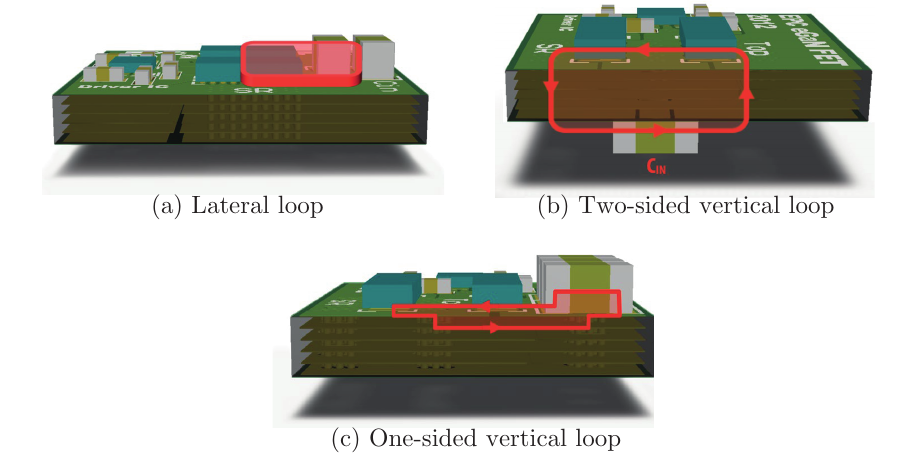}
	\caption{Comparison of three different commutation loop designs \cite{PCBLayout}}
	\label{fig:CommutationLoops}	
\end{figure}

Apart from DC link ceramics one can observe a commutation capacitor bank on the other side of the half-bridges. These capacitors are connected across the half-bridges to DC+ and output terminals A, B and C. They are used to realize a one-sided vertical commutation loop, the advantages of which were explained in \cite{PCBLayout}. Essentially, one-sided vertical commutation loop results in the smallest loop cross-sectional area and, consequently, the smallest loop inductance (Fig. \ref{fig:CommutationLoops}). Small loop inductance, in turn, is needed to reduce the over-voltages, switching losses and a chance of parasitic turn-on/-off, especially during hard-switched transitions. The results in \cite{PCBLayout} experimentally verify those statements.

A second important factor affecting the switching losses and a chance of parasitic turn-on/-off is the gate driver, in particular its turn-on/-off gate resistances, gate capacitance \cite{GaNGIT} and gate loop inductance. The gate driver from \cite{GaNGIT} has been used. Within this gate driver, dedicated capacitance is used to provide a bidirectional gate-source capacitance $C_{gs}$ driving capability even with a unipolar gate driver. This capacitance indirectly adds to the switching losses in a form of gate driver losses, since it has to be charged every switching period. Regarding gate resistance and gate loop inductance, as in a standard gate driver, they affect the charge-discharge time of $C_{gs}$. This time is important for switching loss estimation as it determines the voltage-current overlap in the channel of the device. The shorter the interval the lower the losses. Therefore, to reduce the gate loop inductance the same vertical layout design has been applied, resulting in a one-sided vertical gate loops depicted in Fig. \ref{fig:PowerBoardTop}. The devices also feature a low-inductance Kelvin source connection, thus further reducing loop inductance.

$\bullet$ \textbf{Connectors, control board}

\begin{figure}[H]
	\centering
	\includegraphics[width=0.88\textwidth]{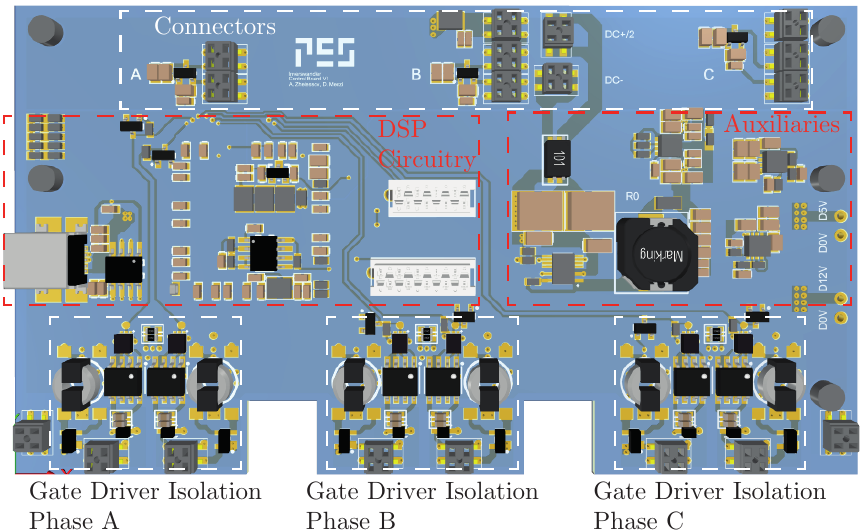}
	\caption{Top view of the control PCB}
	\label{fig:ControlBoardTop}	
\end{figure}

\begin{figure}[h]
	\centering
	\includegraphics[width=0.88\textwidth]{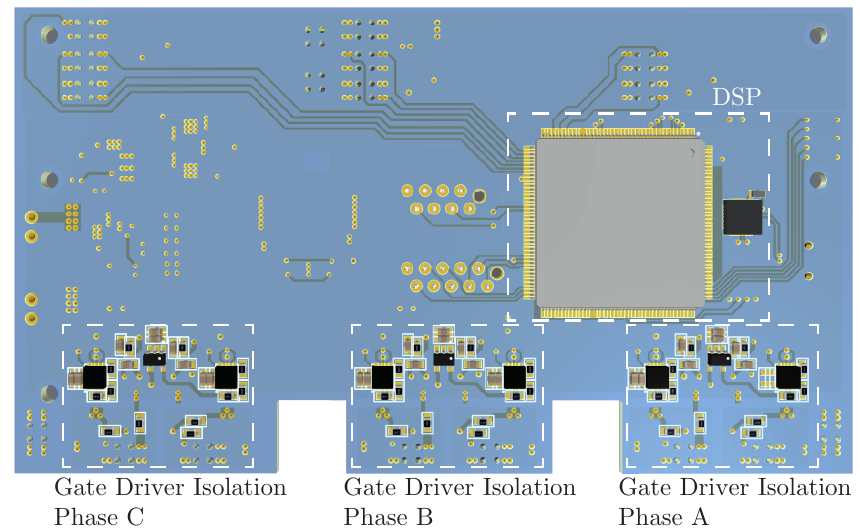}
	\caption{Bottom view of the control PCB}
	\label{fig:ControlBoardBottom}	
\end{figure}

Fig. \ref{fig:PowerBoardTop} features numerous berg-type connectors. There is one such connector per each gate driver and several connectors in measurements and auxiliary areas of the board. Connectors are used to receive the isolated gate signals from / send the measurements and auxiliary power to a separately designed control board on top of the power board. The control board features a DSP, remaining auxiliary supplies as well as gate driver isolation circuits (Fig. \ref{fig:ControlBoardTop}-\ref{fig:ControlBoardBottom}). Finally, the screws attach the heatsink, power and control board together.

\section{Virtual Prototype}

\begin{figure}[h]
	\centering
	\includegraphics[width=\textwidth]{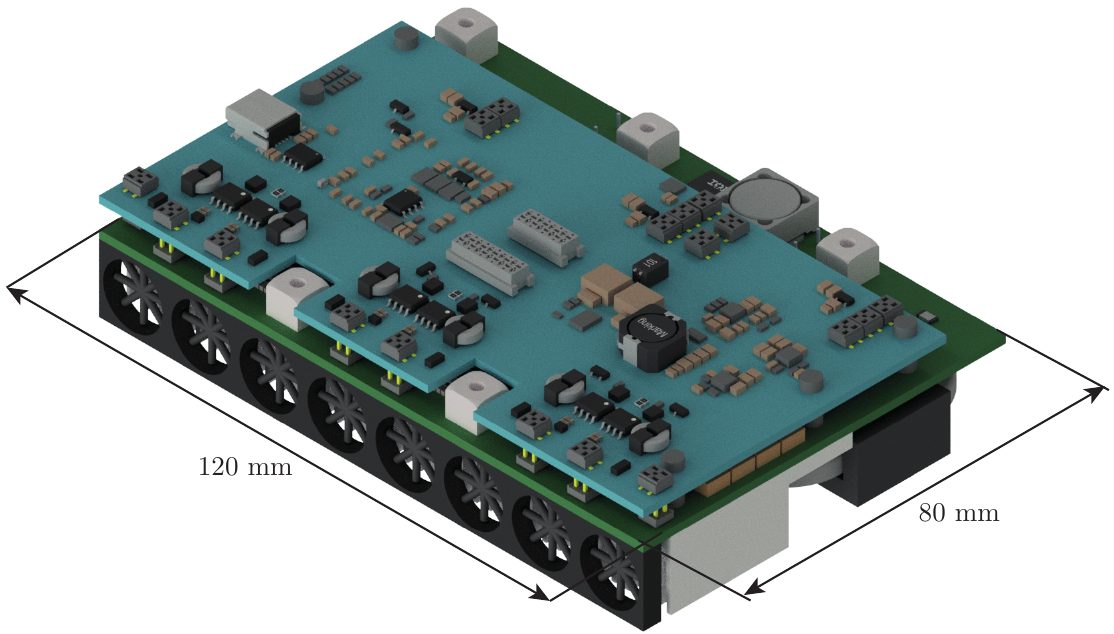}
	\caption{Virtual prototype of the three-phase inverting buck-boost converter. $P_{nom} = 1$kW, $\rho_{exp} = 3.8$kW/Liter, $\eta_{exp} \in [95.5\%..96.4\%]$}
	\label{fig:VirtualPrototype}	
\end{figure}

The last section of the present chapter is dedicated to the complete virtual prototype and its performance evaluation. Fig. \ref{fig:VirtualPrototype} depicts the perspective appearance of the designed converter with lateral dimensions and main performance metrics. With nominal power being 1kW, the expected power density is 3.8kW/Liter, while the efficiency is expected to be in the range of [95.5\%..96.4\%]. Note that the efficiency figure depends on the operating point and that reported value of 95.5\% corresponds to the worst case efficiency occurring at $V_{dc} = 80$V, $V_{o} = 80$V, $P = 1$kW. Overall, Fig. \ref{fig:VirtualPrototype} suggests that the converter has a relatively high fill factor within its boxed shape - a prerequisite and good indicator of a near-maximum power density.

\begin{figure}[h]
	\centering
	\begin{subfigure}[H]{0.45\textwidth}
		\captionsetup{justification=centering}
		\includegraphics[width=\textwidth]{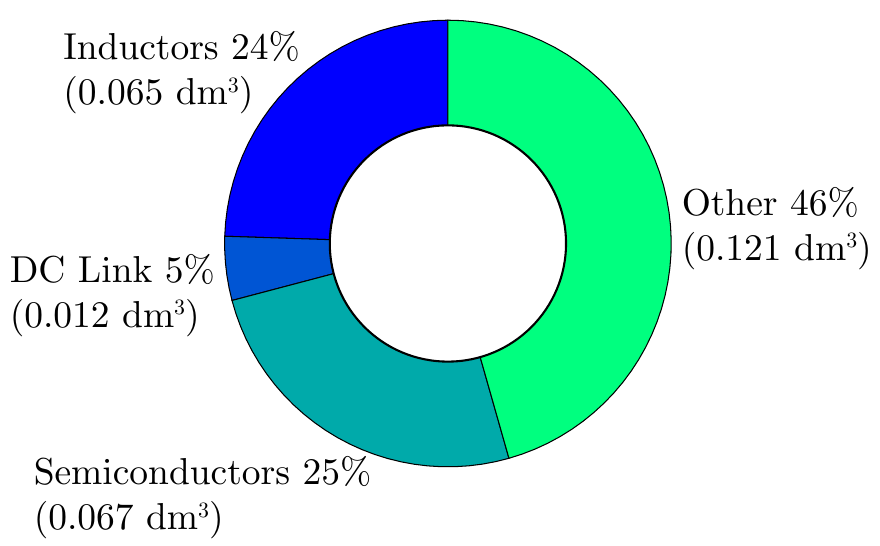}
		\caption{Boxed volume distribution}
		\label{fig:ProtoVolumeDistribution}
	\end{subfigure}
	~
	\begin{subfigure}[H]{0.51\textwidth}
		\includegraphics[width=\textwidth]{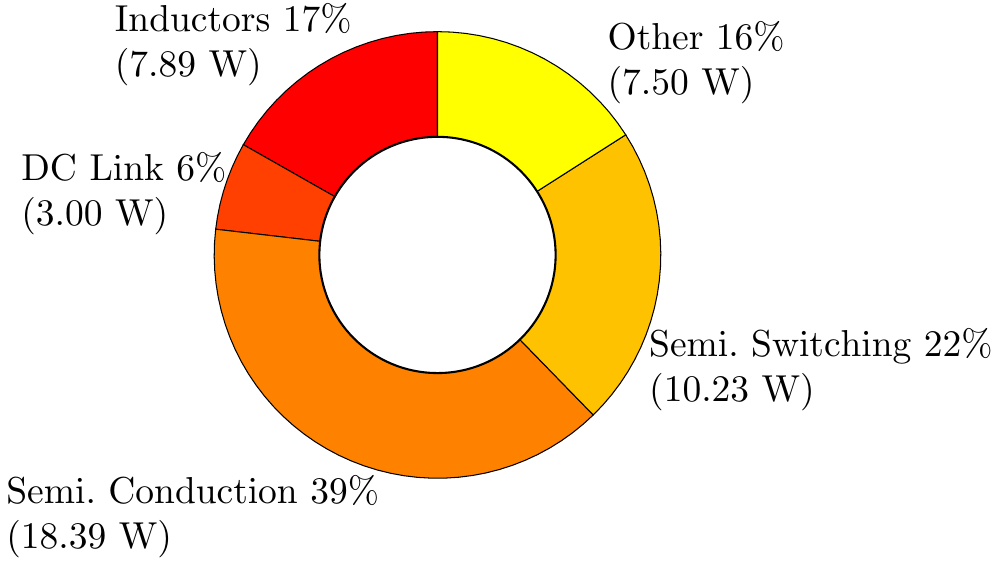}
		\caption{Worst case loss distribution}
		\label{fig:ProtoLossDistribution}
	\end{subfigure}
	\caption{Virtual prototype volume and loss distributions}
	\label{fig:ProtoDistributions}
\end{figure}

Fig. \ref{fig:ProtoDistributions} shows a more precise performance evaluation, namely boxed volume and worst case loss distribution. Regarding volume distribution, it is seen in Fig. \ref{fig:ProtoVolumeDistribution} that the two PCBs account for more that 40\% of total converter volume. The reason is that power board components have a non-negligible height compared to the overall converter height. In these conditions, some safety distance of 5mm has to be kept between power and control PCBs. Although 5mm seem to be a small value, it is already about 20\% of the total converter height. Thus, power board, 5mm air space between the boards and the control board with components on it together account for 40\% of the volume.

\begin{figure}[h]
	\centering
	\includegraphics[width=\textwidth]{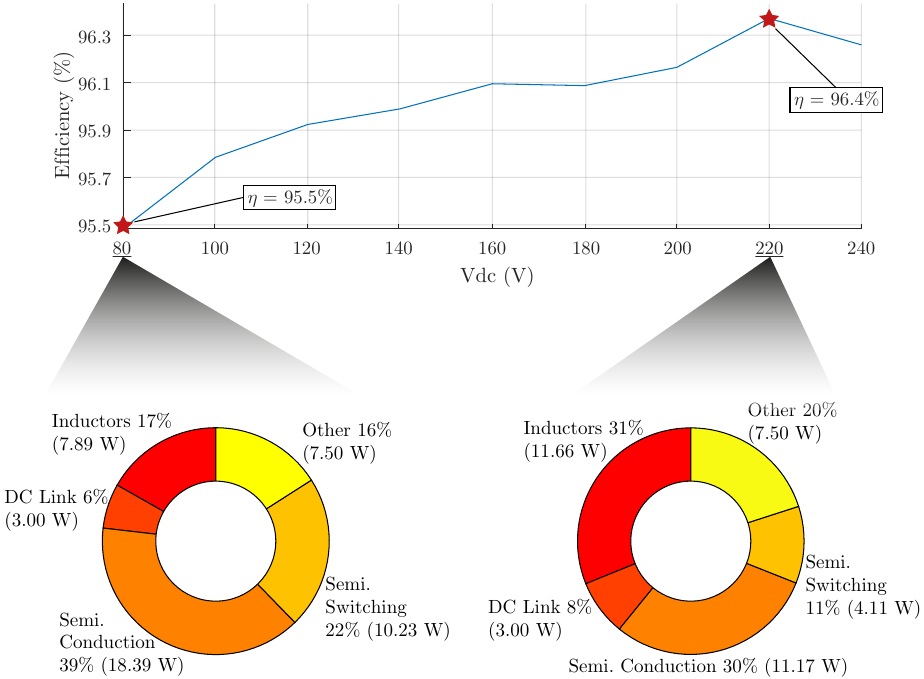}
	\caption{Theoretical prototype efficiency as a function of DC voltage for $\hat{V}_o = 80$V, $P = 1$kW. Theoretical converter loss distributions are shown for the minimum ($\eta_{min} = 95.5$\%) and maximum ($\eta_{max} = 96.4$\%) efficiency points}
	\label{fig:Efficiency_Vdc}	
\end{figure}

Regarding loss distribution, Fig. \ref{fig:ProtoLossDistribution} illustrates the loss distribution corresponding to the worst case operating point. In these conditions, main loss contribution (over 50\%) comes from the semiconductors. This is expected if one recalls Fig. \ref{fig:VdcSweep10u} and discussions therein. Note that as one traverses from 80V to 240V in DC voltage, not only the efficiency grows, but also relative loss distribution changes significantly, with inductor and semiconductor loss contributions rising and dropping, respectively (Fig. \ref{fig:Efficiency_Vdc}).

\afterpage{\blankpage}
\chapter{Conclusion}

\section{Summary}

To sum up, several major outcomes of this thesis can be outlined here:\\

$\bullet$ \ A new converter topology - an three-phase inverting buck-boost converter has been presented and analyzed in this thesis. The topology is of interest because it provides voltage buck-boost capability and bidirectional operation within a single power conversion stage. In addition, the topology is phase modular and has a low number of active components, potentially making it an economically viable option. Introductory chapters presented the topology derivation, fundamental operating concept and three applicable modulation schemes.\\

$\bullet$ \ Two potential applications of the topology have been studied, the first one being a rectifier application for a More Electric Aircraft. Within the scope of this application, modulation candidates were compared and Discontinuous PWM (DPWM) has been selected for further implementation. Control structure has been designed and verified through numerous simulations. Emerging modulation-associated control problems were identified and resolved.\\

$\bullet$ \ The second considered application was an inverter for a Fuel Cell Powered High-Speed Motor Drive. Within the scope of this application, modulation and control were developed, a virtual hardware prototype was designed. The prototype features 1kW power rating, 3.8kW/Liter power density and the efficiency in the range of [95.5\%..96.4\%].\\

$\bullet$ \ Apart from that, a thermal measurement board has been designed to measure the loss map of new generation GaN devices, used in the inverter prototype. The board was designed for a calorimetric measurement setup. It is capable of measuring both hard/soft-switching. Moreover, it provides the opportunity to measure the effect of various DOFs, such as number of parallel devices and gate drive component values.

\section{Outlook}

As an outlook we foresee the assembly, commissioning and testing/measuring of the inverter prototype/measurement board. After those are done one can compare the prototype with existing solutions. The measured loss map can be used for further converter optimization as well as in other projects employing the same semiconductor devices.

\bibliography{bibliography,long}
\appendix
\afterpage{\blankpage}

\chapter{Calorimetric Measurement Setup for a GaN GIT Transistor}
This appendix addresses the calorimetric measurement setup for a new generation GaN transistors. Essentially, employed calorimetric setup and designed measurement board are the same as in \cite{Heller} apart from several minor modifications that will be addressed below.

\begin{figure}[h]
	\centering
	\includegraphics[width=\textwidth]{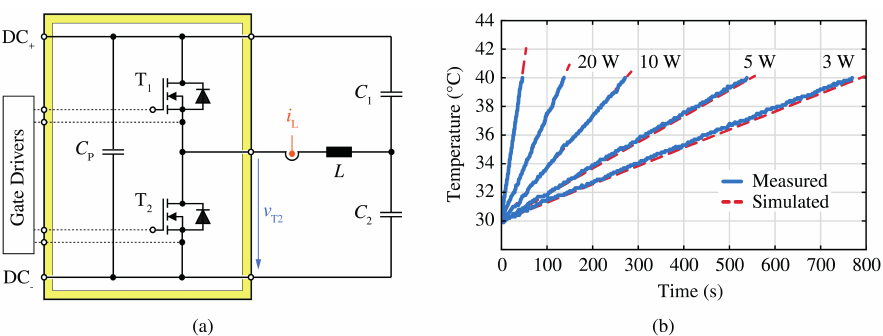}
	\caption{Setup for the soft-switching losses measurement: schematic of the half-bridge operating in triangular current mode (a); and rise of the temperature on the heatsink for different power	levels (b). After setup calibration, different dT/dt allow the calculation of the switching losses \cite{Heller}}
	\label{fig:CalorimetricSetup}
\end{figure}

Generally, calorimetric setup is used to characterize the switching losses of a wide bandgap semiconductors e.g. Gallium Nitride (GaN), because of the enabled high switching speed. The idea behind calorimetric setup is to measure a power related change in a heatsink (brass block) temperature. Then the power is determined with the help of a thermal model \cite{Heller}. To save time, a temperature transient is measured instead of waiting for a steady state to evolve, followed by a subsequent data fitting with a thermal model (Fig. \ref{fig:CalorimetricSetup}b). To obtain a valid measurements with such a setup, several properties should be optimized, including minimum thermal resistance from the devices under test (DUT) to the heatsink, minimum power and gate loop inductances etc. More details can be found in \cite{Heller}.

\begin{figure}[h]
	\centering
	\includegraphics[width=\textwidth]{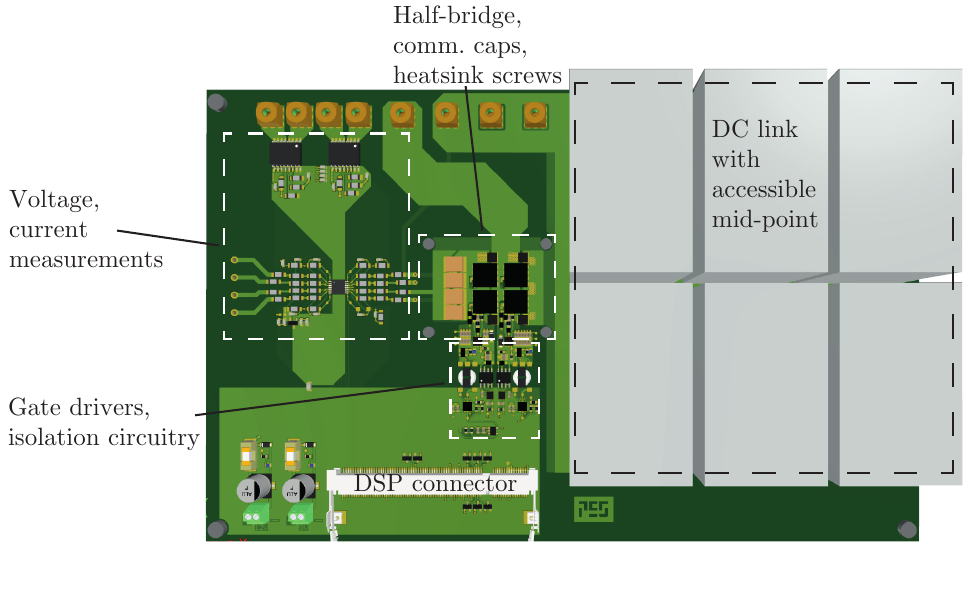}
	\caption{Top view of the designed measurement board}
	\label{fig:MeasBoard}	
\end{figure}

$\bullet$ Measurement board. The designed measurement board is depicted in Fig. \ref{fig:MeasBoard}. It features the DC link with an accessible mid-point (MP), half-bridge with commutation capacitors, gate drivers and isolation circuitry, DSP connector, 5V and 12V power supplies as well as voltage/current measurement circuits and connectors. \textbf{DC link} has two rows of film capacitors with three parallel 20$\mu$F capacitors in each row. One row is connected between DC- and MP, while the other is between MP and DC+. Three respective connectors (DC-, MP, DC+) can be found near the film capacitors (Fig. \ref{fig:MeasBoard}). In \textbf{half-bridge} area one can see two parallel switches with anti-parallel diodes as well as ceramic commutation capacitors. The half-bridge layout features a one-sided vertical commutation loop (Fig. \ref{fig:CommutationLoops}c) as later implemented in the converter prototype. The screws of the brass block were placed towards the corners of the half-bridge area due to space constraints (Fig. \ref{fig:MeasBoard}). Regarding \textbf{gate drivers and isolation circuitry}, those were placed as near as possible to the half-bridge. One reason is the gate loop inductance, which increases with distance between the gate driver and the switches. Another reason is noise reduction over the isolated high-side signal between the digital signal isolator and the gate driver ICs. It was observed that the noise can be superimposed over the isolated high-side signal due to the parasitic capacitances between the trace of the isolated signal and closely-located steady potentials (DC+/DC- for instance). Since high-side isolated gate signal jumps in potential with the switched node, mentioned parasitic capacitances can disturb the signal by injecting the charging-discharging currents to the respective trace if it is long enough. Thus, the distance between the gate driver ICs and the digital signal isolators is minimized. The \textbf{DSP connector} and \textbf{power supplies} are used to connect and power the standard PES DSP board for control purposes. General purpose \textbf{voltage/current measurements} and connectors were added to have a full accessibility and control over the half-bridge.


\begin{table}[H]
	\centering
	\caption{System specifications for a calorimetric measurement setup}
	\label{t:SpecsMeasBoard}
	\begin{tabular}{ | P{1.75cm} || P{2cm} | P{1.5cm} | P{9.2cm} | }
		\hline
		Parameter & Value & Unit & Param. Explanation (Electrical $|$ Thermal $|$ Layout)\\
		\hline\hline
		Device & GaN 600V & - & Semiconductor: Infineon GaN 600V, 55mOhm \cite{IGLD60R070D1}\\
		$N_{par}$ & 2 & - & Number of parallel devices\\	
		$N_{HB}$ & 4 & - & Number of devices per half-bridge\\		
		$R_{ds,on}$ & 100 & mOhm & Drain-source on resistance per device at 100\degree C\\		
		$V_{dc}$ & $0-400$ & V & DC voltage range (Fig. \ref{fig:CalorimetricSetup})\\	
		$C_{dc}$ & $30$ & uF & DC link capacitance (3x20=60uF MP to DC-, 3x20=60uF DC+ to MP, 60/2=30uF DC+ to DC-)\\			
		$\hat{I}_{ss}$ & $40$ & A & Maximum soft-switched inductor current (Fig. \ref{fig:CalorimetricSetup})\\
		$\hat{I}_{hs}$ & $10$ & A & Maximum hard-switched inductor current\\
		$t_{dead}$ & $20$ & ns & Dead time as in \cite{Heller}, can change\\
		$d$ & $0.5$ & - & Half-bridge duty cycle to be used for soft-switching (can be used for hard-switching)\\		
		$\hat{E}_{ss}$ & $3.2$ & uJ & Measured peak soft-switched energy per device at $I_{ds} = 20$A, $V_{dc} = 400$V \cite{Heller}\\
		$\hat{E}_{hs}$ & $40$ & uJ & Measured peak hard-switched energy per device at $I_{ds} = 5$A, $V_{dc} = 400$V \cite{GaNGIT}\\
		\hline
		$R_{jc,pd}$ & $1$ & K/W & Per-device junction-to-case thermal resistance \cite{IGLD60R070D1}\\
		$R_{chs,pd}$ & $4.8$ & K/W & Measured per-device case-to-heatsink th. resist. \cite{OriginCoss}\\		
		$T_{amb}$ & $20..25$ & \degree C & Ambient temperature during measurements\\		
		$T_{j,max}$ & $120$ & \degree C & Maximum tolerable junction temperature\\
		$T_{br}$ & $30-40$ & \degree C & Brass block transient meas. temperature range\\
		$T_{br,rise}$ & $10$ & \degree C & Brass block temperature rise to be measured\\		
		$t_{min}$ & $120$ & sec. & Minimum brass block heating time to be measured\\
		$S\rho$ & $3205$ & kJ/K.m$^3$ & Brass material thermal capacitance per volume\\		
		\hline
		$l$ & $1.7$ & mm & PCB thickness (via length)\\
		$K_{Cu}$ & $385$ & W/K.m & Thermal conductivity of copper\\
		$K_s$ & $60$ & W/K.m & Thermal conductivity of solder\\
		$r_{out}$ & $0.15$ & mm & Outer radius of a single via\\
		$r_{in}$ & $0.10$ & mm & Inner radius of a single via\\		
		$d$ & $0.3$ & mm & Thickness of employed thermal pad\\
		$\lambda_{pad}$ & $17$ & W/K.m & Thermal conductivity of thermal pad material\\
		$A_{pad}$ & $13.6$ & mm\textsuperscript{2} & Base plate area (used by vias)\\		
		$N_{vias}$ & $36$ & - & Number of vias under each device\\		
		\hline
	\end{tabular}
\end{table}

$\bullet$ Setup specifications. Table \ref{t:SpecsMeasBoard} summarizes the system specifications for the calorimetric measurement setup. The specifications were divided into three major groups - electrical, thermal and layout-related. Within each group the relevant parameters, their values, units and explanations were added. For instance, from the first row it can be seen that the setup was designed for GaN 600V, 55mOhm semiconductor devices \cite{IGLD60R070D1}. In addition to the explanations in the Table \ref{t:SpecsMeasBoard}, some further clarifications are made below for the parameters that are slightly different from \cite{Heller}.

For the \textbf{electrical specifications}, first, the $N_{par}$ parameter in the second row indicates that the setup at hand features two parallel devices instead of one, which translates to four devices per half-bridge $N_{HB}$ instead of two (Fig. \ref{fig:MeasBoard}). Second, the maximum soft-switched inductor current $\hat{I}_{ss}$ is now 40A instead of 20A in \cite{Heller} due to the doubled number of parallel devices $N_{par}$. Third, we also want to use the setup for a hard-switching measurements with maximum inductor hard-switched current $\hat{I}_{hs} = 10$A (5A per device) and the expected maximum energy loss per device per transition $\hat{E}_{hs} = 40$uJ \cite{GaNGIT}.

For the \textbf{thermal specifications}, per-device case-to-heatsink thermal resistance $R_{chs,pd} = 4.8$K/W was taken from the measurements in \cite{Heller} as a mere approximation for the designed setup. The validity of such an approximation is verified through thermal resistance calculations below.

For the \textbf{layout specifications}, first, the number of vias under each device $N_{vias}$ was reduced from 81 to 36 compared to \cite{Heller}, due to modifications in the half-bridge layout. The modifications comprise the aforementioned one-sided vertical commutation and gate loops. Second, a thermal pad with different characteristics was selected due to its availability.

Overall, Table \ref{t:SpecsMeasBoard} summarizes the main setup specifications. From these one can derive the remaining necessary parameters, such as hard-/soft-switching frequency limits and the heatsink (brass block) dimensions. Respective derivations and thermal resistance calculations are addressed next.

$\bullet$ $R_{chs,pd}$ calculation. As it was mentioned above, the per-device case-to-heatsink thermal resistance $R_{chs,pd}$ parameter was assumed to be the same for the designed calorimetric setup and the setup in \cite{Heller}. To check this assumption, the expected theoretical $R_{chs,pd}$ can be calculated as:

\begin{equation}
\setlength{\jot}{10pt} 
\label{eq:Rchs,pd}
\begin{split}
&R_{chs,pd} = \frac{1}{N_{par}} \Big(K_s \pi r_{in}^2 + K_{Cu} \pi (r_{out}^2 - r_{in}^2)\Big)^{-1} + \frac{d}{\lambda_{pad} A_{pad}} = 4.08 \text{ K/W}
\end{split}
\end{equation}

Here one assumed that $R_{chs,pd}$ consists of the via-related term and the thermal pad-related term, while the heat spreader contribution was neglected due to its wide area and good thermal conductivity (Fig. \ref{fig:ThermalModel}b). The obtained theoretical $R_{chs,pd} = 4.08$K/W is slightly less than the measured value of $4.8$K/W. However, provided the thermal resistance dependency on various factors, like mechanical contact pressure between the heatsink, thermal pad and heat spreader, effective number of vias used for heat transfer etc., obtained value is in a good match with the assumed $R_{chs,pd}$. As a last note on $R_{chs,pd}$, one should not forget that stated $R_{chs,pd} = 4.8$K/W is a mere assumption and that the measurements of this parameter have to be conducted for the designed measurement board during the calibration process \cite{Heller}.

$\bullet$ Switching frequency limits. Within the framework of calorimetric measurements, the switching frequency is used as a degree of freedom to obtain a lower ratio of conduction/switching losses and thereby increase the measurement accuracy \cite{Heller}. This is why it was not stated in the electrical specifications section of Table \ref{t:SpecsMeasBoard}. Nevertheless, maximum switching frequency limits should be derived and respected in order not to overheat the devices during measurements. 

From the Table \ref{t:SpecsMeasBoard} it is known that the setup was designed to measure both hard- and soft-switching losses with half-bridge DC voltage range of [0..400]V. The desired switched current range for two parallel devices is: soft-switching - [0..40]A peak, hard-switching - [0..10]A peak. These translate to the following per-device ranges: [0..20]A for soft-switching and [0..5]A for hard-switching (assuming equal loss distribution between the deivces). Obviously, the switching frequency limits depend on the operating voltage/current, however as a general limits one can regard the ones obtained for maximum voltage/current ratings (400V, 40A peak soft-switched, 10A DC hard-switched), because at these voltage/current ratings both the switching and conduction losses are at their maximum, leaving no room for the switching frequency to increase if one does not want to overheat the devices.

In addition, from \cite{Heller} it is known that for hard-/soft-switching the triangular / DC current is injected into the switched node, respectively. Using this information and the specifications from Table \ref{t:SpecsMeasBoard}, the maximum switching frequencies can be derived.

The calculations for soft-switching are as follows:

\begin{equation}
\setlength{\jot}{10pt} 
\label{eq:SoS_fsmax}
\begin{split}
& P_{max,pd} = (T_{j,max} - T_{br,max})/(R_{jc,pd} + R_{chs,pd}) = 13.3 \text{ W ($P_{max,pd}$ - maximum dissipation per device)}
\\
& P_{cond} = (\hat{I}_{ss}/\sqrt{3})^2 R_{ds,on}/N_{par} = 26.7 \text{ W ($P_{cond}$ - cond. loss per half-bridge with 4 devices)} \Rightarrow
\\
& P_{cond,pd} = P_{cond}/N_{HB} = 6.67 \text{ W ($P_{cond,pd}$ - cond. loss per device)} \Rightarrow
\\
& P_{sw,pd} = P_{max,pd} - P_{cond,pd} = 6.67 \text{ W ($P_{sw,pd}$ - switching loss margin per device)} \Rightarrow
\\
& f_{sw,max} = P_{sw,pd}/\hat{E}_{ss} = 2 \text{ MHz ($f_{sw,max}$ - maximum switching frequency for soft-switching)}
\end{split}
\end{equation}

The calculations for hard-switching are as follows:

\begin{equation}
\setlength{\jot}{10pt} 
\label{eq:HaS_fsmax}
\begin{split}
& P_{max,pd} = (T_{j,max} - T_{br,max})/(R_{jc,pd} + R_{chs,pd}) = 13.3 \text{ W ($P_{max,pd}$ - maximum dissipation per device)}
\\
& P_{cond} = (\hat{I}_{hs})^2 R_{ds,on}/N_{par} = 5 \text{ W ($P_{cond}$ - cond. loss per half-bridge with 4 devices)} \Rightarrow
\\
& P_{cond,pd} = P_{cond}/N_{HB} = 1.25 \text{ W ($P_{cond,pd}$ - cond. loss per device)} \Rightarrow
\\
& P_{sw,pd} = P_{max,pd} - P_{cond,pd} = 12.1 \text{ W ($P_{sw,pd}$ - switching loss margin per device)} \Rightarrow
\\
& f_{sw,max} = P_{sw,pd}/\hat{E}_{hs} = 300 \text{ kHz ($f_{sw,max}$ - maximum switching frequency for hard-switching)}
\end{split}
\end{equation}

From these calculations it can be seen that for hard-/soft-switching the respective switching frequencies should not exceed 2MHz / 300kHz at maximum voltage/current ratings of the setup. Note that the calculations are only approximate, as they are based on numerous assumptions, such as the datasheet $R_{ds,on}$ value, symmetric loss distribution between the devices, $R_{chs,pd}$ value and $\hat{E}_{ss}$, $\hat{E}_{hs}$ from measurement results of older generation devices.

$\bullet$ Brass block dimensioning. Within the framework of calorimetric measurements, the heatsink (brass block) temperature over time can be approximated as $RC$ exponential curve that starts at initial heatsink temperature and asymptotically converges to the steady state temperature. To reduce the measurement times, \cite{Heller} proposes measuring initial transient, which can be approximated by a straight line (Fig. \ref{fig:CalorimetricSetup}b). In this first approximation, the temperature slope $dT/dt = P/C_{th}$, where $P$ signifies the power injection into heatsink and $C_{th}$ is its thermal capacitance. It was shown in \cite{Heller} that an acceptable measurement accuracy can be achieved if the measurements take at least 2 minutes (120 seconds) and within this time the heatsink temperature rises from 30\degree C to 40\degree C. If one uses the same framework as in \cite{Heller}, the heatsink has to be designed such that its thermal capacitance $C_{th}$ is large enough. With a large $C_{th}$ one can limit the slope $dT/dt = P/C_{th}$ such that even with maximum power injection $P$ into the heatsink, its temperature does not rise from 30\degree C to 40\degree C faster than in 120 seconds. On the other hand, one does not want the $C_{th}$ to be too large either, because in that case the linear transient assumption will not hold true for low-power measurements. Moreover, larger $C_{th}$ will result in a longer measurement times. From these considerations one can derive the following equation for the brass block thermal capacitance $C_{th}$ and volume $V_{br}$:

\begin{equation}
\setlength{\jot}{10pt} 
\label{eq:HeatsinkCapacitance}
\begin{split}
& \frac{N_{HB} P_{max,pd}}{C_{th}} = \frac{(T_{br,max} - T_{br,min})}{t_{min}} \Rightarrow 
\\
& C_{th} = \frac{N_{HB} P_{max,pd} t_{min}}{(T_{br,max} - T_{br,min})} \approx 640 \text{ J/K} \Rightarrow
\\
& V_{br} = \frac{C_{th}}{S\rho} \approx 200 \text{ cm$^3$ ($V_{br}$ - volume of a brass block)}
\end{split}
\end{equation}

The equation (\ref{eq:HeatsinkCapacitance}) estimates the brass block volume to be about 200cm$^3$. Remember from Fig. \ref{fig:MeasBoard} that the screws of the brass block have a predefined location. Since these screws are used to achieve a strong mechanical connection between the brass block and the measurement board, provided the screws are 5/4 centimeters apart, the length/width ($l_{br}/w_{br}$) of the brass block should not be less than 6/5 centimeters, respectively (including some margin for thread drilling). This leaves us with the following height ($h_{br}$) of the brass block: 

\begin{equation}
\label{eq:HeatsinkHeight}
V_{br} = l_{br} \times w_{br} \times h_{br} \Rightarrow h_{br} = \frac{V_{br}}{l_{br} \times w_{br}} = \frac{200\text{ cm$^3$}}{6\text{ cm} \times 5\text{ cm}} = 6.7 \text{ cm}
\end{equation}

Note that the brass block calculations above are approximate as they rely on several assumptions. For example, they depend on the maximum power per-device $P_{max,pd}(R_{chs,pd})$ (\ref{eq:SoS_fsmax}-\ref{eq:HaS_fsmax}), while it is known that the per-device case-to-heatsink thermal resistance $R_{chs,pd}$ was inherited from \cite{Heller}. Therefore, similar calculations have to be performed again once the $R_{chs,pd}$ measurement result is available.

Overall, with brass block dimensions determined, the calorimetric measurement setup is ready for usage. The measurement board has already been ordered. For measurement procedures and further details, please refer to \cite{Heller}.
\listoffigures
\listoftables
\bibliographystyle{IEEEtran}
\includepdf[pages=-]{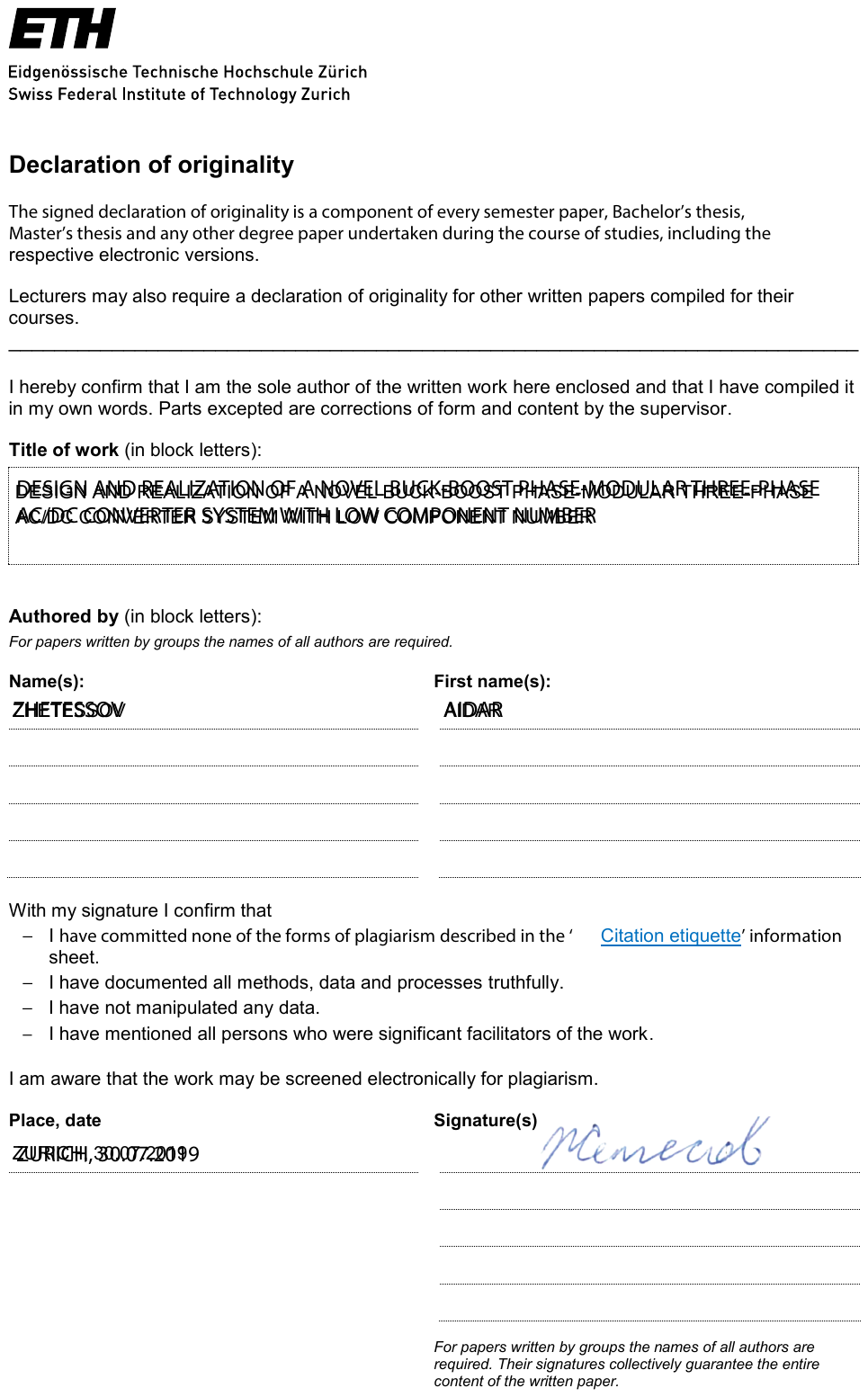}
\end{document}